\documentclass[aps,prl,twocolumn,amsmath,amssymb,superscriptaddress,longbibliography]{revtex4-2}
\usepackage[english]{babel}
\usepackage{xcolor}
\usepackage{amssymb} 
\usepackage{dcolumn}
\usepackage{bm}
\usepackage{graphicx}
\usepackage{amsmath}
\usepackage{braket}

\usepackage{graphicx}        

\graphicspath{{pict/}{}}

\usepackage[normalem]{ulem}

\usepackage{dcolumn}
\usepackage{bm}
\usepackage[pdfstartview=FitH, CJKbookmarks=true, bookmarksnumbered=true, bookmarksopen=true, colorlinks=true, pdfborder=001, citecolor=blue, linkcolor=blue, urlcolor=blue, linktocpage=true] {hyperref}

\newcommand{\ha}{\hat{a}}

\newcommand{\hp}{\hat{\psi}}

\newcommand{\pu}{\phi_{\uparrow}}
\newcommand{\pd}{\phi_{\downarrow}}

\newcommand{\bfr}{\mathbf{r}}
\newcommand{\bmrho}{\bm{\rho}}

\makeatletter

\begin{document}

\title{Dark Superradiance in Cavity-Coupled Polar Molecular Bose-Einstein Condensates}

\author{Yuqi Wang}
\affiliation{Guangdong Provincial Key Laboratory of Quantum Metrology and Sensing $\&$ School of Physics and Astronomy, Sun Yat-Sen University (Zhuhai Campus), Zhuhai 519082, China}
\affiliation{State Key Laboratory of Low Dimensional Quantum Physics, Department of Physics, Tsinghua University, Beijing 100084, China}

\author{Su Yi}
\email{yisu@nbu.edu.cn}
\affiliation{Institute of Fundamental Physics and Quantum Technology $\&$ school of physical science and technology, Ningbo University, Ningbo, 315211, China}
\affiliation{Peng Huanwu Collaborative Center for Research and Education, Beihang University, Beijing 100191, China}

\author{Yuangang Deng}
\email{dengyg3@mail.sysu.edu.cn}
\affiliation{Guangdong Provincial Key Laboratory of Quantum Metrology and Sensing $\&$ School of Physics and Astronomy, Sun Yat-Sen University (Zhuhai Campus), Zhuhai 519082, China}
	
\date{\today}

\begin{abstract}
We propose an experimental scheme to realize phase transition from {\it dark superradiance} to conventional superradiance in a microwave cavity coupled to polar molecules. The competition between cavity-mediated infinite-range repulsions and finite-range attractive dipolar interactions stabilizes a variety of exotic quantum phases, including vortex, vortex anti-vortex pairs, and superradiant phase, all emerging without external driving fields. In vortex phase associated with {\it dark superradiance}, cavity remains in vacuum state while profoundly reshaping the condensate's ground-state wave functions. In particular, the spin configuration locally parallel but globally anti-parallel is a direct consequence of competing for two nonlocal interactions. Beyond Dicke paradigm, dipolar dressing of condensate enables access to an unexplored regime of repulsion-dominated superradiance. A Bogoliubov analysis of low-energy excitation spectrum confirms that the condensate remains stable, avoiding roton-maxon induced collapse even in strongly dipolar regime. 
\end{abstract}
 
\maketitle

{\it Introduction}.---Recent breakthroughs in the production and manipulation of ultracold polar molecules ~\cite{doi:10.1126/science.aau7230, PhysRevLett.124.033401, Schindewolf,PhysRevLett.116.205303,PhysRevLett.114.205302,PhysRevLett.113.255301,PhysRevLett.113.205301} have paved the way for exploring strongly correlated quantum matter governed by long-range dipole-dipole interactions (DDI)~\cite{Chem.Rev.112.4949-5011,Cornish2024,Carr_2009}. Leveraging the rich long-lived rotational states, polar molecules underpin a range of fundamental quantum phenomena, ranging from ultracold controlled chemistry~\cite{,Park2023,PRXQuantum.5.020358},  to quantum computation~\cite{Sawant_2020,Picard2024}, and precise fundamental physics~\cite{PhysRevLett.119.153001,hudson2011improved, doi:10.1126/science.adg4084}. The implementation of collisional shielding techniques~\cite{PhysRevLett.121.163401, doi:10.1126/science.abg9502,Chen2023} has recently enabled the achievement of quantum degenerate and Bose-Einstein condensate (BEC) of polar molecules~\cite{bigagli2024observation} by mitigating two- and three-body losses. These advances pave the way for studying rich dipolar quantum phenomena such as quantum magnetism~\cite{yan2013observation,PhysRevLett.111.185305,Li2023,Christakis2023}, spin liquids~\cite{yao2018quantum}, and supersolids~\cite{PhysRevLett.98.260405,guo2019low,PhysRevLett.122.130405,PhysRevLett.128.195302,PhysRevX.9.021012}. Despite experimental observations of self-bound droplets and supersolids~\cite{guo2019low,PhysRevLett.122.130405,PhysRevLett.128.195302,PhysRevX.9.021012,PhysRevLett.116.215301,PhysRevX.6.041039,schmitt2016self}, most studies have focused on weakly dipolar regime, as strong DDI often induce instabilities associated with roton-maxon softening and phonon instability~\cite{PhysRevLett.90.250403}.

Meanwhile, ultracold atoms inside cavities have become a cornerstone for engineering strong light-matter interactions in controlled environments~\cite{RevModPhys.85.553,RevModPhys.95.035002,Ritsch_2021,RevModPhys.91.025005}. Cavity-mediated interactions (CMI), characterized by infinite range coupling, have facilitated the realization of dynamical spin-orbit coupling~\cite{PhysRevLett.112.143007,PhysRevLett.121.163601,PhysRevLett.123.160404}, self-organized crystalline orders~\cite{doi:10.1126/science.abd4385,Baumann2010,doi:10.1126/science.1220314}, and quantum simulators~\cite{,PhysRevResearch.5.013002,Leonard2017,leonard2017monitoring,PhysRevLett.132.093402}. However, cavity superradiance has been primarily explored in the regime of attractive CMI, facilitating Dicke phase transition via roton-mode softening under discrete ${\cal Z}_2$ symmetry breaking~\cite{doi:10.1126/science.abd4385,Baumann2010,doi:10.1126/science.1220314}. Notably, dynamical phase transitions induced by driven field and spin-exchange CMI have been observed~\cite{muniz2020exploring}, while ferromagnetic to paramagnetic phase transition displays symmetric behavior responsing of CMI sign. Despite these significant advances, the many-body ground state triggered by repulsive CMI remains unexplored, primarily due to the absence of a lower energy bound on blue sideband of cavity. So far, most efforts have treated CMI~\cite{RevModPhys.85.553,RevModPhys.95.035002,Ritsch_2021} and DDI~\cite{Chem.Rev.112.4949-5011,Cornish2024,Carr_2009} separately, with focused on competition with short-range collisional interactions~\cite{landig2016quantum}. The recent realization of polar molecule BEC~\cite{bigagli2024observation} offers a compelling opportunity to explore the interplay between finite-range DDI and infinite-range repulsive CMI. Understanding how these competing nonlocal interactions reshape fundamental quantum phenomena may unlock new regimes of strongly correlated quantum matter physics~\cite{Li2023,Christakis2023} and establish a versatile platform for studying nonequilibrium quantum dynamics~\cite{defenu2024out,young2024observing,PhysRevLett.134.083603}.


\begin{figure}[ptb]
	\includegraphics[width=0.89\columnwidth]{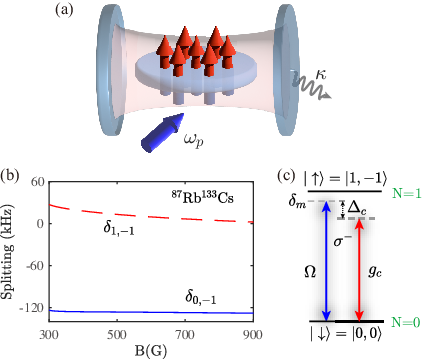} 
	\caption{(a) Scheme for creating {\it dark superradiance} in cavity-coupled polar molecules. (b) $B$ dependences of hyperfine splittings and (c) relevant energy level.}\label{scheme} 
\end{figure}

In this Letter, we address this gap by proposing a cavity-coupled pseudospin-$1/2$ model for pancake-shaped polar molecules, which enables tunable infinite-range CMI and finite-range DDI. We show that, depending on the interplay between attractive DDI and repulsive CMI, three distinct phases emerge: vortex (V), vortex anti-vortex pair (VP), and superradiance (SR) phases. Unlike the extensively studied Dicke-type superradiance driven by attractive CMI, we uncover a novel {\it dark superradiance} to superradiance phase transition in the regime of repulsive CMI. This {\it dark superradiant} phase is characterized by vanishing photon number and zero CMI energy, and it hosts emergent spin vortex with spontaneous broken chiral symmetry. Strikingly, the mechanism underlying vortex formation goes beyond conventional paradigms such as spin-orbit coupling that exchanges spin and orbital angular momenta in dipolar quantum gases~\cite{Deng2015,PhysRevLett.97.020401,PhysRevLett.100.170403} or artificial gauge field engineering in neutral ultracold atoms~\cite{lin2009synthetic}. To confirm the symmetry-breaking nature of these phases, we analyze the low-energy Bogoliubov excitation spectra and identify distinct collective modes associated with each phase. Notably, our proposal offers advantage that dipolar condensate, governed purely by spin-exchange DDI, remains stable without roton-maxon induced collapse. Additionally, the introduction of the novel concept of {\it dark superradiance} offers a new perspective on superradiance and provides unique platform for exploring emergent many-body phenomena in DDI-dressed condensates coupled to cavities.

{\it Model}.---We consider a BEC of $N$ polar molecules in ${}^{1}\Sigma(\nu=0)$ confined within a high-finesse microwave cavity [Fig.~\ref{scheme}(a)]. The internal state of each molecule is described in uncoupled basis $|M_1M_2NM_N\rangle$, where $M_N$, $M_1$ and $M_2$ are the projections of the rotational angular momentum ${\mathbf{N}}$, two nuclear spins ${\mathbf{I}}_1$ and ${\mathbf{I}}_2$ along the quantization axis. By applying a sufficiently strong magnetic field $\mathbf{B}$ along $z$-axis (quantization axis), the nuclear Zeeman effect dominates over the complex hyperfine interactions, rendering $M_1$ and $M_2$ good quantum numbers. By focusing on the lowest Zeeman sublevels of nuclear spins ($M_i=I_i$), the relevant Hilbert space consists of four states: $|N,M_N\rangle=|0,0\rangle, |1,0\rangle, |1,\pm1\rangle$, within the lowest two rotational manifolds ($N=0$ and $1$) due to the anharmonicity of rotational spectrum.

Figure.~\ref{scheme}(b) displays the hyperfine splittings $\delta_{0,-1}$ and $\delta_{1,-1}$ for $N=1$ manifold as a function of $\mathbf{B}$ for a ${}^{87}{\rm Rb}{}^{133}{\rm Cs}$ molecule. Remarkably, the typical hyperfine splitting (tens of kHz) significantly exceeds the DDI energy scale ($0.47$ kHz) at the density ($2\times 10^{12} {\rm cm}^{-3}$) for such ${}^{87}{\rm Rb}{}^{133}{\rm Cs}$ molecule, as observed in recent dipolar molecular BEC~\cite{bigagli2024observation}. This energetic separation enables the definition of a pseudospin-$1/2$ model with $|\uparrow\rangle=|1,-1\rangle$ and $|\downarrow\rangle=|0,0\rangle$, where $|\uparrow\rangle$ is well-isolated from $|1,0\rangle$ and $|1,1\rangle$ states. As illustrated in Fig.~\ref{scheme}(c), the transition $|\uparrow\rangle\leftrightarrow|\downarrow\rangle$ is coupled to a far-resonant $\sigma^{-}$-polarized microwave cavity and a resonant classical pump field, characterized by spatially uniform Rabi frequency $g_c$ and $\Omega$, respectively. Unlike optical cavity superradiance~\cite{doi:10.1126/science.abd4385,Baumann2010,doi:10.1126/science.1220314,Leonard2017,leonard2017monitoring,PhysRevResearch.5.013002}, the microwave wave length (GHz-scale frequency) is much larger than condensate size of molecules. 

In the large light-cavity detuning limit, $\left|\Delta_c\right| \gg\{g_c,\Omega\}$, many-body Hamiltonian for rotating polar molecules after adiabatically eliminating cavity field is given by~\cite{SM}
\begin{align}
\hat {\cal H}= & \sum_{\sigma\sigma'}\int d{\mathbf r}\hat\psi_{\sigma}^{\dag}({\mathbf r})[\hat{h}_{\sigma\sigma'}]\hat\psi_{\sigma'}({\mathbf r})  + \chi \int d\mathbf{r}d\mathbf{r^\prime} \hat{S}_+(\mathbf{r})\hat{S}_-(\mathbf{r^\prime}), \nonumber \\
&+  \frac{d^{2}}{4\pi\epsilon_{0}}\sqrt{\frac{4\pi}{45}}\int
\frac{d {\bf r}_{1} d {\bf r}_{2}}{|{\mathbf R}|^3}
Y_{20}(\hat{\mathbf R})\hat{S}_+(\mathbf{r})\hat{S}_-(\mathbf{r^\prime}), \nonumber \\ 
&+\sum_{\sigma \sigma'}\frac{2\pi \hbar^2 a_{\sigma\sigma'}}{m}\int d{\bf r}\hat{\psi}^\dag_{\sigma}({\bf r})\hat{\psi}^\dag_{\sigma'}({\bf r})\hat{\psi}_{\sigma'}({\bf r})\hat{\psi}_{\sigma}({\bf r}),  \label{Ham}
\end{align}
where $\hp_{\sigma=\uparrow,\downarrow}(\bfr)$ are the field operators for spin-$\sigma$ molecules, $\hat{S}_+(\mathbf{r})=\hat{S}_-^\dagger(\mathbf{r})=\hat{\psi}_\uparrow^\dagger(\mathbf{r})\hat{\psi}_\downarrow(\mathbf{r})$ is the spin operator, and $\hat{h}={{\mathbf p}^{2}}/{2m} + V({\mathbf r}) + \delta_m \sigma_z/2 + \Omega \sigma_x$ is the single-particle Hamiltonian with $\sigma_{x,z}$ being Pauli matrices, $\delta_m$ being the light-molecule detuning, $V_{b}$ being the trapping potential, and $m$ being the mass of polar molecules. 

The second term describes the spin-exchange CMI with $\chi={g_c^2{\Delta}_c}/({{\Delta}_c^2 +\kappa^2})$ and $\kappa=(2\pi)2\,{\rm kHz}$ being the cavity decay rate~\cite{PhysRevLett.107.240501}. Interestingly, $\chi$ is primarily governed by atom-cavity coupling $g_c$ and can be highly tuned via light-cavity detuning $\Delta_c$, while remaining independent of external pump field $\Omega$~\cite{norcia2018cavity}. The third term accounts for contact interactions, where $a_{\sigma\sigma'}$ denotes $s$-wave scattering lengths for intraspecies ($\sigma=\sigma'$) and interspecies ($\sigma\neq\sigma'$) molecular interactions. The final term represents the intrinsic DDI, where $d$ is permanent electric dipole moment, $\epsilon_{0}$ is electric permittivity of vacuum, $Y_{20}({\hat{\bf R}})$ is spherical harmonics, and  $\hat{\mathbf R}={\mathbf R}/|{\mathbf R}|$ is an unit vector. Importantly, the spin-exchange interaction for polar molecules is attractive in $xy$ plane and repulsive along the $z$-axis, in contrast to atomic dipolar counterpart, which also includes density-density interactions~\cite{PhysRevLett.108.125301}. Furthermore, DDI can be tuned to be repulsive in $xy$ plane by replacing $|\uparrow\rangle$ to $|1,0\rangle$ states. The sign of DDI plays a crucial role in realizing {\it dark superradiance} and vortex structure in molecular condensate. 

\begin{figure}[ptb]
\includegraphics[width=0.95\columnwidth]{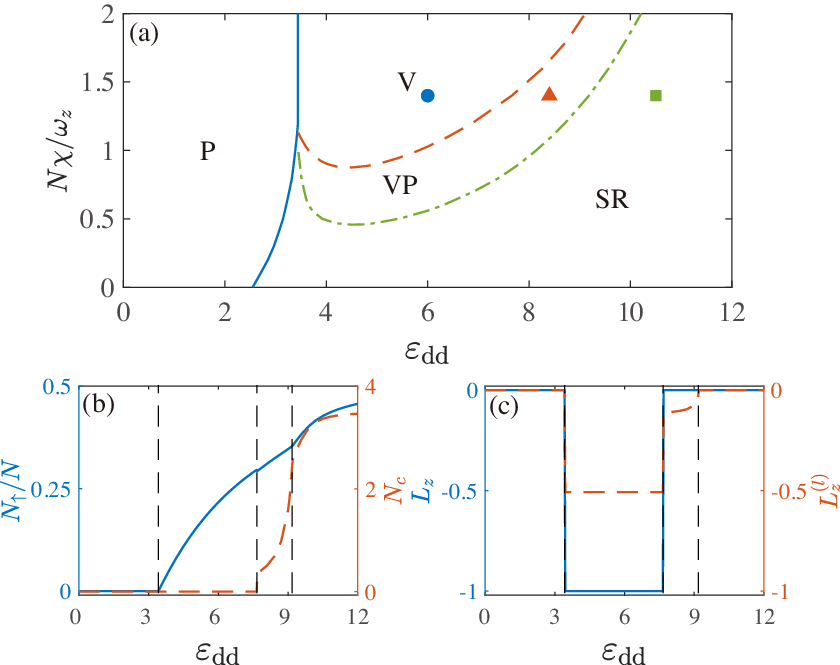} 
\caption{Phase diagram on $\varepsilon_{\rm dd}$-$\chi$ parameter plane for $\Omega=0$. The $\varepsilon_{\rm dd}$ dependence of (b) $N_{\uparrow}$ and $N_c$, (c) $L_z$ and $L_z^{(l)}$ for $N\chi/\omega_z=1.4$. Phase boundaries are indicated by vertical lines. The markers' in (a) correspond to $\varepsilon_{\rm dd}$=6, 8.4, and 10.5 with $N\chi/\omega_z=1.4$, respectively. The phases boundaries are indicated by vertical dotted lines in (b) and (c).}\label{phase} 
\end{figure}

{\it Ground state properties}.---We now study quantum phases of the condensate, arising from the interplay between spin-exchange DDI and CMI. We consider a pancake-shaped condensate with $N=10^4$ molecules, confined in a harmonic trap with frequencies $(\omega_{\rho},\omega_z)=2\pi\times(0.1,1)\,{\rm kHz}$, corresponding to a characteristic length $a_z=\sqrt{\hbar/(m\omega_z)}$. The ground-state wave functions are obtained by numerically solving the imaginary-time Gross-Pitaevskii equation with $\psi_\sigma \equiv \langle \hat{\psi}_\sigma \rangle$~\cite{SM}.  We take the parameters $\delta_m=(2\pi)1\,{\rm kHz}$ and $\Delta_c=(2\pi)1\,{\rm MHz}$, and $a_{\sigma\sigma'}=100~a_{\rm B}$ with $a_{\rm B}$ being the Bohr radius. By introduce the dipolar length $a_{dd}=md^2/(12\pi\hbar^2\epsilon_{0})$, the two free parameters reduce to $\varepsilon_{\rm dd}= a_{dd}/ a_{\sigma\sigma'}$ and $\chi$. Notably, unlike tuning short-range collisional interactions via Feshbach resonance-which often suffer from large two-body losses~\cite{PhysRevX.9.011051}, $\varepsilon_{\rm dd}$ can be highly tuned by applying microwave fields~\cite{Deng2015}. The steady-state intracavity photon number, self-consistently determined from the ground-state molecular wave functions, satisfies
\begin{align}
N_c=g_c^2   \int d\mathbf{r}{S}_+(\mathbf{r}) d\mathbf{r'}{S}_-(\mathbf{r'}) /({{\Delta}_c^2 +\kappa^2}), \label{cavity}
\end{align}
which can be treated as an order parameter for characterizing the superradiance of cavity.

Figure~\ref{phase}(a) presents the phase diagram in the $\varepsilon_{\rm dd}$-$\chi$ parameter plane. The region labeled P denotes polarized phase, while vortex (V) and vortex anti-vortex pairs (VP) represent distinct vortex states. A superradiant (SR) phase is identified by a nonzero $N_c$. Unlike Dicke superradiance, which required a critical driving strength~\cite{doi:10.1126/science.abd4385,Baumann2010,doi:10.1126/science.1220314}, these exotic many-body phases for cavity-coupled rotating polar molecules occur in an entirely different quantum regime, characterized by repulsive CMI ($\chi>0$) and the absence of driven field $\Omega=0$. 

\begin{figure}[ptb]
\includegraphics[width=0.95\columnwidth]{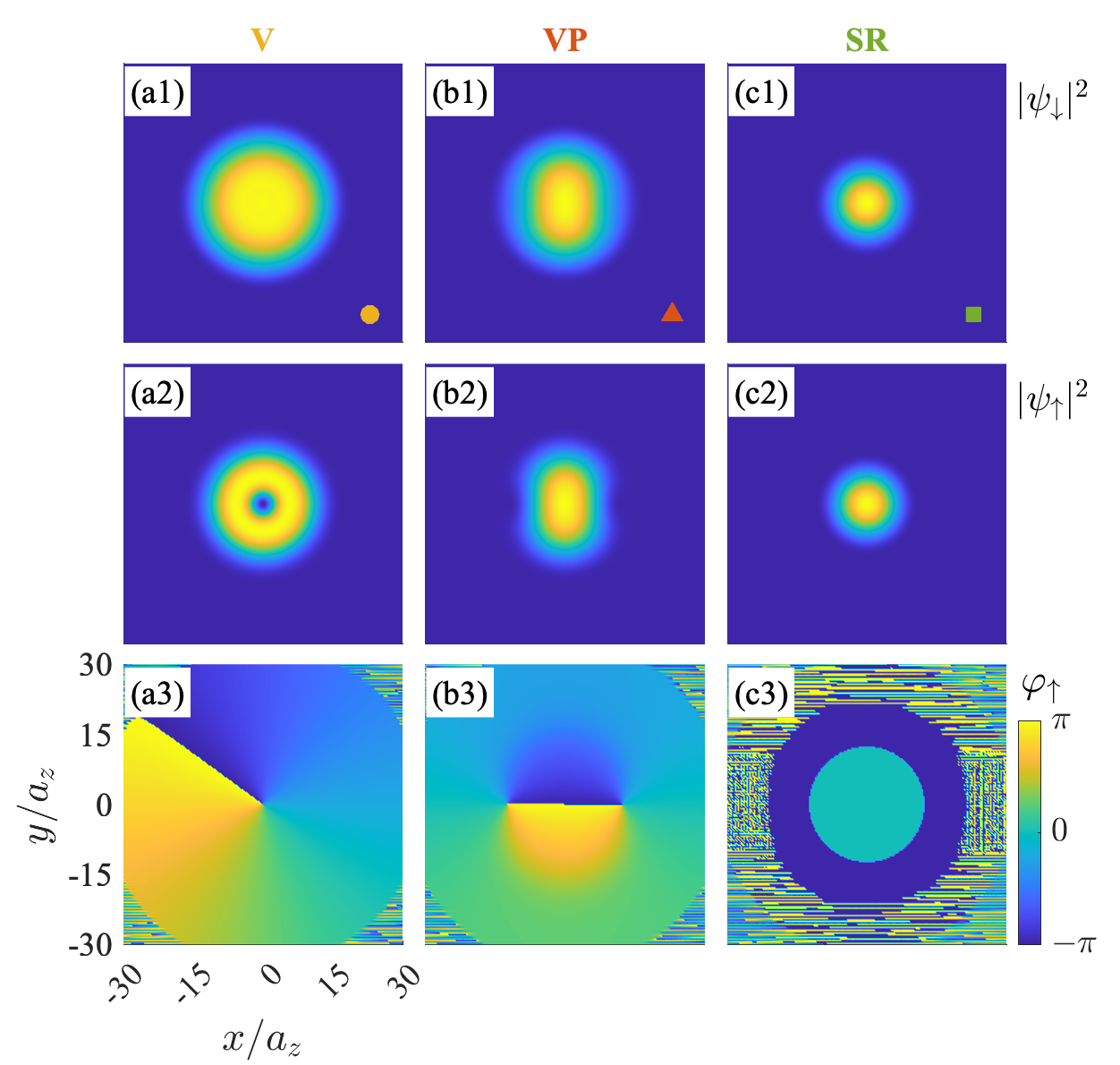} 
\caption{The condensate wave functions for V (columns 1) with $\varepsilon_{\rm dd}=6$, VP (columns 2) with $\varepsilon_{\rm dd}=8.4$, and SR  (columns 3) with $\varepsilon_{\rm dd}=10.5$, respectively. Row 1 denotes density profile ${|\psi}_\downarrow|^2$. Rows 2 and 3 denote density profile ${|\psi}_\uparrow|^2$ and relative phase $\varphi_{\uparrow}$. The other parameter is $N\chi/\omega_z=1.4$.}\label{denprof} 
\end{figure}

For conventional dipolar quantum gases with $\varepsilon_{\rm dd}>1$~\cite{guo2019low,PhysRevLett.122.130405,PhysRevLett.128.195302,PhysRevX.9.021012,PhysRevLett.116.215301,PhysRevX.6.041039,schmitt2016self}, self-bound droplet and supersolid phases appear in the mean-field collapse regime. In contrast to early approaches employing coherent-state ansatzes, where repulsive quantum fluctuations stabilize dipolar condensates~\cite{PhysRevA.84.041604}, recent investigations have explored stabilization of self-bound dipolar droplets using generalized Gaussian-state ansatzes or quantum Monte Carlo simulation~\cite{PhysRevResearch.2.043074,PhysRevLett.130.183001,jin2024bose}. Interestingly, in our model that DDI solely hosts spin-exchange interactions, the condensate remains stable within the mean-field framework even for $\varepsilon_{\rm dd}\gg1$. This guarantees the validity of quantum many-body ground states obtained via Gross-Pitaevskii equation. In fact, the contribution of  repulsive quantum fluctuations is negligible in this regime, exerting minimal influence on the ground state structure.  

Figure~\ref{phase}(b) shows $\varepsilon_{\rm dd}$ dependence of the molecule number $N_{\uparrow}= \int d\mathbf{r}|\langle \hat{\psi}_\uparrow(\mathbf{r})\rangle|^2$ and steady-state photon number $N_c$. Clearly, both $N_{\uparrow}$ and $N_c$ are zero in P phase. In V phase, $N_{\uparrow}$ becomes finite while $N_c$ remains zero. To distinguish V and VP phases, we calculate the average orbital angular momentum of $|\uparrow\rangle$ state over the entire trap and over the left half-plane $L_{z}=\int d{\bf r} \psi_{\uparrow}^*{\hat{L}_z}\psi_{\uparrow}/N_{\uparrow}$ and $L_{z}^{(l)}=\int_{x\leq0} d{\bf r} \psi_{\uparrow}^*{\hat{L}_z}\psi_{\uparrow}/N_{\uparrow}$, where ${\hat L}_z=-i\hbar(x\partial_y - y\partial_x)$ is the $z$ component of orbital angular momentum. Different to V phase with $L_z=-1$, the VP phase corresponds to $L_z=0$ and $L_z^{(l)} \neq 0$, as shown in Fig.~\ref{phase}(c). The fact that $|L_z^{(l)}|$ is much smaller than $1/2$ suggests that vortex-anti-vortex cores are displaced from the trap center. In SR phase, both $L_z$ and $L_z^{(l)}$ vanish while $N_c$ becomes finite. Notably, within SR phase, the populations of the two molecular components become comparable, causing $N_c$ to saturate as $\varepsilon_{\rm dd}$ increases.

\begin{figure}[ptb]
\includegraphics[width=0.95\columnwidth]{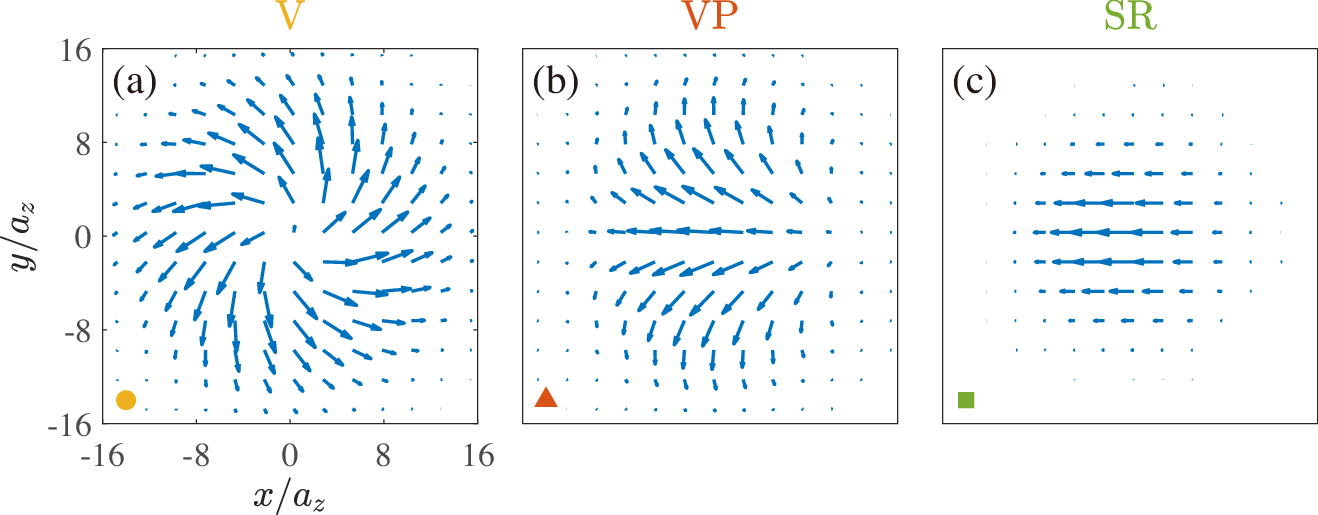} 
\caption{Spin distribution for typical points in (a) V, (b) VP and (c) SR phases with $N\chi/\omega_z=1.4$ and $\varepsilon_{\rm dd}$=6, 8.4, and 10.5, respectively.}\label{spindist} 
\end{figure}

For weak dipolar interactions, the molecules predominantly occupy $|\downarrow\rangle$ state due to the large positive light-molecule detuning $\delta_m$, resulting in $N_{\uparrow}=0$ and vacuum cavity state with $N_c=0$. Notably, the condensate remains in P phase for arbitrary $\chi$ ($>0$) when $\varepsilon_{\rm dd}<2.55$ in our simulations. This is consistent with experimental observation of ferromagnetic phase for cold atoms in cavity~\cite{muniz2020exploring}. Indeed, P phase is robust against variations in $\delta$ when $\varepsilon_{\rm dd}=0$, since energy contribution from repulsive CMI $E_{\chi}=\hbar N_c\Delta_c$ is always non-negative. Under the driven field $\Omega>0$, the rich self-organized density patterns observed for non-dipolar gases emerge only in attractive CMI regime with $\chi<0$ at red cavity-light detuning $\Delta_c <0$~\cite{Ritsch_2021}. As $\varepsilon_{\rm dd}$ increases, molecules gradually occupy spin-$\uparrow$ state to minimize the attractive DDI energy. Due to the competition between two long-range spin-exchange interactions, three distinct phases including unconventional {\it dark superradiance} are emerged. 

Figure~\ref{denprof} shows the typical density profiles $|{\psi}_\sigma|^2$ and corresponding phase distributions $\varphi_{\sigma}={\rm arg}({\psi}_\sigma)$ of condensate wave functions across different phases. In our simulations, spin-$\downarrow$ component remains highly populated due to positive $\delta_m$, leading to a uniform phase profile for $\varphi_{\downarrow}$.  Any phase structure developing in high-density spin state would induce a significant kinetic energy cost, rendering such configuration energetically unfavorable. Owing to interplay between finite-range attractive DDI and infinite-range CMI, a single-quantized vortex state carrying an orbital angular momentum of $\hbar$ forms in the less populated spin-$\uparrow$, accompanied by a structureless density profile for highly populated state. Notably, V phase corresponds to {\it dark superradiance} with $N_c=0$, where pseudospin density $S_+(\bmrho)$ exhibits odd parity and spontaneous chiral (${\cal Z}_2$) symmetry breaking. It is crucial to emphasize that the mechanism responsible for V phase is qualitatively distinct from previously studied spin vortex for dipolar condensates, which typically originate from spin-orbit coupling  facilitating spin and orbital angular momenta exchange~\cite{Deng2015,PhysRevLett.97.020401,PhysRevLett.100.170403}, artificial gauge field~\cite{lin2009synthetic}, or mechanical rotation in BEC~\cite{Ketterle2001,Cornell2003,Cornell2004}. 

\begin{figure}[ptb]
\includegraphics[width=0.9\columnwidth]{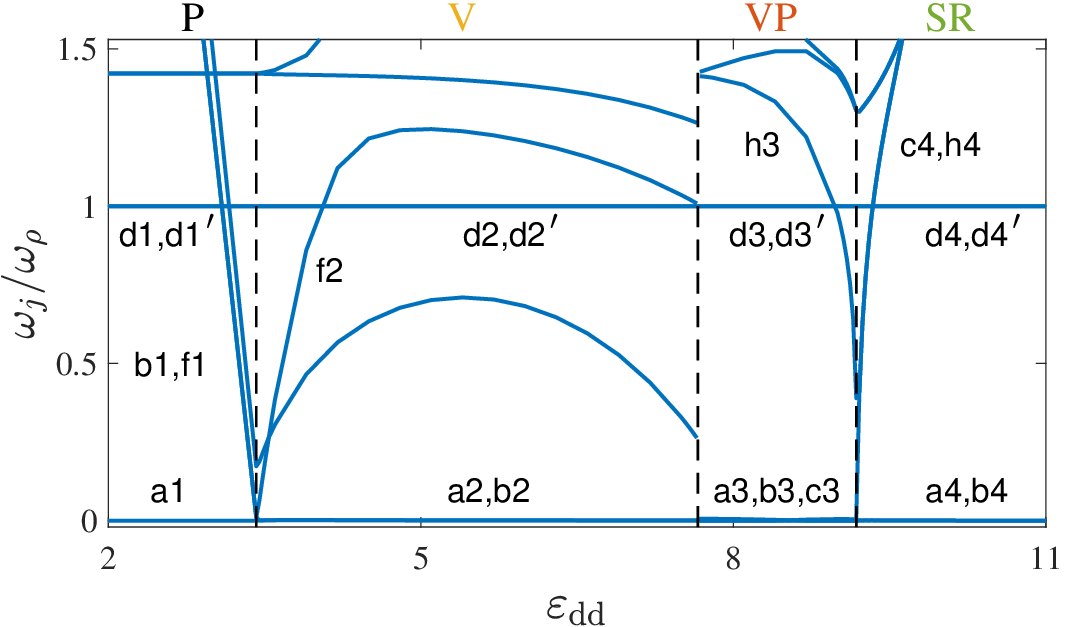} 
\caption{Low energy spectrum as $\varepsilon_{\rm dd}$ varies for $N\chi/\omega_z=1.4$. The vertical dashed lines indicate phase boundaries.}\label{bdg2} 
\end{figure}

This novel vortex generation mechanism arises from competition between repulsive CMI and attractive DDI. Although the spin-exchange CMI does not directly contribute to the ground-state energy, i.e., $E_{\chi}=0$, the V phase only emerges when $\chi$ exceeds a critical threshold. Distinguish from conventional superradiance, this phenomenon occurs in vacuum cavity and is spontaneously selected by the system through energy competition. This behavior is reminiscent of the concept of {\it dark state polaritons} in three-level electromagnetically induced transparency~\cite{PhysRevLett.84.5094}, where quantum superposition states form without photon absorption or emission, effectively decoupling from the excited state due to destructive interference.

More remarkably, we observe the V-VP transition upon further increasing $\varepsilon_{\rm dd}$. As shown in the second column of Fig.~\ref{denprof}, the quantized vortex splits into a vortex-anti-vortex pair, indicating spontaneous axial symmetry breaking. Different to V phase, characterized by $|L_z|=1$ and $|L_z^{(l)}|=0.5$, VP phase contains two vortex cores symmetrically located about origin, each carrying opposite orbital angular moment, resulting in a net total angular momentum $L_z=0$. Since these vortices are displaced from the high-density region of condensate, the local orbital angular momentum satisfies $|L_z^{(l)}|<0.2$, although phase profiles still exhibit a characteristic $2\pi$ winding. The boundary of this phase transition is clearly identified by behavior of photon number $N_c$, as shown in Fig.~\ref{phase}(b). Different to the second-order Dicke phase transition~\cite{Baumann2010}, the transition from vacuum cavity to superradiant phase is first order, marked by a discontinuous jump of $N_c$ from zero to a finite value. This transition is driven by an energetic trade-off: repulsive CMI energy increases with $N_c$, while attractive DDI energy decreases.

For sufficiently large ratio of $\varepsilon_{\rm dd}\omega_z/N\chi$, the populations of two molecular spin become approximately equal [Fig.~\ref{phase}(b)]. In this regime, both CMI energy and photon number $N_c$ saturate, indicating maximum overlap for spin-$1/2$ molecule. In competing to $E_{\chi}$, the attractive DDI dominates over repulsive CMI in determining the ground state structure. Consequently, a transition to SR phase occurs with manifesting finite photon number $N_c$, favoring a structureless density and trivial phase for condensate wavefunction. As shown in the third column of Fig~\ref{denprof}, the condensate in SR phase shrinks in the trap center, minimizing both attractive DDI and trap energy, while preserving rotational symmetry, thereby eliminating any anisotropic features. 

The emergence of vortices across different phases can be further understood from the perspective of spin textures. Figure~\ref{spindist} illustrates the planar spin distribution ${\bf s}({\bmrho})=(S_x, S_y)$ for each quantum phase. The spin-exchange interactions mediated by finite-range attraction of DDI and  infinite-range repulsion of CMI exhibit competing characteristics. This competition leads to a frustration effect, where local spins tend to align parallel (favored by DDI) but are driven to be anti-parallel globally (due to CMI). In V phase, where DDI and CMI are comparable, the planar spin distribution forms a spin vortex with $2\pi$ phase winding to minimize repulsive CMI energy. As for VP phase, two local spin vortices carry opposite orbital angular momenta, resulting in a net winding number of zero. This reflects the fine balance between two competing nonlocal interactions, which leads to spontaneous axial symmetry breaking. When $\varepsilon_{\rm dd}\omega_z/N\chi$ becomes, the transverse spin components ${\bf s}(\bmrho)$ tend to form a parallel spin configuration, as shown in Fig.~\ref{spindist}(c), consistent with the structureless SR phase.

{\it Low-energy excitation}.---To gain deeper insight of quantum phases, we study the collective excitations of condensate by solving the Bogoliubov-de Gennes (BdG) equations~\cite{SM}
\begin{align}\label{bdg}
\hbar\omega_j f_j(\bmrho) = \Sigma^z\int d\bmrho' \mathcal{H}_{\rm BdG}(\bmrho,\bmrho') f_j(\bmrho),
\end{align}
where $f_j\equiv\big(u_{\uparrow,j}, u_{\downarrow,j}, v_{\uparrow,j}, v_{\downarrow,j}\big)^T$ represents the Bogoliubov mode functions and $\Sigma^z={\rm diag}(1,1,-1,-1)$ is the Pauli-$z$ matrix in the Nambu space. By diagonalizing BdG Hamiltonian which preserves the inherent particle-hole symmetry, we obtain the Bogoliubov excitation spectrum $\omega_j$. 

Figure~\ref{bdg2} presents the low-energy excitation spectrum as function of $\varepsilon_{\rm dd}$ for $N\chi/\omega_z=1.4$. For brevity, the excitations in P, V, VP and SR phases are labeled by Arabic numbers 1, 2, 3 and 4, respectively. Across all phases, two degenerate dipole modes $d$ and $d'$, independent of $\varepsilon_{\rm dd}$, remain fixed at trap frequency $\omega_{\bmrho}$,  corresponding to center-of-mass motion in transverse plane. As expected in P phase, a single zero-energy Goldstone mode (a1) emerges due to spontaneous breaking of internal ${\cal U}(1)$ gauge symmetry of spin-$\downarrow$ superfluid. In contrast, V, VP, and SR phases exhibit macroscopical occupation of both spin components, thereby supporting two zero-energy Goldstone modes (a2, b2 in V, a3, b3 in VP, and a4, b4 in SR), reflecting breaking of two independent ${\cal U}(1)$ symmetries. Moreover, VP phase hosts a third gapless mode (c3) arising from broken axial symmetry continuously. Interestingly, in P phase, two degenerate excitation modes (b1 and f1) soften as $\varepsilon_{\rm dd}$ increases. As the system transitions to V phase, one of these evolves into Goldstone mode (b2), while the other becomes gapped Higgs mode (f2). A similar transition occurs across SR-VP phase boundary: modes c4 and h4 in SR evolve into Goldstone mode c3 and Higgs mode h3 in VP as $\varepsilon_{\rm dd}$ decreases. 

Remarkably, condensate dominated by spin-exchange DDI remains stable even for strong dipolar regime ($\varepsilon_{\rm dd}\gg1$). This stability arises from lacking density-density DDI responsible for phonon instability~\cite{Deng2015}. In contrast to magnetic dipolar gases, where roton-maxon softening can induce collapse when $|\varepsilon_{\rm dd}|>1$~\cite{PhysRevLett.90.250403}, the pure spin-exchange DDI ensures a stable ground state within mean-field framework. Our proposal enable access to an unexplored strong dipolar regime, providing a novel platform to explore exotic quantum states governed by competing nonlocal interactions~\cite{micheli2006toolbox,gregory2021robust,PhysRevLett.104.125301}.

{\it Conclusion}.---Leveraging recent advance of dipolar molecules, we propose an experimental scheme for cavity-coupled polar molecule governed by two distinct long-range interactions. This setup bridges cavity QED systems with polar molecules, opening access to a previously unexplored interaction regime. We uncover a novel {\it dark superradiance} vortex phase emerging from competition between repulsive CMI and attractive DDI. This phase, characterized by exact cavity vacuum and vanishing CMI energy, fundamentally differs from conventional vortex phases requiring synthetic or intrinsic spin-orbit couplings. Additionally, we demonstrate the first order transition from VP to SR phases triggered by strong DDI and accompanied by nonzero photon population. 

 A key experimental advantage of our scheme lies in the inherent leakage of cavity, which  enables precise, nondemolition measurements of quantum phase transitions, overcoming longstanding challenges in detecting polar molecules  due to their complex internal structure. The distinct phases are further identified via low-energy Bogoliubov spectrum, revealing spontaneous chiral or axial symmetry breaking. Importantly, the condensate remains stable against roton-maxon collapse even for strong DDI ($\varepsilon_{\rm dd}\gg1$), opening new handles on designing of versatile quantum simulators~\cite{doi:10.1126/science.adf8999, doi:10.1126/science.adf4272,You25} and precision metrology~\cite{luo2024momentum,PhysRevLett.132.093402}. Finally, our approach naturally extends to optical cavities coupled to polar molecules via Raman transitions between rotational levels, where self-organized quantum phases with spatially periodic crystalline orders~\cite{RevModPhys.85.553,RevModPhys.95.035002,Ritsch_2021,RevModPhys.91.025005} enriched by anisotropic DDI offer an exciting avenue for future exploration.


{\em Acknowledgments}.---This work was supported by the NSFC (Grants No. 12274473 and No. 12135018), by the National Key Research and Development Program of China (Grant No. 2021YFA0718304), by the Strategic Priority Research Program of CAS (Grant No. XDB28000000).

%

\newpage{}

\begin{widetext}
\begin{center}
	\textbf{\large Supplementary Materials: Dark Superradiance in Cavity-Coupled Polar Molecular Bose-Einstein Condensates}
\end{center}

\setcounter{equation}{0} \setcounter{figure}{0} \setcounter{table}{0} %
\renewcommand{\theequation}{S\arabic{equation}} \renewcommand{\thefigure}{S%
	\arabic{figure}} \renewcommand{\bibnumfmt}[1]{[S#1]}

\subsection{Cavity-molecule Hamiltonian with controlled hyperfine structures}\label{effham}
We consider ultracold polar molecules prepared in the $^{1}\Sigma(v=0)$ electronic ground state and subjected to a bias magnetic field ${\mathbf B}=B\hat{\mathbf z}$. Each molecule possesses three angular momentum degrees of freedom: rotational angular momentum ${\mathbf N}$ and two nuclear spins ${\mathbf I}_{1}$ and ${\mathbf I}_{2}$. The internal states are conveniently described in the uncoupled basis $|M_{1}M_{2}NM_{N}\rangle$, where $M_{N}$ and $M_{i}$ denote the projections of ${\mathbf N}$ and ${\mathbf I}_{i}$ onto the quantization axis, respectively. The internal Hamiltonian for a bialkali $^{1}\Sigma$ molecules reads~\cite{Huston2008}
\begin{align}
\hat{H}_{\rm in}= \hat{H}_{\rm rot} + \hat{H}_Z + \hat{H}_{\rm hf},\label{smhin},
\end{align}
comprising rotational, Zeeman, and hyperfine contributions. The dominant intrinsic energy scale is set by the rotational term, $\hat H_{\rm rot}=B_{v}{\mathbf N}^{2}$, where $B_{v}$ is the rotational constant, typically on the order GHz. The Zeeman term takes the form, 
\begin{align}
\hat{H}_{Z}&=- g_r \mu_N {\mathbf N}\cdot {\mathbf B} -
\sum_{i=1}^2 g_i \mu_N {\mathbf I}_i\cdot {\mathbf
B}(1-\sigma_i),
\end{align}
where $\mu_N$ is the nuclear magnetic moment, $g_r$ is the rotational $g$-factor of the molecule, $g_i$ denotes the nuclear $g$-factor for the $i$th nucleus, and $\sigma_{i}$ is the nuclear shielding parameter. 

\begin{table}[h]
\tabcolsep 1pt \caption{Molecular parameters for bialkali polar molecules.  Subscripts 1 and 2 refer to the less electronegative atom and to the more electronegative one~\cite{Huston2008,ran2010hyperfine,PhysRevA.96.042506}.} \vspace*{-12pt}
\begin{center}
\def\temptablewidth{0.78\textwidth}
{\rule{\temptablewidth}{1.5pt}}
\begin{tabular*}{\temptablewidth}{@{\extracolsep{\fill}}lccc}
Molecule & $^7$Li$^{133}$Cs &  $^{23}$Na$^{87}$Cs & $^{87}$Rb$^{133}$Cs\\
\hline
$I_{1}$             & $3/2$        & $3/2$    & $3/2$\\
$I_{2}$             & $7/2$       & $7/2$    & $7/2$\\
$g_{1}$             & $2.171$   & $1.478$  & $1.834$\\
$g_{2}$             & $0.738$   & $0.738$  & $0.738$\\
$(eqQ)_{1}$\;(kHz)  & $18.5$       & $-97$   & $-872$\\
$(eqQ)_{2}$\;(kHz)  & $188$      & $150$  & $51$ \\
$\sigma_{1}$\;(ppm) & $108.2$     & $639.2$   & $3531$ \\
$\sigma_{2}$\;(ppm) & $6242.5$    & $6278.7$   & $6367$  \\
$c_1$\;(Hz)         & $32$        & $14.2$   & $98.4$\\
$c_{2}$\;(Hz)       & $3014$     & $854.5$  & $194.1$\\
$c_3$\;(Hz)         & $140$      & $105.6$   & $192.4$\\
$c_4$\;(Hz)         & $1610$    & $3941.8$  & $17345.4$\\
$g_r$               & $0.0106$   & $-$ & $0.0062$\\
$d$\;(Debye)        & $5.52$     & $4.75$   & $1.25$
\end{tabular*}\label{tabl}
{\rule{\temptablewidth}{1pt}}
\end{center}
\end{table}

The nuclear hyperfine interaction comprise four distinct contributions: the nuclear electric quadrupole interaction $\hat H_{Q}$, the nuclear spin-rotation interaction $\hat H_{IN}$, the tensor nuclear spin-spin interactions $\hat H_{t}$, and the scalar $\hat H_{\rm sc}$ nuclear spin-spin interactions. The total hyperfine Hamiltonian is given by
\begin{align}
\hat{H}_{\rm hf}=&\;\hat H_{Q}+\hat H_{IN}+\hat H_{t}+\hat H_{\rm sc}\nonumber\\
=&\;\sum_{i=1}^{2}\frac{\sqrt{6}(eQ_{i}q_{i})}{4I_{i}(2I_{i}-1)}T^{(2)}({\mathbf C})\cdot T^{(2)}({\mathbf I}_{i},{\mathbf I}_{i})+\sum_{i=1}^2c_i{\mathbf N}\cdot {\mathbf I}_i -c_{3}\sqrt{6}\,T^{(2)}({\mathbf C})\cdot T^{(2)}({\mathbf I}_{1},{\mathbf I}_{2})+c_{4}{\mathbf I}_{1}\cdot{\mathbf I}_{2}, \label{smhhf}
\end{align}
where $T^{(2)}({\mathbf C})$ denotes the second order unnormalized spherical harmonic with components $T_{q}^{(2)}({\mathbf C})\equiv C_{q}^{(2)}(\theta,\varphi)=\sqrt{\frac{4\pi}{5}}Y_{2,q}(\theta,\varphi)$ with $(\theta,\varphi)$ being the spherical coordinate and $T^{(2)}({\mathbf I}_{i},{\mathbf I}_{j})$  is the rank-$2$ spherical tensor operator constructed from the vector operators ${\mathbf I}_{i}$ and ${\mathbf I}_{j}$. Here, $eQ_{i}$ is the electric quadrupole moment of nucleus $i$, $q_{i}$ is the corresponding the negative of the electric field gradient at nucleus $i$, $c_{i}$ represents the strength of the nuclear spin-rotation coupling for the $i$th nucleus, and $c_{3}$ and $c_{4}$ are, respectively, the strengths of the nuclear tensor and scalar spin-spin interaction. Explicitly, the matrix elements of each hyperfine interaction in the uncoupled basis can be expressed analytically as follows~\cite{,Deng2015}
\begin{align}
\langle M_{1}M_{2}NM_{N}|\hat H_{Q}|M_{1}'M_{2}'N'M_{N}'\rangle=&\sum_{i=1,2}\frac{(eQq)_{i}}{4}\delta_{M_{\bar i}M_{\bar i}'}\sum_{p}(-1)^{p-M_{N}+I_{i}-M_{i}}\sqrt{(2N+1)(2N'+1)}\nonumber\\
&\times\begin{pmatrix}N&2&N'\\-M_{N}&p&M_{N}'\end{pmatrix}\begin{pmatrix}I_{i}&2&I_{i}\\-M_{i}&-p&M_{i}'\end{pmatrix}\begin{pmatrix}N&2&N'\\0&0&0\end{pmatrix}\begin{pmatrix}I_{i}&2&I_{i}\\-I_{i}&0&I_{i}\end{pmatrix}^{-1},\label{smmathq}\\
\langle M_{1}M_{2}NM_{N}|\hat H_{IN}|M_{1}'M_{2}'N'M_{N}'\rangle=&\;\delta_{NN'}\sum_{q}(-1)^{q+N-M_{N}}\sqrt{N(N+1)(2N+1)}\begin{pmatrix}N&1&N\\-M_{N}&q&M_{N}'\end{pmatrix}\nonumber\\
&\times\sum_{i=1,2}c_{i}(-1)^{I_{i}-M_{i}}\delta_{M_{\bar i}M_{\bar i}'}\sqrt{I_{i}(I_{i}+1)(2I_{i}+1)}\begin{pmatrix}I_{i}&1&I_{i}\\-M_{i}&-q&M_{i}'\end{pmatrix},\label{smmathin}\\
\langle M_{1}M_{2}NM_{N}|\hat H_{t}|M_{1}'M_{2}'N'M_{N}'\rangle=&-c_{3}\sqrt{6}\sqrt{I_{1}(I_{1}+1)(2I_{1}+1)}\sqrt{I_{2}(I_{2}+1)(2I_{2}+1)}\sqrt{(2N+1)(2N'+1)}\nonumber\\
&\times\begin{pmatrix}N&2&N'\\0&0&0\end{pmatrix}\sum_{p}(-1)^{p-M_{N}+I_{1}-M_{1}+I_{2}-M_{2}}\begin{pmatrix}N&2&N'\\-M_{N}&p&M_{N}'\end{pmatrix}\nonumber\\
&\times\sum_{m}\langle 1,m;1,-p-m|2,-p\rangle\begin{pmatrix}I_{1}&1&I_{1}\\-M_{1}&m&M_{1}'\end{pmatrix}\begin{pmatrix}I_{2}&1&I_{2}\\-M_{2}&-p-m&M_{2}'\end{pmatrix},\label{smmatht}\\
\langle M_{1}M_{2}NM_{N}|\hat H_{\rm sc}|M_{1}'M_{2}'N'M_{N}'\rangle=&\;c_{4}\delta_{NN'}\delta_{M_{N}M_{N}'}\sqrt{I_{1}(I_{1}+1)(2I_{1}+1)}\sqrt{I_{2}(I_{2}+1)(2I_{2}+1)}\nonumber\\
&\times(-1)^{I_{1}-M_{1}+I_{2}-M_{2}}\sum_{p}(-1)^{p}\begin{pmatrix}I_{1}&1&I_{2}\\-M_{1}&p&M_{1}'\end{pmatrix}\begin{pmatrix}I_{2}&1&I_{2}\\-M_{2}&-p&M_{2}'\end{pmatrix},\label{smmathsc}
\end{align}
where $\bar i=3-i$ denotes the nuclear index complementary to $i$.

\begin{figure}[tbp]
\includegraphics[width=0.75\columnwidth]{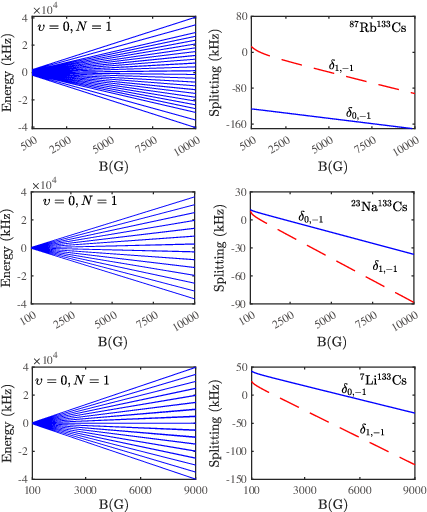}
\caption{(color online). Zeeman and hyperfine splittings of the lowest nuclear Zeeman levels with $M_i=I_i$ (columns 2) for polar molecules in the  $N=1$ rotational manifold. Column 1 shows the Zeeman splittings and column 2 shows the hyperfine splittings as functions of the external magnetic field for ${}^{87}{\rm Rb}{}^{133}{\rm Cs}$ (rows 1), ${}^{23}{\rm Na}{}^{87}{\rm Cs}$ (rows 2), and ${}^{7}{\rm Li}{}^{133}{\rm Cs}$ (rows 3), respectively.}\label{splitting}
\end{figure}

Due to the anharmonic nature of the rotational spectrum, given by $E_{\rm rot} = B_{v}N(N+1)$, we restrict our attention to the two lowest rotational states, $N=0$ and $1$, which are separated by an energy gap of $2B_{v}$. The first excited manifold contains three hyperfine sublevels corresponding to $|1,M_N\rangle$ with $M_N = 0, \pm1$. Although the nuclear hyperfine interaction $\hat{H}_{\rm hf}$ couples different internal states, it's effect can be effectively suppressed by a sufficiently strong external magnetic field via the Zeeman interaction $\hat H_{Z}$, which couples ${\mathbf B}$ to both the rotational angular momentum ${\mathbf N}$ and the nuclear spins ${\mathbf I}_{i}$. In this high-field regime, the nuclear Zeeman interaction dominates over the hyperfine interaction, thereby rendering $M_{1}$ and $M_{2}$ become good quantum numbers. 

Focusing on the lowest-energy nuclear Zeeman levels ($M_{i}=I_{i}$) in the $N=0$ and $1$ manifolds, the internal Hilbert space reduces to four relevant states: $|N,M_{N}\rangle=|0,0\rangle$, $|1,0\rangle$, and $|1,\pm1\rangle$. In this reduced basis, both the rotational Hamiltonian $\hat {H}_{\rm rot}$ and the Zeeman Hamiltonian $\hat{H}_Z$ remain diagonal in the uncoupled basis $\{|M_{1}M_{2}NM_{N}\rangle\}$. Moreover, it can be further verified that the hyperfine interaction is also diagonal in this reduced four-level Hilbert space, implying that each state is characterized by a well-defined rotational projection quantum number $M_{N}$.

Figure~\ref{splitting} shows the magnetic-field dependence of the Zeeman and hyperfine splittings of the lowest nuclear Zeeman levels with $M_i=I_i$ (columns 2) for polar molecules in the  $N=1$ rotational manifold. As can be seen, the typical hyperfine splittings $\delta_{0,-1}=E_{|1,0\rangle}-E_{|1,-1\rangle}$ and $\delta_{1,-1}=E_{|1,1\rangle}-E_{|1,-1\rangle}$, lie in the range of several tens of kHz for magnetic field in the range of $300$-$900\,$G for ${}^{87}{\rm Rb}{}^{133}{\rm Cs}$ molecule. These splittings ensure that the $|1,-1\rangle$ state is energetically well-isolated from the $|1,0\rangle$ and  $|1,1\rangle$ states.

We consider an ensemble of $N$ bosonic ultracold bialkali polar molecules prepared in the rovibrational ground state ${X}^{1}\Sigma(\nu=0)$, confined in a high-finesse microwave cavity. The internal state of each molecule is described by the uncoupled basis $|M_1M_2NM_N\rangle$, where $M_N$, $M_1$ and $M_2$ are the projections of the rotational angular momentum $\hat{\mathbf{N}}$, two nuclear spins $\hat{\mathbf{I}}_1$ and $\hat{\mathbf{I}}_2$ along the quantization axis. The molecules are illuminated by a transverse $\sigma^{-}$-polarized microwave field with frequency $\omega_p$ resonant with the $|0,0\rangle\leftrightarrow|1,-1\rangle$ transition. This drive induces Rabi frequency with strength $\Omega$ and light-molecule detuning $\delta_m=2B_{v}/\hbar-\omega_{p}$, typically on the order of $1\,$kHz. Assuming all molecules are initially prepared in the $|0,0\rangle$ state, the levels $|1,0\rangle$ and  $|1,1\rangle$ states becomes well-separated from $|1,-1\rangle$ under the conditions $|\delta_m|\ll|\delta_{0,-1}|$ and $|\delta_m|\ll|\delta_{1,-1}|$. The microwave cavity locked at frequency $\omega_c$,with a light-cavity detuning $\Delta_c=\omega_p-\omega_c$ from the probe field, supports a $\sigma^{-}$-polarized mode that resonantly drives the $|0,0\rangle\leftrightarrow|1,-1\rangle$ transition with coupling strength $g_c$ originating from the coherent Bragg scattering.  

By introducing a rotating frame defined by the unitary transformation, ${\cal U} = \exp[-i\omega_p (\hat{a}^\dag \hat{a} + \sum_{q=0,\pm1}\hat{\psi}^\dag_{1q}\hat{\psi}_{1q})t]$, the time-independent single-particle Hamiltonian describing the cavity-molecule interaction becomes
\begin{align} 
{\hat { h}} = - \hbar \Delta_c \hat{a}^\dag \hat{a} + \hbar \delta_m |1,-1\rangle\langle 1,-1| + \hbar\left(\Omega|1,-1\rangle\langle 0,0| + g_c\hat{a} |1,-1\rangle\langle 0,0| +{\rm H.c.}\right) , 
 \label{single}
\end{align}
where $\hat{a}$ is the annihilation operator of the cavity photon. Given the spatial homogeneity of the cavity and pump fields on the scale of the molecular condensate, the Rabi frequencies $\Omega$ and $g_c$ are treated as position-independent. For short-hand notation, 
 we shall denote $|1,-1\rangle$ and $|0,0\rangle$ as $|\uparrow\rangle$ and $|\downarrow\rangle$, respectively. In the second-quantized form,  we rewrite the Hamiltonian (\ref{single}) in terms of the annihilation operators $\hat \psi_{\sigma}$ for spin-$\sigma$ molecule as  
\begin{eqnarray}
\hat {\cal H}_{0}=- \hbar \Delta_c \hat{a}^\dag \hat{a} +\hbar \delta_m \int d{\mathbf r} \hat\psi^{\dag}_{\uparrow}({\mathbf r}) \hat\psi_{\uparrow}({\mathbf r}) + \hbar (\Omega+g_c \hat{a})  \int d{\mathbf r} \hat\psi^{\dag}_{\uparrow}({\mathbf r}) \hat\psi_{\downarrow}({\mathbf r})  + {\rm H.c.}.\label{hamh0}
\end{eqnarray}
The Heisenberg equations of motion in the rotating frame are then given by
\begin{align}
i\dot{\hat \psi}_{\uparrow} &=  \delta_m {\hat \psi}_{\uparrow} + (\Omega+g_c\hat{a})\hat\psi_{\downarrow}, \nonumber \\
i\dot{\hat \psi}_{\downarrow} &=  (\Omega+g_c\hat{a}^\dag)\hat\psi_{\uparrow}, \nonumber \\
i\dot{\hat a}&= (-\Delta_c-i\kappa) \hat a + g_c \int d{\mathbf r} \hat\psi^{\dag}_{\downarrow}({\mathbf r}) \hat\psi_{\uparrow}({\mathbf r}),
\end{align}
with $\kappa$ being the cavity decay rate. 

In order to gain some physical insight, it is appropriate to adiabatically eliminate the cavity field in the far dispersive regime with $|\Delta_c/\kappa|\gg 1$. The cavity field quickly reaching a steady state is much faster than the external motion. The steady-state equation of motion for the cavity field can be formally solved, yielding  
\begin{align} \label{steady}
\hat a= \frac{g{\hat{\Xi}}}{{\Delta}_c +
i\kappa},
\end{align}%
where ${\hat{\Xi}}$ is the introduced parameter defined as, ${\hat{\Xi}}  = \int d {\mathbf r} 
\hat{\psi}_\downarrow^\dag({\mathbf r})\hat{\psi}_\uparrow({\mathbf r})$. Thus the steady-state intracavity photon number is given by
\begin{align}
N_s= \langle \hat a^\dag\hat a \rangle=\frac{g_c^2 \langle\hat{\Xi}^\dag  \hat{\Xi}\rangle }{{\Delta}_c^2 +
\kappa^2}  \approx\frac{g_c^2 \langle\hat{\Xi}^\dag\rangle \langle  \hat{\Xi}\rangle }{{\Delta}_c^2 +
\kappa^2} 
\end{align}%
where the molecular fields are assuming as the coherent states.

To extract the effective cavity-mediated long-range interactions of atomic fields, we eliminate the cavity mode by substituting the steady-state solution Eq.~(\ref{steady}) into the Hamiltonian. This yields a long-range spin-exchange interaction of the form 
\begin{align}
\hat{\cal {H}}_{\rm {eff}} =\chi \int d {\bf r} d {\bf r'}
\hat{\psi}_\uparrow^\dag ({\mathbf r})\hat{\psi}_\downarrow^\dag({\mathbf r'}) \hat{\psi}_\uparrow ({\mathbf r'})\hat{\psi}_\downarrow({\mathbf r})=\chi \int d\mathbf{r}d\mathbf{r^\prime} \hat{S}_+(\mathbf{r})\hat{S}_-(\mathbf{r^\prime}) \label{many}
\end{align}
where $\chi=\frac{g_c^2{\Delta}_c}{{\Delta}_c^2 +
\kappa^2}$ ($V_I>0$) is the tunable strength of the cavity-meditated  two-body interaction and $S_+(\mathbf{r})=S_-^\dagger(\mathbf{r})=\hat{\psi}_\uparrow^\dagger(\mathbf{r})\hat{\psi}_\downarrow(\mathbf{r})$ is the spin operator.  We remark that the controllable cavity-mediated long-range spin-exchange interaction between spin-$\uparrow$ and spin-$\downarrow$ molecules conserves the number of the molecules in each spin state and dominates over the two-body contact interaction. 

For completeness, we include the $s$-wave contact interaction
\begin{eqnarray}
\hat {\cal H}_{\rm con}=\sum_{\sigma\sigma'}\frac{2\pi\hbar^{2} a_{\sigma\sigma'}}{m} \int
d{\mathbf r}\hat\psi_{\sigma}^{\dag}({\mathbf r})
\hat\psi_{\sigma'}^{\dag}({\mathbf r})\hat\psi_{\sigma'}({\mathbf
r}) \hat\psi_{\sigma}({\mathbf r}),\label{hcol}
\end{eqnarray} 
where $a_{\sigma\sigma'}$ are the $s$-wave scattering lengths between spin-$\sigma$ and -$\sigma'$ molecules. For simplicity, we take the typical values of $a_{\uparrow\uparrow}=a_{\downarrow\downarrow}=a_{\uparrow\downarrow}=100a_{B}$ with $a_{B}$ being the Bohr radius. The resulting contact interaction energy is on the order of tens of Hz. Importantly, we remark that the emergent spin structure is insensitive to the specific values of $a_{\sigma\sigma'}$ as $\hat{\cal H}_{\rm con}$ conserves spin populations.

\subsection{Derivation of dipolar interaction for polar molecules}\label{appc}

The electric DDI between two polar molecules with dipole operators $d\hat {\mathbf d}_{1}$ and $d\hat {\mathbf d}_{2}$, is given by
\begin{align}
V_{\rm dd}({\mathbf R}) &=\frac{g_{d}}{|{\mathbf R}|^{3}}\left[\hat{\mathbf d}_{1}\cdot \hat{\mathbf d}_{2}-3(\hat{\mathbf d}_{1}\cdot\hat{\mathbf R})\,(\hat{\mathbf d}_{2}\cdot\hat{\mathbf R})\right], \nonumber \\
&= -\frac{g_{d}}{|{\mathbf R}|^{3}}\sqrt{\frac{24\pi}{5}} 
\sum_{\mu=-2}^{\mu=2}(-1)^{\mu} 
Y_{2\mu}^*(\hat{\mathbf R})(\hat{\bf d}_1\otimes \hat{\bf  d}_2)^2_{m},\nonumber \\
&= -\frac{g_{d}}{|{\mathbf R}|^{3}}\sqrt{\frac{24\pi}{5}} 
\sum_{\mu=-2}^{\mu=2}(-1)^{\mu} 
Y_{2\mu}^*(\hat{\mathbf R})\Sigma_{2\mu},
\end{align}
where $g_{d}=d^{2}/(4\pi\epsilon_{0})$ is the DDI strength with $\epsilon_{0}$ being the electric permittivity of vacuum, ${\mathbf R}$ is the vector connecting the two molecules, and $\hat{\mathbf R}={\mathbf R}/|{\mathbf R}|$. $Y_{2\mu}({\hat{R}})$ is a spherical harmonics of rank-2 given by
\begin{align}
Y_{20}({\hat{R}}) &= \sqrt{5\over {16\pi}}\,(3\cos^2\theta-1), \nonumber \\
Y_{2\pm1}({\hat{R}}) &= \mp\sqrt{15\over {8\pi}}\,\cos\theta\sin\theta e^{\pm i\varphi}, \nonumber \\
Y_{2\pm2}({\hat{R}}) &= \frac{1}{2}\sqrt{15\over {8\pi}}\,\sin\theta^2e^{\pm 2 i\varphi}, \nonumber
\end{align}
with $(\theta,\varphi)$ being the polar and azimuthal angles. The tensor $(\hat{\bf d}_1\otimes \hat{\bf  d}_2)^2_{m}$ is a rank-2 spherical tensor formed from the two molecular hyperfine spin vector operators
$\Sigma_{2\mu}$ defined as
\begin{align}
\Sigma_{2,0} &=(\hat{\bf d}_1\otimes \hat{\bf  d}_2)^2_{0}= \frac{1}{\sqrt{6}}(\hat{d}_{1-}\hat{d}_{2+} + 2\hat{d}_{1z}\hat{d}_{2z}+\hat{d}_{1+}\hat{d}_{2-}), \nonumber \\
\Sigma_{2,\pm1} &=(\hat{\bf d}_1\otimes \hat{\bf  d}_2)^2_{\pm1}= \frac{1}{\sqrt{2}}(\hat{d}_{1z}\hat{d}_{2\pm} + \hat{d}_{1\pm}\hat{d}_{2z}), \nonumber \\
\Sigma_{2,\pm2} &=(\hat{\bf d}_1\otimes \hat{\bf  d}_2)^2_{\pm2}= \hat{d}_{1\pm}\hat{d}_{2\pm}, \nonumber \label{spherical}
\end{align}
where the spin ladder operators are defined by $ \hat{d}_{\pm}= \mp( \hat{d}_{x}\pm i \hat{d}_{y})/\sqrt{2}$. 

According to the Wigner-Echart theorem, the matrix elements of the dipole moment operator ${\mathbf d}$ in the rotational state basis $|NM_{N} are given by
\rangle$:
\begin{align}
\langle NM_N|\hat{d}_q|N'M_N'\rangle &= (-1)^{2N-M_N}d\sqrt{(2N+1)(2N'+1)}\begin{pmatrix}
N & 1 & N' \\
-M_{N}  & q  & M_{N}' \end{pmatrix}
\begin{pmatrix}
N & 1 & N' \\
0  & 0 & 0
\end{pmatrix},
\end{align}
In the Hilbert space $\{|0,0\rangle, |1,0\rangle, |1,\pm1\rangle\}$, $\Sigma_{2,\mu}$ can be written out explicitly in the second-quantized form as
\begin{eqnarray}
\hat{d}_{+} &=& -\frac{1}{\sqrt{3}}(\hat{\psi}^\dag_{00}\hat{\psi}_{1-1}-\hat{\psi}^\dag_{11}\hat{\psi}_{00}),\nonumber\\
\hat{d}_{-} &=&
-\frac{1}{\sqrt{3}}(\hat{\psi}^\dag_{00}\hat{\psi}_{11} -
\hat{\psi}^\dag_{1-1}\hat{\psi}_{00}),\nonumber\\
\hat{d}_{z}
&=&\frac{1}{\sqrt{3}}(\hat{\psi}^\dag_{00}\hat{\psi}_{10}+\hat{\psi}^\dag_{10}\hat{\psi}_{00}).
\end{eqnarray}

To reveal the dipolar processes explicitly, we rewrite the DDI Hamiltonian as
\begin{align}
\hat {\cal H}_{\rm dd}=&\; \frac{g_d}{2}\sqrt{\frac{16\pi}{45}}\int
\frac{d {\bf r}_{1} d {\bf r}_{2}}{|{\mathbf R}|^3}
\left\{Y_{20}(\hat{\mathbf R})\left[\hat{\psi}_{00}^\dag({\mathbf
r}_{1})\hat{\psi}_{11}^\dag({\mathbf r}_{2})\hat{\psi}_{00}({\mathbf
r}_{2})\hat{\psi}_{11}({\mathbf r}_{1}) +\hat{\psi}_{00}^\dag({\mathbf
r}_{1})\hat{\psi}_{1-1}^\dag({\mathbf r}_{2})\hat{\psi}_{00}({\mathbf
r}_{2})\hat{\psi}_{1-1}({\mathbf r}_{1})\right.\right.\nonumber\\
& \qquad\qquad\qquad\qquad\quad\;\left.- 2\hat{\psi}_{00}^\dag({\mathbf r}_{1})\hat{\psi}_{10}^\dag({\mathbf
r}_{2})\hat{\psi}_{00}({\mathbf r}_{2})\hat{\psi}_{1,0}({\mathbf
r}_{1})\right]-Y_{20}(\hat{\mathbf R})\left[\hat{\psi}_{00}^\dag({\mathbf
r}_{1})\hat{\psi}_{00}^\dag({\mathbf r}_{2})\hat{\psi}_{1-1}({\mathbf
r}_{2})\hat{\psi}_{11}({\mathbf r}_{1})\right.\nonumber \\
& \qquad\qquad\qquad\qquad\quad\;\left.\left.+ \hat{\psi}_{00}^\dag({\mathbf r}_{1})\hat{\psi}_{00}^\dag({\mathbf
r}_{2})\hat{\psi}_{10}({\mathbf r}_{2})\hat{\psi}_{10}({\mathbf r}_{1}) +
{\rm H.c.}\right] \right\}\nonumber\\
& -\frac{g_d}{2}\sqrt{\frac{16\pi}{15}}\int \frac{d {\bf r}_{1} d {\bf r}_{2}}{|{\mathbf R}|^3}
\left\{Y_{2-1}(\hat{\mathbf R})\left[\hat{\psi}_{00}^\dag({\mathbf
r}_{1})\hat{\psi}_{00}^\dag({\mathbf r}_{2})\hat{\psi}_{1-1}({\mathbf
r}_{2})\hat{\psi}_{10}({\mathbf r}_{1}) +\hat{\psi}_{00}^\dag({\mathbf
r}_{1})\hat{\psi}_{10}^\dag({\mathbf r}_{2})\hat{\psi}_{00}({\mathbf
r}_{2})\hat{\psi}_{1-1}({\mathbf r}_{1})\right.\right.\nonumber \\
&\qquad\qquad\qquad\qquad\qquad\;\left.\left.
-\hat{\psi}_{00}^\dag({\mathbf r}_{1})\hat{\psi}_{11}^\dag({\mathbf r}_{2})\hat{\psi}_{00}({\mathbf r}_{2})\hat{\psi}_{10}({\mathbf r}_{1}) - \hat{\psi}_{11}^\dag({\mathbf r}_{1})\hat{\psi}_{10}^\dag({\mathbf
r}_{2})\hat{\psi}_{00}({\mathbf r}_{2})\hat{\psi}_{00}({\mathbf r}_{1})\right] +
{\rm H.c.}\right\}\nonumber\\
& -\frac{g_d}{2}\sqrt{\frac{8\pi}{15}}\int\frac{d {\bf r}_{1} d {\bf r}_{2}}{|{\mathbf R}|^3}
\left\{Y_{2-2}(\hat{\mathbf R})\left[\hat{\psi}_{00}^\dag({\mathbf
r}_{1})\hat{\psi}_{00}^\dag({\mathbf r}_{2})\hat{\psi}_{1,-1}({\mathbf r}_{2})\hat{\psi}_{1,-1}({\mathbf r}_{1}) -2\hat{\psi}_{11}^\dag({\mathbf r}_{1})\hat{\psi}_{00}^\dag({\mathbf
r}_{2})\hat{\psi}_{1-1}({\mathbf r}_{2})\hat{\psi}_{00}({\mathbf r}_{1})\right.\right. \nonumber \\
&\qquad\qquad\qquad\qquad\quad\;\;\;
\left.\left.+\hat{\psi}_{11}^\dag({\mathbf r}_{1})\hat{\psi}_{11}^\dag({\mathbf r}_{2})\hat{\psi}_{00}({\mathbf r}_{2})\hat{\psi}_{00}({\mathbf r}_{1})\right] + {\rm H.c.}\right\},\label{smhdd0}
\end{align}
where we have arranged all terms according to the components of the spherical harmonics. From Eq. (\ref{smhdd0}), it is apparent that the DDI conserves the total (rotational + orbital) angular momentum. Under a microwave field, we perform a unitary transformation to the rotating frame, leading to the time-dependent DDI Hamiltonian:
\begin{align}
\hat {\cal H}_{\rm dd}\rightarrow &\;{\cal U}^{\dag}\hat{\cal H}_{\rm dd}{\cal U}\nonumber\\
=&\; \frac{g_d}{2}\sqrt{\frac{16\pi}{45}}\int
\frac{d {\bf r}_{1} d {\bf r}_{2}}{|{\mathbf R}|^3}
\left\{Y_{20}(\hat{\mathbf R})\left[\hat{\psi}_{00}^\dag({\mathbf
r}_{1})\hat{\psi}_{11}^\dag({\mathbf r}_{2})\hat{\psi}_{00}({\mathbf
r}_{2})\hat{\psi}_{11}({\mathbf r}_{1}) +\hat{\psi}_{00}^\dag({\mathbf
r}_{1})\hat{\psi}_{1-1}^\dag({\mathbf r}_{2})\hat{\psi}_{00}({\mathbf
r}_{2})\hat{\psi}_{1-1}({\mathbf r}_{1})\right.\right.\nonumber\\
& \qquad\qquad\qquad\qquad\quad\;\left.- 2\hat{\psi}_{00}^\dag({\mathbf r}_{1})\hat{\psi}_{1,0}^\dag({\mathbf
r}_{2})\hat{\psi}_{00}({\mathbf r}_{2})\hat{\psi}_{10}({\mathbf
r}_{1})\right]-Y_{20}(\hat{\mathbf R})\left[\hat{\psi}_{00}^\dag({\mathbf
r}_{1})\hat{\psi}_{00}^\dag({\mathbf r}_{2})\hat{\psi}_{1-1}({\mathbf
r}_{2})\hat{\psi}_{11}({\mathbf r}_{1})e^{-2i\omega_{p}t}\right.\nonumber \\
& \qquad\qquad\qquad\qquad\quad\;\left.\left.+ \hat{\psi}_{00}^\dag({\mathbf r}_{1})\hat{\psi}_{00}^\dag({\mathbf
r}_{2})\hat{\psi}_{10}({\mathbf r}_{2})\hat{\psi}_{10}({\mathbf r}_{1})e^{-2i\omega_{p}t} +
{\rm H.c.}\right] \right\}\nonumber\\
& -\frac{g_d}{2}\sqrt{\frac{16\pi}{15}}\int \frac{d {\bf r}_{1} d {\bf r}_{2}}{|{\mathbf R}|^3}
\left\{Y_{2-1}(\hat{\mathbf R})\left[\hat{\psi}_{00}^\dag({\mathbf
r}_{1})\hat{\psi}_{00}^\dag({\mathbf r}_{2})\hat{\psi}_{1-1}({\mathbf
r}_{2})\hat{\psi}_{10}({\mathbf r}_{1})e^{-2i\omega_{\rm mw}t} +\hat{\psi}_{00}^\dag({\mathbf
r}_{1})\hat{\psi}_{10}^\dag({\mathbf r}_{2})\hat{\psi}_{00}({\mathbf
r}_{2})\hat{\psi}_{1-1}({\mathbf r}_{1})\right.\right.\nonumber \\
&\qquad\qquad\qquad\qquad\qquad\;\left.\left.
-\hat{\psi}_{00}^\dag({\mathbf r}_{1})\hat{\psi}_{11}^\dag({\mathbf r}_{2})\hat{\psi}_{00}({\mathbf r}_{2})\hat{\psi}_{10}({\mathbf r}_{1}) - \hat{\psi}_{11}^\dag({\mathbf r}_{1})\hat{\psi}_{10}^\dag({\mathbf
r}_{2})\hat{\psi}_{00}({\mathbf r}_{2})\hat{\psi}_{00}({\mathbf r}_{1})e^{2i\omega_{p}t}\right] +
{\rm H.c.}\right\}\nonumber\\
& -\frac{g_d}{2}\sqrt{\frac{8\pi}{15}}\int\frac{d {\bf r}_{1} d {\bf r}_{2}}{|{\mathbf R}|^3}
\left\{Y_{2-2}(\hat{\mathbf R})\left[\hat{\psi}_{00}^\dag({\mathbf
r}_{1})\hat{\psi}_{00}^\dag({\mathbf r}_{2})\hat{\psi}_{1-1}({\mathbf r}_{2})\hat{\psi}_{1-1}({\mathbf r}_{1})e^{-2i\omega_{p}t} -2\hat{\psi}_{11}^\dag({\mathbf r}_{1})\hat{\psi}_{00}^\dag({\mathbf
r}_{2})\hat{\psi}_{1-1}({\mathbf r}_{2})\hat{\psi}_{00}({\mathbf r}_{1})\right.\right. \nonumber \\
&\qquad\qquad\qquad\qquad\quad\;\;\;
\left.\left.+\hat{\psi}_{11}^\dag({\mathbf r}_{1})\hat{\psi}_{11}^\dag({\mathbf r}_{2})\hat{\psi}_{00}({\mathbf r}_{2})\hat{\psi}_{00}({\mathbf r}_{1})e^{2i\omega_{p}t}\right] + {\rm H.c.}\right\},\label{smhdd1}
\end{align}
Under the rotating-wave approximation, the time-dependent terms with higher frequencies (of order of GHz) can be safely neglected since the typical energy scale for DDI interaction is around $0.47\,$kHz in ${}^{87}{\rm Rb}{}^{133}{\rm Cs}$ molecule at the typical experimental density ($2\times 10^{12} {\rm cm}^{-3}$) for dipolar molecular BEC~\cite{bigagli2024observation}. Thus, the effective DDI that is time averaged over a period of $2\pi/\omega_{p}$ is
\begin{align} 
\hat {\cal H}_{\rm dd}\simeq&\; \frac{g_d}{2}\sqrt{\frac{16\pi}{45}}\int
\frac{d {\bf r}_{1} d {\bf r}_{2}}{|{\mathbf R}|^3}
\left\{Y_{20}(\hat{\mathbf R})\left[\hat{\psi}_{00}^\dag({\mathbf
r}_{1})\hat{\psi}_{11}^\dag({\mathbf r}_{2})\hat{\psi}_{00}({\mathbf
r}_{2})\hat{\psi}_{11}({\mathbf r}_{1}) +\hat{\psi}_{00}^\dag({\mathbf
r}_{1})\hat{\psi}_{1-1}^\dag({\mathbf r}_{2})\hat{\psi}_{00}({\mathbf
r}_{2})\hat{\psi}_{1-1}({\mathbf r}_{1})\right.\right.\nonumber\\
& \qquad\qquad\qquad\qquad\quad\;\left.\left.- 2\hat{\psi}_{00}^\dag({\mathbf r}_{1})\hat{\psi}_{1,0}^\dag({\mathbf
r}_{2})\hat{\psi}_{00}({\mathbf r}_{2})\hat{\psi}_{10}({\mathbf
r}_{1})\right] \right\}\nonumber\\
& -\frac{g_d}{2}\sqrt{\frac{16\pi}{15}}\int \frac{d {\bf r}_{1} d {\bf r}_{2}}{|{\mathbf R}|^3}
\left\{Y_{2-1}(\hat{\mathbf R})\left[\hat{\psi}_{00}^\dag({\mathbf
r}_{1})\hat{\psi}_{10}^\dag({\mathbf r}_{2})\hat{\psi}_{00}({\mathbf
r}_{2})\hat{\psi}_{1-1}({\mathbf r}_{1})
-\hat{\psi}_{00}^\dag({\mathbf r}_{1})\hat{\psi}_{11}^\dag({\mathbf r}_{2})\hat{\psi}_{00}({\mathbf r}_{2})\hat{\psi}_{10}({\mathbf r}_{1})\right] +
{\rm H.c.}\right\}\nonumber\\
& -\frac{g_d}{2}\sqrt{\frac{8\pi}{15}}\int\frac{d {\bf r}_{1} d {\bf r}_{2}}{|{\mathbf R}|^3}
\left[-2Y_{2-2}(\hat{\mathbf R})\hat{\psi}_{11}^\dag({\mathbf r}_{1})\hat{\psi}_{00}^\dag({\mathbf
r}_{2})\hat{\psi}_{1-1}({\mathbf r}_{2})\hat{\psi}_{00}({\mathbf r}_{1})+ {\rm H.c.}\right],\label{smhdd2}
\end{align}
For large enough hyperfine splitting, with the assumption that level $|1,-1\rangle$ becomes well-separated from $|1,0\rangle$ and  $|1,1\rangle$ states during the time scale considered here. As a result, we may simply drop all terms containing $\hat\psi_{10}$ and $\hat\psi_{11}$ in $\hat{\cal H}_{\rm dd}$, which eventually leads to the effective DDI Hamiltonian, 
\begin{align} 
\hat {\cal H}_{\rm dd}\simeq&\; \frac{g_d}{2}\sqrt{\frac{16\pi}{45}}\int
\frac{d {\bf r}_{1} d {\bf r}_{2}}{|{\mathbf R}|^3}
Y_{20}(\hat{\mathbf R})\hat{\psi}_{00}^\dag({\mathbf
r}_{1})\hat{\psi}_{1-1}^\dag({\mathbf r}_{2})\hat{\psi}_{00}({\mathbf
r}_{2})\hat{\psi}_{1-1}({\mathbf r}_{1}), \nonumber\\
&=\; \frac{g_d}{2}\sqrt{\frac{16\pi}{45}}\int
\frac{d {\bf r}_{1} d {\bf r}_{2}}{|{\mathbf R}|^3}
Y_{20}(\hat{\mathbf R})\hat{\psi}_{\downarrow}^\dag({\mathbf
r}_{1})\hat{\psi}_{\uparrow}^\dag({\mathbf r}_{2})\hat{\psi}_{\downarrow}({\mathbf
r}_{2})\hat{\psi}_{\uparrow}({\mathbf r}_{1}).
\label{smhdd3}
\end{align}
Clearly, $\hat {\cal H}_{\rm dd}$ represents the dipolar spin-exchange interaction between spin-$\uparrow$ and -$\downarrow$ molecules. 

\subsection{Gross-Pitaevskii equations for spin-half polar molecules}

After integrating out cavity field in the far-dispersive regime, the many-body Hamiltonian for spin-$1/2$ rotating polar molecules takes the form
\begin{align}
\hat {\cal H}= & \sum_{\sigma\sigma'}\int d{\mathbf r}\hat\psi_{\sigma}^{\dag}({\mathbf r})[\hat{h}_{\sigma\sigma'}]\hat\psi_{\sigma'}({\mathbf r})  +\frac{1}{2}\sum_{\sigma \sigma'}g_{\sigma\sigma'}\int d{\bf r}\hat{\psi}^\dag_{\sigma}({\bf r})\hat{\psi}^\dag_{\sigma'}({\bf r})\hat{\psi}_{\sigma'}({\bf r})\hat{\psi}_{\sigma}({\bf r}) \nonumber \\
& +
\chi \int d {\bf r} d {\bf r'}
\hat{\psi}_\uparrow^\dag ({\mathbf r})\hat{\psi}^\dag_\downarrow({\mathbf r'}) \hat{\psi}_\uparrow ({\mathbf r'})\hat{\psi}_\downarrow({\mathbf r}) +  \frac{g_d}{2}\sqrt{\frac{16\pi}{45}}\int
\frac{d {\bf r}_{1} d {\bf r}_{2}}{|{\mathbf R}|^3}
Y_{20}(\hat{\mathbf R})\hat{\psi}_{\downarrow}^\dag({\mathbf
r}_{1})\hat{\psi}_{\uparrow}^\dag({\mathbf r}_{2})\hat{\psi}_{\downarrow}({\mathbf
r}_{2})\hat{\psi}_{\uparrow}({\mathbf r}_{1}),  \label{SM_Ham}
\end{align}
where $g_{{\sigma\sigma}}={4\pi\hbar^{2}a_{{\sigma\sigma}}}/{m}$ denotes the short-range interaction strength and $g_{d}=d^{2}/(4\pi\epsilon_{0})$ characterizes the DDI strength. Notably, the Hamiltonian $\hat {\cal H}$ incorporates both the dipolar interaction and the cavity-mediated spin-exchange interaction. The competition between these two long-range interactions plays a key role in determining the ground-state properties of the system. 

For simplicity, we consider polar molecules confined in the quasi-two-dimensional (quasi-2D) harmonic potential $V(\bfr)=m(\omega_x^2x^2+\omega_y^2y^2+\omega_z^2z^2)/2$ with the trapping frequencies $(\omega_x, \omega_y, \omega_z) = (\omega_{\rho}, \omega_{\rho}, \omega_z)$. In the limit $\omega_z/\omega_{\rho} \gg 1$, the molecular motion along the $z$-axis is effectively frozen to the lowest harmonic oscillator level. For the collisional interaction, we assume equal $s$-wave scattering lengths such that $g_{\uparrow\uparrow}=g_{\downarrow\downarrow}=g_{\uparrow\downarrow}$. Under quasi-2D geometry, the field operators can be decomposed as 
\begin{align}
\hp_{\sigma}(\bfr)=\hat{\phi}_{\sigma}(\bmrho)\exp[-z^2/(2a_z^2)]/(\pi a_z^2)^{1/4}, \label{SM_2D}
\end{align}
where $\bmrho\equiv(x,y)$ and $a_z=\sqrt{\hbar/(m\omega_z)}$ is the harmonic oscillator length along the $z$-direction. After integrating out the $z$ variable and substitute Eq.~\eqref{SM_2D} into the 3D Hamiltonian in Eq.~\eqref{SM_Ham}, we obtain the effective 2D Hamiltonian for spin-$1/2$ molecular condensate
\begin{align}\label{Hami}
\hat {\cal H}_1 &= \sum_{\sigma\sigma'}\int d{\mathbf \rho}\hat\phi_{\sigma}^{\dag}({\mathbf \rho})[\hat{h}_{\sigma\sigma'}]\hat\phi_{\sigma'}({\mathbf r}) + \frac{1}{2}\sum_{\sigma \sigma'} \int d\bmrho \frac{g_{\sigma \sigma'}}{\sqrt{2\pi}a_z}\hat{\phi}_{\sigma}^{\dag}(\bmrho)\hat{\phi}_{\sigma'}^{\dag}(\bmrho)\hat{\phi}_{\sigma'}(\bmrho)\hat{\phi}_{\sigma}(\bmrho),
 \notag\\
&\quad + \chi\int d\bmrho d\bmrho' \pd^{\dag}(\bmrho)\pu^{\dag}(\bmrho')\pd(\bmrho')\pu(\bmrho) - \frac{1}{6}\int d\bmrho d\bmrho' \pd^{\dag}(\bmrho)\pu^{\dag}(\bmrho') U_{\rm dd}(\bmrho-\bmrho') \pd(\bmrho')\pu(\bmrho),
\end{align}
where the effective quasi-2D dipolar interaction is given by $U_{\rm dd}(\bmrho)=\mathcal{F}^{-1}[\widetilde{U}_{\rm dd}] z$ with $\mathcal{F}^{-1}[\cdot]$ denoting the inverse Fourier transform. In the 2D momentum space, the effective quasi-2D dipolar interaction reads 
\begin{align}
\widetilde{U}_{\rm dd}(\mathbf{k}_{\rho})=\frac{4\pi g_{\rm d}}{3\sqrt{2\pi}a_z}D\left( \frac{|\mathbf{k}_{\rho}|a_z}{\sqrt{2}} \right),
\end{align}
where $\mathbf{k}_{\rho}=(k_x,k_y)$~\cite{Fischer2006} and $D(x)=2-3\sqrt{\pi}xe^{x^2}{\rm erfc}(x)$ with ${\rm erfc}(x)$ being the complementary error function. 

To explore the ground-state phases of the molecular condensate, we solve the cavity-coupled Gross-Pitaevskii equations self-consistently with the steady-state solution for the cavity photon field. In the mean-field approach, the molecular field operators $\psi_\sigma$ are replaced by the condensate wave function $\psi_\sigma(\bmrho) \equiv \langle \hat{\psi}_\sigma(\bmrho) \rangle$. The steady-state cavity photon number is then self-consistently determined by the ground-state molecular wavefunctions as
\begin{align}
N_c =\langle \ha^\dag \ha \rangle =\frac{g_c^2} {{\Delta}_c^2 +\kappa^2} \int d\bmrho \int d\bmrho'S_{+}(\bmrho) S_{-}(\bmrho'),
\end{align}
where $\hat{a}$ is the annihilation operator for the microwave cavity field. Notably, the photon number relates to the cavity-induced interaction energy via $N_c=E_{\chi}/\hbar\Delta_c$, corresponding to the cavity-induced interactions $E_{\chi} = \chi \left|\int d\bmrho S_{+}(\bmrho) \right|^2$. 

To obtain the ground state of the condensate wave function, we numerically minimize the free energy functional by evolving the system in imaginary time. The imaginary-time Gross-Pitaevskii equations for the spin-$1/2$ molecular condensate take the form
\begin{subequations}
\begin{align}
-\frac{\partial}{\partial\tau}\pu(\bmrho) 
&= \Big[ h_{\uparrow\uparrow}(\bmrho) + g_{s} n(\bmrho) \Big]\pu(\bmrho) + \Omega \pd(\bmrho)+ \int d\bmrho' \Big[ \chi - \frac{1}{6}U_{\rm dd}(\bmrho-\bmrho') \Big]\pd^*(\bmrho')\pu(\bmrho')\pd(\bmrho), \\
-\frac{\partial}{\partial\tau}\pd(\bmrho)
&= \Big[  h_{\downarrow\downarrow}(\bmrho) +g_{s} n(\bmrho) \Big]\pd(\bmrho) +  \Omega\pu(\bmrho) + \int d\bmrho' \Big[ \chi - \frac{1}{6}U_{\rm dd}(\bmrho-\bmrho') \Big]\pu^*(\bmrho')\pd(\bmrho')\pu(\bmrho),
\end{align}
\end{subequations}
where $\tau$ is the imaginary time and $n(\bmrho)$ is the total 2D condensate density. The single-particle Hamiltonian are defined as: $h_{\uparrow\uparrow}(\bmrho)={{\mathbf p}^{2}}/{2m} + V(\bmrho) + \delta_m/2$ and $h_{\downarrow\downarrow}(\bmrho)={{\mathbf p}^{2}}/{2m} + V(\bmrho) - \delta_m/2$. The effective 2D contact interaction strength is given by $g_s=g_{\uparrow\uparrow}/{\sqrt{2\pi}a_z}$.

\subsection{Bogoliubov-de-Gennes equations}

To investigate the collective excitations of the condensate, we analyze the Bogoliubov-de Gennes (BdG) equations. We consider small perturbations around the ground-state wavefunctions $(\phi_{\uparrow},\phi_{\downarrow})$ of the form:
\begin{align}\label{Bogotrans}
	\psi_{\sigma}=e^{-i\mu t/\hbar}\Big[ \phi_{\sigma} + \sum_j\lambda_j\big( u_{\sigma,j}e^{-i\omega_j t} + v_{\sigma,j}^*e^{i\omega_j t} \big) \Big]\exp[-z^2/(2a_z^2)]/(\pi a_z^2)^{1/4},
\end{align}
where $\mu$ is the chemical potential of the system,  $\omega_j$ are quasi-particle energies, and $\lambda_j$ is the real perturbative parameter.  The Bogoliubov amplitudes $u_{\sigma,j}$ and $v_{\sigma,j}$ satisfy the orthonormality conditions:
\begin{subequations}
	\begin{align}
		&\sum_{\sigma}\int d\bmrho \Big[ u_{\sigma,j}^*(\bmrho)u_{\sigma,l}(\bmrho) - v_{\sigma,j}^*(\bmrho)v_{\sigma,l}(\bmrho) \Big]=\delta_{jl}, \\
		&\sum_{\sigma}\int d\bmrho \Big[ v_{\sigma,j}(\bmrho)u_{\sigma,l}(\bmrho) - u_{\sigma,j}(\bmrho)v_{\sigma,l}(\bmrho) \Big]=0,
	\end{align}
\end{subequations}
and the completeness relations:
\begin{subequations}
	\begin{align}
		&\sum_{j}\Big[ u_{\sigma,j}(\bmrho)u_{\sigma',j}^*(\bmrho') - v_{\sigma,j}^*(\bmrho)v_{\sigma',j}(\bmrho') \Big]=\delta_{\sigma\sigma'}\delta(\bmrho-\bmrho'), \\
		&\sum_{j}\Big[ u_{\sigma,j}(\bmrho)v_{\sigma',j}^*(\bmrho') - v_{\sigma,j}^*(\bmrho)u_{\sigma',j}(\bmrho') \Big]=0.
	\end{align}
\end{subequations}

Substituting Eq.~\eqref{Bogotrans} into the time-dependent real-time GPE and retaining terms linear in $\lambda_j$, we derive the BdG equations:
\begin{subequations}
\begin{align}
i\frac{\partial}{\partial t}\pu(\bmrho)
&= \Big[ h_{\uparrow\uparrow}(\bmrho) + g_{s} n(\bmrho) \Big]\pu(\bmrho) + \Omega \pd(\bmrho)+ \int d\bmrho' \Big[ \chi - \frac{1}{6}U_{\rm dd}(\bmrho-\bmrho') \Big]\pd^*(\bmrho')\pu(\bmrho')\pd(\bmrho), \\
i\frac{\partial}{\partial t}\pd(\bmrho)
&= \Big[  h_{\downarrow\downarrow}(\bmrho) + g_{s} n(\bmrho) \Big]\pd(\bmrho) +  \Omega\pu(\bmrho) + \int d\bmrho' \Big[ \chi - \frac{1}{6}U_{\rm dd}(\bmrho-\bmrho') \Big]\pu^*(\bmrho')\pd(\bmrho')\pu(\bmrho),
\end{align}
\end{subequations}
to the linear order of $\lambda_j$ and split the positive and negative frequency terms, one obtains:
\begin{subequations}
\begin{align}
(\mu+\omega_j) u_{\uparrow,j}(\bmrho) &=  \Big[h_{\uparrow\uparrow}(\bmrho) + 2g_{s}n_{\uparrow}(\bmrho) + g_{s}n_{\downarrow}(\bmrho)\Big]u_{\uparrow,j}(\bmrho) + \int d\bmrho'\Big[\chi - \frac{1}{6}U_{\rm dd}(\bmrho-\bmrho')\Big] \phi^*_{\downarrow}(\bmrho')\phi_{\downarrow}(\bmrho) u_{\uparrow,j}(\bmrho') \notag\\
&\quad+ \bigg\{ \Omega+g_{s}\phi_{\uparrow}(\bmrho)\phi_{\downarrow}^*(\bmrho) + \int d\bmrho'' \Big[ \chi - \frac{1}{6}U_{\rm dd}(\bmrho-\bmrho'') \Big]\phi_{\uparrow}(\bmrho'')\phi_{\downarrow}^*(\bmrho'') \bigg\} u_{\downarrow,j}(\bmrho) \notag\\
&\quad+ g_{s}\phi^2_{\uparrow}(\bmrho) v_{\uparrow,j}(\bmrho)
+ g_{s}\phi_{\uparrow}(\bmrho)\phi_{\downarrow}(\bmrho) v_{\downarrow,j}(\bmrho) + \int d\bmrho'\Big[\chi - \frac{1}{6}U_{\rm dd}(\bmrho-\bmrho')\Big] \phi_{\downarrow}(\bmrho)\phi_{\uparrow}(\bmrho') v_{\downarrow,j}(\bmrho'), \\
(\mu+\omega_j) u_{\downarrow,j}(\bmrho) &= \bigg\{\Omega+ g_{s}\phi_{\uparrow}^*(\bmrho)\phi_{\downarrow}(\bmrho) + \int d\bmrho'' \Big[ \chi - \frac{1}{6}U_{\rm dd}(\bmrho-\bmrho'') \Big]\phi_{\uparrow}^*(\bmrho'')\phi_{\downarrow}(\bmrho'') \bigg\} u_{\uparrow,j}(\bmrho) + g_{s}\phi^2_{\downarrow}(\bmrho) v_{\downarrow,j}(\bmrho) \notag\\
&\quad+ \Big[h_{\downarrow\downarrow}(\bmrho) + 2g_{s}n_{\downarrow}(\bmrho) + g_{s}n_{\uparrow}(\bmrho)\Big] u_{\downarrow,j}(\bmrho) + \int d\bmrho'\Big[\chi - \frac{1}{6}U_{\rm dd}(\bmrho-\bmrho')\Big] \phi^*_{\uparrow}(\bmrho')\phi_{\uparrow}(\bmrho) u_{\downarrow,j}(\bmrho') \notag\\
&\quad+ g_{s}\phi_{\uparrow}(\bmrho)\phi_{\downarrow}(\bmrho) v_{\uparrow,j}(\bmrho) + \int d\bmrho'\Big[\chi - \frac{1}{6}U_{\rm dd}(\bmrho-\bmrho')\Big] \phi_{\uparrow}(\bmrho)\phi_{\downarrow}(\bmrho') v_{\uparrow,j}(\bmrho'), \\
(\mu-\omega_j) v_{\uparrow,j}(\bmrho) &= g_{s}\phi^{*2}_{\uparrow}(\bmrho)u_{\uparrow,j}(\bmrho) 
+ g_{s}\phi_{\uparrow}^*(\bmrho)\phi_{\downarrow}^*(\bmrho)u_{\downarrow,j}(\bmrho) + \int d\bmrho'\Big[\chi - \frac{1}{6}U_{\rm dd}(\bmrho-\bmrho')\Big] \phi_{\downarrow}^*(\bmrho)\phi_{\uparrow}^*(\bmrho') u_{\downarrow,j}(\bmrho') \notag\\
&\quad+ \Big[h_{\uparrow\uparrow}(\bmrho)  + 2g_{d}n_{\uparrow}(\bmrho) + g_{s}n_{\downarrow}(\bmrho)\Big] v_{\uparrow,j}(\bmrho) + \int d\bmrho'\Big[\chi - \frac{1}{6}U_{\rm dd}(\bmrho-\bmrho')\Big] \phi_{\downarrow}(\bmrho')\phi^*_{\downarrow}(\bmrho) v_{\uparrow,j}(\bmrho') \notag\\
&\quad+ \bigg\{ \Omega + g_{s}\phi^*_{\uparrow}(\bmrho)\phi_{\downarrow}(\bmrho) + \int d\bmrho'' \Big[ \chi - \frac{1}{6}U_{\rm dd}(\bmrho-\bmrho'') \Big]\phi_{\uparrow}^*(\bmrho'')\phi_{\downarrow}(\bmrho'') \bigg\} v_{\downarrow,j}(\bmrho), \\
(\mu-\omega_j) v_{\downarrow,j}(\bmrho) &= g_{s}\phi_{\uparrow}^*(\bmrho)\phi_{\downarrow}^*(\bmrho) u_{\uparrow,j}(\bmrho) + \int d\bmrho'\Big[\chi - \frac{1}{6}U_{\rm dd}(\bmrho-\bmrho')\Big] \phi_{\uparrow}^*(\bmrho)\phi_{\downarrow}^*(\bmrho') u_{\uparrow,j}(\bmrho') 
+ g_{s}\phi^{*2}_{\downarrow}(\bmrho) u_{\downarrow,j}(\bmrho) \notag\\
&\quad+ \Big[h_{\downarrow\downarrow}(\bmrho)+ 2g_{s}n_{\downarrow}(\bmrho) + g_{s}n_{\uparrow}(\bmrho)\Big] v_{\downarrow,j}(\bmrho) + \int d\bmrho'\Big[\chi - \frac{1}{6}U_{\rm dd}(\bmrho-\bmrho')\Big] \phi_{\uparrow}(\bmrho')\phi^*_{\uparrow}(\bmrho) v_{\downarrow,j}(\bmrho') \notag\\
&\quad+ \bigg\{ \Omega+ g_{s}\phi_{\uparrow}(\bmrho)\phi_{\downarrow}^*(\bmrho) + \int d\bmrho'' \Big[ \chi - \frac{1}{6}U_{\rm dd}(\bmrho-\bmrho'') \Big]\phi_{\uparrow}(\bmrho'')\phi_{\downarrow}^*(\bmrho'') \bigg\} v_{\uparrow,j}(\bmrho).
\end{align}
\end{subequations}

In a more compact form,  the BdG equations can be written as
\begin{align}
\hbar\omega_j f_j(\bmrho) = \Sigma^z\int d\bmrho' \mathcal{H}(\bmrho,\bmrho') f_j(\bmrho),
\end{align}
where $f_j\equiv\big(u_{\uparrow,j}, u_{\downarrow,j}, v_{\uparrow,j}, v_{\downarrow,j}\big)^T$ and $\Sigma^z={\rm diag}(1,1,-1,-1)=\sigma_z\otimes I$ denotes the Bogoliubov mode functions and Pauli $z$ matrix in the Nambu space, respectively. Explicitly, $\mathcal{H}$ has the form
\begin{align}
	\mathcal{H} = \begin{pmatrix}
		\mathcal{E}_{\uparrow\uparrow} & \mathcal{E}_{\uparrow\downarrow} & \Delta_{\uparrow\uparrow} & \Delta_{\uparrow\downarrow} \\
		\mathcal{E}_{\downarrow\uparrow} & \mathcal{E}_{\downarrow\downarrow} & \Delta_{\downarrow\uparrow} & \Delta_{\downarrow\downarrow} \\
		\Delta^*_{\uparrow\uparrow} & \Delta^*_{\uparrow\downarrow} & \mathcal{E}^*_{\uparrow\uparrow} & \mathcal{E}^*_{\uparrow\downarrow} \\
		\Delta^*_{\downarrow\uparrow} & \Delta^*_{\downarrow\downarrow} & \mathcal{E}^*_{\downarrow\uparrow} & \mathcal{E}^*_{\downarrow\downarrow}
	\end{pmatrix},
\end{align}
with 

\begin{subequations}
\begin{align}
\mathcal{E}_{\uparrow\uparrow}(\bmrho,\bmrho') &= \Big[h_{\uparrow\uparrow}(\bmrho) + 2g_{s}n_{\uparrow}(\bmrho) + g_{s}n_{\downarrow}(\bmrho))\Big]\delta(\bmrho-\bmrho') + \Big[\chi - \frac{1}{6}U_{\rm dd}(\bmrho-\bmrho')\Big] \phi^*_{\downarrow}(\bmrho')\phi_{\downarrow}(\bmrho), \\
\mathcal{E}_{\uparrow\downarrow}(\bmrho,\bmrho') &= \bigg\{ \Omega + g_{s}\phi_{\uparrow}(\bmrho)\phi_{\downarrow}^*(\bmrho) + \int d\bmrho'' \Big[ \chi - \frac{1}{6}U_{\rm dd}(\bmrho-\bmrho'') \Big]\phi_{\uparrow}(\bmrho'')\phi_{\downarrow}^*(\bmrho'') \bigg\}\delta(\bmrho-\bmrho'), \\
\mathcal{E}_{\downarrow\downarrow}(\bmrho,\bmrho') &= \Big[h_{\downarrow\downarrow}(\bmrho) + 2g_{s}n_{\downarrow}(\bmrho) + g_{s}n_{\uparrow}(\bmrho)\Big]\delta(\bmrho-\bmrho') + \Big[\chi - \frac{1}{6}U_{\rm dd}(\bmrho-\bmrho')\Big] \phi^*_{\uparrow}(\bmrho')\phi_{\uparrow}(\bmrho), \\
\mathcal{E}_{\downarrow\uparrow}(\bmrho,\bmrho') &= \mathcal{E}^*_{\uparrow\downarrow}(\bmrho',\bmrho), \\
\Delta_{\uparrow\uparrow}(\bmrho,\bmrho') &= g_{s}\phi^2_{\uparrow}(\bmrho)\delta(\bmrho-\bmrho'), \\
\Delta_{\uparrow\downarrow}(\bmrho,\bmrho') &= g_{s}\phi_{\uparrow}(\bmrho)\phi_{\downarrow}(\bmrho)\delta(\bmrho-\bmrho') + \Big[\chi - \frac{1}{6}U_{\rm dd}(\bmrho-\bmrho')\Big] \phi_{\downarrow}(\bmrho)\phi_{\uparrow}(\bmrho'), \\
\Delta_{\downarrow\downarrow}(\bmrho,\bmrho') &= g_{s}\phi^2_{\downarrow}(\bmrho)\delta(\bmrho-\bmrho'), \\
\Delta_{\downarrow\uparrow}(\bmrho,\bmrho') &= \Delta_{\uparrow\downarrow}(\bmrho',\bmrho).
\end{align}
\end{subequations}


%

\end{widetext}


\begin{thebibliography}{82}%
\makeatletter
\providecommand \@ifxundefined [1]{%
 \@ifx{#1\undefined}
}%
\providecommand \@ifnum [1]{%
 \ifnum #1\expandafter \@firstoftwo
 \else \expandafter \@secondoftwo
 \fi
}%
\providecommand \@ifx [1]{%
 \ifx #1\expandafter \@firstoftwo
 \else \expandafter \@secondoftwo
 \fi
}%
\providecommand \natexlab [1]{#1}%
\providecommand \enquote  [1]{``#1''}%
\providecommand \bibnamefont  [1]{#1}%
\providecommand \bibfnamefont [1]{#1}%
\providecommand \citenamefont [1]{#1}%
\providecommand \href@noop [0]{\@secondoftwo}%
\providecommand \href [0]{\begingroup \@sanitize@url \@href}%
\providecommand \@href[1]{\@@startlink{#1}\@@href}%
\providecommand \@@href[1]{\endgroup#1\@@endlink}%
\providecommand \@sanitize@url [0]{\catcode `\\12\catcode `\$12\catcode
  `\&12\catcode `\#12\catcode `\^12\catcode `\_12\catcode `\%12\relax}%
\providecommand \@@startlink[1]{}%
\providecommand \@@endlink[0]{}%
\providecommand \url  [0]{\begingroup\@sanitize@url \@url }%
\providecommand \@url [1]{\endgroup\@href {#1}{\urlprefix }}%
\providecommand \urlprefix  [0]{URL }%
\providecommand \Eprint [0]{\href }%
\providecommand \doibase [0]{https://doi.org/}%
\providecommand \selectlanguage [0]{\@gobble}%
\providecommand \bibinfo  [0]{\@secondoftwo}%
\providecommand \bibfield  [0]{\@secondoftwo}%
\providecommand \translation [1]{[#1]}%
\providecommand \BibitemOpen [0]{}%
\providecommand \bibitemStop [0]{}%
\providecommand \bibitemNoStop [0]{.\EOS\space}%
\providecommand \EOS [0]{\spacefactor3000\relax}%
\providecommand \BibitemShut  [1]{\csname bibitem#1\endcsname}%
\let\auto@bib@innerbib\@empty
\bibitem [{\citenamefont {Marco}\ \emph {et~al.}(2019)\citenamefont {Marco},
  \citenamefont {Valtolina}, \citenamefont {Matsuda}, \citenamefont {Tobias},
  \citenamefont {Covey},\ and\ \citenamefont
  {Ye}}]{doi:10.1126/science.aau7230}%
  \BibitemOpen
  \bibfield  {author} {\bibinfo {author} {\bibfnamefont {L.~D.}\ \bibnamefont
  {Marco}}, \bibinfo {author} {\bibfnamefont {G.}~\bibnamefont {Valtolina}},
  \bibinfo {author} {\bibfnamefont {K.}~\bibnamefont {Matsuda}}, \bibinfo
  {author} {\bibfnamefont {W.~G.}\ \bibnamefont {Tobias}}, \bibinfo {author}
  {\bibfnamefont {J.~P.}\ \bibnamefont {Covey}},\ and\ \bibinfo {author}
  {\bibfnamefont {J.}~\bibnamefont {Ye}},\ }\bibfield  {title} {\bibinfo
  {title} {A degenerate fermi gas of polar molecules},\ }\href
  {https://doi.org/10.1126/science.aau7230} {\bibfield  {journal} {\bibinfo
  {journal} {Science}\ }\textbf {\bibinfo {volume} {363}},\ \bibinfo {pages}
  {853} (\bibinfo {year} {2019})}\BibitemShut {NoStop}%
\bibitem [{\citenamefont {Tobias}\ \emph {et~al.}(2020)\citenamefont {Tobias},
  \citenamefont {Matsuda}, \citenamefont {Valtolina}, \citenamefont {De~Marco},
  \citenamefont {Li},\ and\ \citenamefont {Ye}}]{PhysRevLett.124.033401}%
  \BibitemOpen
  \bibfield  {author} {\bibinfo {author} {\bibfnamefont {W.~G.}\ \bibnamefont
  {Tobias}}, \bibinfo {author} {\bibfnamefont {K.}~\bibnamefont {Matsuda}},
  \bibinfo {author} {\bibfnamefont {G.}~\bibnamefont {Valtolina}}, \bibinfo
  {author} {\bibfnamefont {L.}~\bibnamefont {De~Marco}}, \bibinfo {author}
  {\bibfnamefont {J.-R.}\ \bibnamefont {Li}},\ and\ \bibinfo {author}
  {\bibfnamefont {J.}~\bibnamefont {Ye}},\ }\bibfield  {title} {\bibinfo
  {title} {Thermalization and sub-poissonian density fluctuations in a
  degenerate molecular fermi gas},\ }\href
  {https://doi.org/10.1103/PhysRevLett.124.033401} {\bibfield  {journal}
  {\bibinfo  {journal} {Phys. Rev. Lett.}\ }\textbf {\bibinfo {volume} {124}},\
  \bibinfo {pages} {033401} (\bibinfo {year} {2020})}\BibitemShut {NoStop}%
\bibitem [{\citenamefont {Schindewolf}\ \emph {et~al.}(2022)\citenamefont
  {Schindewolf}, \citenamefont {Bause}, \citenamefont {Chen}, \citenamefont
  {Duda}, \citenamefont {Karman}, \citenamefont {Bloch},\ and\ \citenamefont
  {Luo}}]{Schindewolf}%
  \BibitemOpen
  \bibfield  {author} {\bibinfo {author} {\bibfnamefont {A.}~\bibnamefont
  {Schindewolf}}, \bibinfo {author} {\bibfnamefont {R.}~\bibnamefont {Bause}},
  \bibinfo {author} {\bibfnamefont {X.-Y.}\ \bibnamefont {Chen}}, \bibinfo
  {author} {\bibfnamefont {M.}~\bibnamefont {Duda}}, \bibinfo {author}
  {\bibfnamefont {T.}~\bibnamefont {Karman}}, \bibinfo {author} {\bibfnamefont
  {I.}~\bibnamefont {Bloch}},\ and\ \bibinfo {author} {\bibfnamefont {X.-Y.}\
  \bibnamefont {Luo}},\ }\bibfield  {title} {\bibinfo {title} {Evaporation of
  microwave-shielded polar molecules to quantum degeneracy},\ }\href
  {https://doi.org/10.1038/s41586-022-04900-0} {\bibfield  {journal} {\bibinfo
  {journal} {Nature (London)}\ }\textbf {\bibinfo {volume} {607}},\ \bibinfo
  {pages} {677} (\bibinfo {year} {2022})}\BibitemShut {NoStop}%
\bibitem [{\citenamefont {Guo}\ \emph {et~al.}(2016)\citenamefont {Guo},
  \citenamefont {Zhu}, \citenamefont {Lu}, \citenamefont {Ye}, \citenamefont
  {Wang}, \citenamefont {Vexiau}, \citenamefont {Bouloufa-Maafa}, \citenamefont
  {Qu\'em\'ener}, \citenamefont {Dulieu},\ and\ \citenamefont
  {Wang}}]{PhysRevLett.116.205303}%
  \BibitemOpen
  \bibfield  {author} {\bibinfo {author} {\bibfnamefont {M.}~\bibnamefont
  {Guo}}, \bibinfo {author} {\bibfnamefont {B.}~\bibnamefont {Zhu}}, \bibinfo
  {author} {\bibfnamefont {B.}~\bibnamefont {Lu}}, \bibinfo {author}
  {\bibfnamefont {X.}~\bibnamefont {Ye}}, \bibinfo {author} {\bibfnamefont
  {F.}~\bibnamefont {Wang}}, \bibinfo {author} {\bibfnamefont {R.}~\bibnamefont
  {Vexiau}}, \bibinfo {author} {\bibfnamefont {N.}~\bibnamefont
  {Bouloufa-Maafa}}, \bibinfo {author} {\bibfnamefont {G.}~\bibnamefont
  {Qu\'em\'ener}}, \bibinfo {author} {\bibfnamefont {O.}~\bibnamefont
  {Dulieu}},\ and\ \bibinfo {author} {\bibfnamefont {D.}~\bibnamefont {Wang}},\
  }\bibfield  {title} {\bibinfo {title} {Creation of an ultracold gas of
  ground-state dipolar $^{23}\mathrm{Na}^{87}\mathrm{Rb}$ molecules},\ }\href
  {https://doi.org/10.1103/PhysRevLett.116.205303} {\bibfield  {journal}
  {\bibinfo  {journal} {Phys. Rev. Lett.}\ }\textbf {\bibinfo {volume} {116}},\
  \bibinfo {pages} {205303} (\bibinfo {year} {2016})}\BibitemShut {NoStop}%
\bibitem [{\citenamefont {Park}\ \emph {et~al.}(2015)\citenamefont {Park},
  \citenamefont {Will},\ and\ \citenamefont
  {Zwierlein}}]{PhysRevLett.114.205302}%
  \BibitemOpen
  \bibfield  {author} {\bibinfo {author} {\bibfnamefont {J.~W.}\ \bibnamefont
  {Park}}, \bibinfo {author} {\bibfnamefont {S.~A.}\ \bibnamefont {Will}},\
  and\ \bibinfo {author} {\bibfnamefont {M.~W.}\ \bibnamefont {Zwierlein}},\
  }\bibfield  {title} {\bibinfo {title} {Ultracold dipolar gas of fermionic
  $^{23}\mathrm{Na}^{40}\mathrm{K}$ molecules in their absolute ground state},\
  }\href {https://doi.org/10.1103/PhysRevLett.114.205302} {\bibfield  {journal}
  {\bibinfo  {journal} {Phys. Rev. Lett.}\ }\textbf {\bibinfo {volume} {114}},\
  \bibinfo {pages} {205302} (\bibinfo {year} {2015})}\BibitemShut {NoStop}%
\bibitem [{\citenamefont {Molony}\ \emph {et~al.}(2014)\citenamefont {Molony},
  \citenamefont {Gregory}, \citenamefont {Ji}, \citenamefont {Lu},
  \citenamefont {K\"oppinger}, \citenamefont {Le~Sueur}, \citenamefont
  {Blackley}, \citenamefont {Hutson},\ and\ \citenamefont
  {Cornish}}]{PhysRevLett.113.255301}%
  \BibitemOpen
  \bibfield  {author} {\bibinfo {author} {\bibfnamefont {P.~K.}\ \bibnamefont
  {Molony}}, \bibinfo {author} {\bibfnamefont {P.~D.}\ \bibnamefont {Gregory}},
  \bibinfo {author} {\bibfnamefont {Z.}~\bibnamefont {Ji}}, \bibinfo {author}
  {\bibfnamefont {B.}~\bibnamefont {Lu}}, \bibinfo {author} {\bibfnamefont
  {M.~P.}\ \bibnamefont {K\"oppinger}}, \bibinfo {author} {\bibfnamefont
  {C.~R.}\ \bibnamefont {Le~Sueur}}, \bibinfo {author} {\bibfnamefont {C.~L.}\
  \bibnamefont {Blackley}}, \bibinfo {author} {\bibfnamefont {J.~M.}\
  \bibnamefont {Hutson}},\ and\ \bibinfo {author} {\bibfnamefont {S.~L.}\
  \bibnamefont {Cornish}},\ }\bibfield  {title} {\bibinfo {title} {Creation of
  ultracold $^{87}\mathrm{Rb}^{133}\mathrm{Cs}$ molecules in the rovibrational
  ground state},\ }\href {https://doi.org/10.1103/PhysRevLett.113.255301}
  {\bibfield  {journal} {\bibinfo  {journal} {Phys. Rev. Lett.}\ }\textbf
  {\bibinfo {volume} {113}},\ \bibinfo {pages} {255301} (\bibinfo {year}
  {2014})}\BibitemShut {NoStop}%
\bibitem [{\citenamefont {Takekoshi}\ \emph {et~al.}(2014)\citenamefont
  {Takekoshi}, \citenamefont {Reichs\"ollner}, \citenamefont {Schindewolf},
  \citenamefont {Hutson}, \citenamefont {Le~Sueur}, \citenamefont {Dulieu},
  \citenamefont {Ferlaino}, \citenamefont {Grimm},\ and\ \citenamefont
  {N\"agerl}}]{PhysRevLett.113.205301}%
  \BibitemOpen
  \bibfield  {author} {\bibinfo {author} {\bibfnamefont {T.}~\bibnamefont
  {Takekoshi}}, \bibinfo {author} {\bibfnamefont {L.}~\bibnamefont
  {Reichs\"ollner}}, \bibinfo {author} {\bibfnamefont {A.}~\bibnamefont
  {Schindewolf}}, \bibinfo {author} {\bibfnamefont {J.~M.}\ \bibnamefont
  {Hutson}}, \bibinfo {author} {\bibfnamefont {C.~R.}\ \bibnamefont
  {Le~Sueur}}, \bibinfo {author} {\bibfnamefont {O.}~\bibnamefont {Dulieu}},
  \bibinfo {author} {\bibfnamefont {F.}~\bibnamefont {Ferlaino}}, \bibinfo
  {author} {\bibfnamefont {R.}~\bibnamefont {Grimm}},\ and\ \bibinfo {author}
  {\bibfnamefont {H.-C.}\ \bibnamefont {N\"agerl}},\ }\bibfield  {title}
  {\bibinfo {title} {Ultracold dense samples of dipolar rbcs molecules in the
  rovibrational and hyperfine ground state},\ }\href
  {https://doi.org/10.1103/PhysRevLett.113.205301} {\bibfield  {journal}
  {\bibinfo  {journal} {Phys. Rev. Lett.}\ }\textbf {\bibinfo {volume} {113}},\
  \bibinfo {pages} {205301} (\bibinfo {year} {2014})}\BibitemShut {NoStop}%
\bibitem [{\citenamefont {Qu\'{e}m\'{e}ner}\ and\ \citenamefont
  {Julienne}(2012)}]{Chem.Rev.112.4949-5011}%
  \BibitemOpen
  \bibfield  {author} {\bibinfo {author} {\bibfnamefont {G.}~\bibnamefont
  {Qu\'{e}m\'{e}ner}}\ and\ \bibinfo {author} {\bibfnamefont {P.~S.}\
  \bibnamefont {Julienne}},\ }\bibfield  {title} {\bibinfo {title} {Ultracold
  molecules under control!},\ }\href {https://doi.org/10.1021/cr300092g}
  {\bibfield  {journal} {\bibinfo  {journal} {Chem. Rev.}\ }\textbf {\bibinfo
  {volume} {112}},\ \bibinfo {pages} {4949} (\bibinfo {year}
  {2012})}\BibitemShut {NoStop}%
\bibitem [{\citenamefont {Cornish}\ \emph {et~al.}(2024)\citenamefont
  {Cornish}, \citenamefont {Tarbutt},\ and\ \citenamefont
  {Hazzard}}]{Cornish2024}%
  \BibitemOpen
  \bibfield  {author} {\bibinfo {author} {\bibfnamefont {S.}~\bibnamefont
  {Cornish}}, \bibinfo {author} {\bibfnamefont {M.}~\bibnamefont {Tarbutt}},\
  and\ \bibinfo {author} {\bibfnamefont {K.}~\bibnamefont {Hazzard}},\
  }\bibfield  {title} {\bibinfo {title} {Quantum computation and quantum
  simulation with ultracold molecules},\ }\href
  {https://doi.org/10.1038/s41567-024-02453-9} {\bibfield  {journal} {\bibinfo
  {journal} {Nat. Phys.}\ }\textbf {\bibinfo {volume} {20}},\ \bibinfo {pages}
  {730} (\bibinfo {year} {2024})}\BibitemShut {NoStop}%
\bibitem [{\citenamefont {Carr}\ \emph {et~al.}(2009)\citenamefont {Carr},
  \citenamefont {DeMille}, \citenamefont {Krems},\ and\ \citenamefont
  {Ye}}]{Carr_2009}%
  \BibitemOpen
  \bibfield  {author} {\bibinfo {author} {\bibfnamefont {L.~D.}\ \bibnamefont
  {Carr}}, \bibinfo {author} {\bibfnamefont {D.}~\bibnamefont {DeMille}},
  \bibinfo {author} {\bibfnamefont {R.~V.}\ \bibnamefont {Krems}},\ and\
  \bibinfo {author} {\bibfnamefont {J.}~\bibnamefont {Ye}},\ }\bibfield
  {title} {\bibinfo {title} {Cold and ultracold molecules: science, technology
  and applications},\ }\href {https://doi.org/10.1088/1367-2630/11/5/055049}
  {\bibfield  {journal} {\bibinfo  {journal} {New J. Phys.}\ }\textbf {\bibinfo
  {volume} {11}},\ \bibinfo {pages} {055049} (\bibinfo {year}
  {2009})}\BibitemShut {NoStop}%
\bibitem [{\citenamefont {Park}\ \emph {et~al.}(2023)\citenamefont {Park},
  \citenamefont {Lu}, \citenamefont {Jamison}, \citenamefont {Tscherbul},\ and\
  \citenamefont {Ketterle}}]{Park2023}%
  \BibitemOpen
  \bibfield  {author} {\bibinfo {author} {\bibfnamefont {J.~J.}\ \bibnamefont
  {Park}}, \bibinfo {author} {\bibfnamefont {Y.-K.}\ \bibnamefont {Lu}},
  \bibinfo {author} {\bibfnamefont {A.~O.}\ \bibnamefont {Jamison}}, \bibinfo
  {author} {\bibfnamefont {T.~V.}\ \bibnamefont {Tscherbul}},\ and\ \bibinfo
  {author} {\bibfnamefont {W.}~\bibnamefont {Ketterle}},\ }\bibfield  {title}
  {\bibinfo {title} {A feshbach resonance in collisions between triplet
  ground-state molecules},\ }\href {https://doi.org/10.1038/s41586-022-05635-8}
  {\bibfield  {journal} {\bibinfo  {journal} {Nature (London)}\ }\textbf
  {\bibinfo {volume} {614}},\ \bibinfo {pages} {54} (\bibinfo {year}
  {2023})}\BibitemShut {NoStop}%
\bibitem [{\citenamefont {Finelli}\ \emph {et~al.}(2024)\citenamefont
  {Finelli}, \citenamefont {Ciamei}, \citenamefont {Restivo}, \citenamefont
  {Schemmer}, \citenamefont {Cosco}, \citenamefont {Inguscio}, \citenamefont
  {Trenkwalder}, \citenamefont {Zaremba-Kopczyk}, \citenamefont {Gronowski},
  \citenamefont {Tomza},\ and\ \citenamefont {Zaccanti}}]{PRXQuantum.5.020358}%
  \BibitemOpen
  \bibfield  {author} {\bibinfo {author} {\bibfnamefont {S.}~\bibnamefont
  {Finelli}}, \bibinfo {author} {\bibfnamefont {A.}~\bibnamefont {Ciamei}},
  \bibinfo {author} {\bibfnamefont {B.}~\bibnamefont {Restivo}}, \bibinfo
  {author} {\bibfnamefont {M.}~\bibnamefont {Schemmer}}, \bibinfo {author}
  {\bibfnamefont {A.}~\bibnamefont {Cosco}}, \bibinfo {author} {\bibfnamefont
  {M.}~\bibnamefont {Inguscio}}, \bibinfo {author} {\bibfnamefont
  {A.}~\bibnamefont {Trenkwalder}}, \bibinfo {author} {\bibfnamefont
  {K.}~\bibnamefont {Zaremba-Kopczyk}}, \bibinfo {author} {\bibfnamefont
  {M.}~\bibnamefont {Gronowski}}, \bibinfo {author} {\bibfnamefont
  {M.}~\bibnamefont {Tomza}},\ and\ \bibinfo {author} {\bibfnamefont
  {M.}~\bibnamefont {Zaccanti}},\ }\bibfield  {title} {\bibinfo {title}
  {Ultracold $\mathrm{Li}\mathrm{Cr}$: A new pathway to quantum gases of
  paramagnetic polar molecules},\ }\href
  {https://doi.org/10.1103/PRXQuantum.5.020358} {\bibfield  {journal} {\bibinfo
   {journal} {PRX Quantum}\ }\textbf {\bibinfo {volume} {5}},\ \bibinfo {pages}
  {020358} (\bibinfo {year} {2024})}\BibitemShut {NoStop}%
\bibitem [{\citenamefont {Sawant}\ \emph {et~al.}(2020)\citenamefont {Sawant},
  \citenamefont {Blackmore}, \citenamefont {Gregory}, \citenamefont
  {Mur-Petit}, \citenamefont {Jaksch}, \citenamefont {Aldegunde}, \citenamefont
  {Hutson}, \citenamefont {Tarbutt},\ and\ \citenamefont
  {Cornish}}]{Sawant_2020}%
  \BibitemOpen
  \bibfield  {author} {\bibinfo {author} {\bibfnamefont {R.}~\bibnamefont
  {Sawant}}, \bibinfo {author} {\bibfnamefont {J.~A.}\ \bibnamefont
  {Blackmore}}, \bibinfo {author} {\bibfnamefont {P.~D.}\ \bibnamefont
  {Gregory}}, \bibinfo {author} {\bibfnamefont {J.}~\bibnamefont {Mur-Petit}},
  \bibinfo {author} {\bibfnamefont {D.}~\bibnamefont {Jaksch}}, \bibinfo
  {author} {\bibfnamefont {J.}~\bibnamefont {Aldegunde}}, \bibinfo {author}
  {\bibfnamefont {J.~M.}\ \bibnamefont {Hutson}}, \bibinfo {author}
  {\bibfnamefont {M.~R.}\ \bibnamefont {Tarbutt}},\ and\ \bibinfo {author}
  {\bibfnamefont {S.~L.}\ \bibnamefont {Cornish}},\ }\bibfield  {title}
  {\bibinfo {title} {Ultracold polar molecules as qudits},\ }\href
  {https://doi.org/10.1088/1367-2630/ab60f4} {\bibfield  {journal} {\bibinfo
  {journal} {New J. Phys.}\ }\textbf {\bibinfo {volume} {22}},\ \bibinfo
  {pages} {013027} (\bibinfo {year} {2020})}\BibitemShut {NoStop}%
\bibitem [{\citenamefont {Picard}\ \emph {et~al.}(2024)\citenamefont {Picard},
  \citenamefont {Park}, \citenamefont {Patenotte}, \citenamefont
  {Gebretsadkan}, \citenamefont {Wellnitz}, \citenamefont {Rey},\ and\
  \citenamefont {Ni}}]{Picard2024}%
  \BibitemOpen
  \bibfield  {author} {\bibinfo {author} {\bibfnamefont {L.~R.~B.}\
  \bibnamefont {Picard}}, \bibinfo {author} {\bibfnamefont {A.~J.}\
  \bibnamefont {Park}}, \bibinfo {author} {\bibfnamefont {G.~E.}\ \bibnamefont
  {Patenotte}}, \bibinfo {author} {\bibfnamefont {S.}~\bibnamefont
  {Gebretsadkan}}, \bibinfo {author} {\bibfnamefont {D.}~\bibnamefont
  {Wellnitz}}, \bibinfo {author} {\bibfnamefont {A.~M.}\ \bibnamefont {Rey}},\
  and\ \bibinfo {author} {\bibfnamefont {K.-K.}\ \bibnamefont {Ni}},\
  }\bibfield  {title} {\bibinfo {title} {Entanglement and iswap gate between
  molecular qubits},\ }\bibfield  {journal} {\bibinfo  {journal} {Nature
  (London)}\ }\href {https://doi.org/10.1038/s41586-024-08177-3}
  {10.1038/s41586-024-08177-3} (\bibinfo {year} {2024})\BibitemShut {NoStop}%
\bibitem [{\citenamefont {Cairncross}\ \emph {et~al.}(2017)\citenamefont
  {Cairncross}, \citenamefont {Gresh}, \citenamefont {Grau}, \citenamefont
  {Cossel}, \citenamefont {Roussy}, \citenamefont {Ni}, \citenamefont {Zhou},
  \citenamefont {Ye},\ and\ \citenamefont {Cornell}}]{PhysRevLett.119.153001}%
  \BibitemOpen
  \bibfield  {author} {\bibinfo {author} {\bibfnamefont {W.~B.}\ \bibnamefont
  {Cairncross}}, \bibinfo {author} {\bibfnamefont {D.~N.}\ \bibnamefont
  {Gresh}}, \bibinfo {author} {\bibfnamefont {M.}~\bibnamefont {Grau}},
  \bibinfo {author} {\bibfnamefont {K.~C.}\ \bibnamefont {Cossel}}, \bibinfo
  {author} {\bibfnamefont {T.~S.}\ \bibnamefont {Roussy}}, \bibinfo {author}
  {\bibfnamefont {Y.}~\bibnamefont {Ni}}, \bibinfo {author} {\bibfnamefont
  {Y.}~\bibnamefont {Zhou}}, \bibinfo {author} {\bibfnamefont {J.}~\bibnamefont
  {Ye}},\ and\ \bibinfo {author} {\bibfnamefont {E.~A.}\ \bibnamefont
  {Cornell}},\ }\bibfield  {title} {\bibinfo {title} {Precision measurement of
  the electron's electric dipole moment using trapped molecular ions},\ }\href
  {https://doi.org/10.1103/PhysRevLett.119.153001} {\bibfield  {journal}
  {\bibinfo  {journal} {Phys. Rev. Lett.}\ }\textbf {\bibinfo {volume} {119}},\
  \bibinfo {pages} {153001} (\bibinfo {year} {2017})}\BibitemShut {NoStop}%
\bibitem [{\citenamefont {Hudson}\ \emph {et~al.}(2011)\citenamefont {Hudson},
  \citenamefont {Kara}, \citenamefont {Smallman}, \citenamefont {Sauer},
  \citenamefont {Tarbutt},\ and\ \citenamefont {Hinds}}]{hudson2011improved}%
  \BibitemOpen
  \bibfield  {author} {\bibinfo {author} {\bibfnamefont {J.~J.}\ \bibnamefont
  {Hudson}}, \bibinfo {author} {\bibfnamefont {D.~M.}\ \bibnamefont {Kara}},
  \bibinfo {author} {\bibfnamefont {I.}~\bibnamefont {Smallman}}, \bibinfo
  {author} {\bibfnamefont {B.~E.}\ \bibnamefont {Sauer}}, \bibinfo {author}
  {\bibfnamefont {M.~R.}\ \bibnamefont {Tarbutt}},\ and\ \bibinfo {author}
  {\bibfnamefont {E.~A.}\ \bibnamefont {Hinds}},\ }\bibfield  {title} {\bibinfo
  {title} {Improved measurement of the shape of the electron},\ }\href@noop {}
  {\bibfield  {journal} {\bibinfo  {journal} {Nature}\ }\textbf {\bibinfo
  {volume} {473}},\ \bibinfo {pages} {493} (\bibinfo {year}
  {2011})}\BibitemShut {NoStop}%
\bibitem [{\citenamefont {Roussy}\ \emph {et~al.}(2023)\citenamefont {Roussy},
  \citenamefont {Caldwell}, \citenamefont {Wright}, \citenamefont {Cairncross},
  \citenamefont {Shagam}, \citenamefont {Ng}, \citenamefont {Schlossberger},
  \citenamefont {Park}, \citenamefont {Wang}, \citenamefont {Ye},\ and\
  \citenamefont {Cornell}}]{doi:10.1126/science.adg4084}%
  \BibitemOpen
  \bibfield  {author} {\bibinfo {author} {\bibfnamefont {T.~S.}\ \bibnamefont
  {Roussy}}, \bibinfo {author} {\bibfnamefont {L.}~\bibnamefont {Caldwell}},
  \bibinfo {author} {\bibfnamefont {T.}~\bibnamefont {Wright}}, \bibinfo
  {author} {\bibfnamefont {W.~B.}\ \bibnamefont {Cairncross}}, \bibinfo
  {author} {\bibfnamefont {Y.}~\bibnamefont {Shagam}}, \bibinfo {author}
  {\bibfnamefont {K.~B.}\ \bibnamefont {Ng}}, \bibinfo {author} {\bibfnamefont
  {N.}~\bibnamefont {Schlossberger}}, \bibinfo {author} {\bibfnamefont {S.~Y.}\
  \bibnamefont {Park}}, \bibinfo {author} {\bibfnamefont {A.}~\bibnamefont
  {Wang}}, \bibinfo {author} {\bibfnamefont {J.}~\bibnamefont {Ye}},\ and\
  \bibinfo {author} {\bibfnamefont {E.~A.}\ \bibnamefont {Cornell}},\
  }\bibfield  {title} {\bibinfo {title} {An improved bound on the electron's
  electric dipole moment},\ }\href {https://doi.org/10.1126/science.adg4084}
  {\bibfield  {journal} {\bibinfo  {journal} {Science}\ }\textbf {\bibinfo
  {volume} {381}},\ \bibinfo {pages} {46} (\bibinfo {year} {2023})}\BibitemShut
  {NoStop}%
\bibitem [{\citenamefont {Karman}\ and\ \citenamefont
  {Hutson}(2018)}]{PhysRevLett.121.163401}%
  \BibitemOpen
  \bibfield  {author} {\bibinfo {author} {\bibfnamefont {T.}~\bibnamefont
  {Karman}}\ and\ \bibinfo {author} {\bibfnamefont {J.~M.}\ \bibnamefont
  {Hutson}},\ }\bibfield  {title} {\bibinfo {title} {Microwave shielding of
  ultracold polar molecules},\ }\href
  {https://doi.org/10.1103/PhysRevLett.121.163401} {\bibfield  {journal}
  {\bibinfo  {journal} {Phys. Rev. Lett.}\ }\textbf {\bibinfo {volume} {121}},\
  \bibinfo {pages} {163401} (\bibinfo {year} {2018})}\BibitemShut {NoStop}%
\bibitem [{\citenamefont {Anderegg}\ \emph {et~al.}(2021)\citenamefont
  {Anderegg}, \citenamefont {Burchesky}, \citenamefont {Bao}, \citenamefont
  {Yu}, \citenamefont {Karman}, \citenamefont {Chae}, \citenamefont {Ni},
  \citenamefont {Ketterle},\ and\ \citenamefont
  {Doyle}}]{doi:10.1126/science.abg9502}%
  \BibitemOpen
  \bibfield  {author} {\bibinfo {author} {\bibfnamefont {L.}~\bibnamefont
  {Anderegg}}, \bibinfo {author} {\bibfnamefont {S.}~\bibnamefont {Burchesky}},
  \bibinfo {author} {\bibfnamefont {Y.-C.}\ \bibnamefont {Bao}}, \bibinfo
  {author} {\bibfnamefont {S.~S.}\ \bibnamefont {Yu}}, \bibinfo {author}
  {\bibfnamefont {T.}~\bibnamefont {Karman}}, \bibinfo {author} {\bibfnamefont
  {E.}~\bibnamefont {Chae}}, \bibinfo {author} {\bibfnamefont {K.-K.}\
  \bibnamefont {Ni}}, \bibinfo {author} {\bibfnamefont {W.}~\bibnamefont
  {Ketterle}},\ and\ \bibinfo {author} {\bibfnamefont {J.~M.}\ \bibnamefont
  {Doyle}},\ }\bibfield  {title} {\bibinfo {title} {Observation of microwave
  shielding of ultracold molecules},\ }\href
  {https://doi.org/10.1126/science.abg9502} {\bibfield  {journal} {\bibinfo
  {journal} {Science}\ }\textbf {\bibinfo {volume} {373}},\ \bibinfo {pages}
  {779} (\bibinfo {year} {2021})}\BibitemShut {NoStop}%
\bibitem [{\citenamefont {Chen}\ \emph {et~al.}(2023)\citenamefont {Chen},
  \citenamefont {Schindewolf}, \citenamefont {Eppelt}, \citenamefont {Bause},
  \citenamefont {Duda}, \citenamefont {Biswas}, \citenamefont {Karman},
  \citenamefont {Hilker}, \citenamefont {Bloch},\ and\ \citenamefont
  {Luo}}]{Chen2023}%
  \BibitemOpen
  \bibfield  {author} {\bibinfo {author} {\bibfnamefont {X.-Y.}\ \bibnamefont
  {Chen}}, \bibinfo {author} {\bibfnamefont {A.}~\bibnamefont {Schindewolf}},
  \bibinfo {author} {\bibfnamefont {S.}~\bibnamefont {Eppelt}}, \bibinfo
  {author} {\bibfnamefont {R.}~\bibnamefont {Bause}}, \bibinfo {author}
  {\bibfnamefont {M.}~\bibnamefont {Duda}}, \bibinfo {author} {\bibfnamefont
  {S.}~\bibnamefont {Biswas}}, \bibinfo {author} {\bibfnamefont
  {T.}~\bibnamefont {Karman}}, \bibinfo {author} {\bibfnamefont
  {T.}~\bibnamefont {Hilker}}, \bibinfo {author} {\bibfnamefont
  {I.}~\bibnamefont {Bloch}},\ and\ \bibinfo {author} {\bibfnamefont {X.-Y.}\
  \bibnamefont {Luo}},\ }\bibfield  {title} {\bibinfo {title} {Field-linked
  resonances of polar molecules},\ }\href
  {https://doi.org/10.1038/s41586-022-05651-8} {\bibfield  {journal} {\bibinfo
  {journal} {Nature (London)}\ }\textbf {\bibinfo {volume} {614}},\ \bibinfo
  {pages} {59} (\bibinfo {year} {2023})}\BibitemShut {NoStop}%
\bibitem [{\citenamefont {Bigagli}\ \emph {et~al.}(2024)\citenamefont
  {Bigagli}, \citenamefont {Yuan}, \citenamefont {Zhang}, \citenamefont
  {Bulatovic}, \citenamefont {Karman}, \citenamefont {Stevenson},\ and\
  \citenamefont {Will}}]{bigagli2024observation}%
  \BibitemOpen
  \bibfield  {author} {\bibinfo {author} {\bibfnamefont {N.}~\bibnamefont
  {Bigagli}}, \bibinfo {author} {\bibfnamefont {W.}~\bibnamefont {Yuan}},
  \bibinfo {author} {\bibfnamefont {S.}~\bibnamefont {Zhang}}, \bibinfo
  {author} {\bibfnamefont {B.}~\bibnamefont {Bulatovic}}, \bibinfo {author}
  {\bibfnamefont {T.}~\bibnamefont {Karman}}, \bibinfo {author} {\bibfnamefont
  {I.}~\bibnamefont {Stevenson}},\ and\ \bibinfo {author} {\bibfnamefont
  {S.}~\bibnamefont {Will}},\ }\bibfield  {title} {\bibinfo {title}
  {Observation of bose-einstein condensation of dipolar molecules},\ }\href
  {https://doi.org/10.1038/s41586-024-07492-z} {\bibfield  {journal} {\bibinfo
  {journal} {Nature (London)}\ }\textbf {\bibinfo {volume} {631}},\ \bibinfo
  {pages} {289} (\bibinfo {year} {2024})}\BibitemShut {NoStop}%
\bibitem [{\citenamefont {Yan}\ \emph {et~al.}(2013)\citenamefont {Yan},
  \citenamefont {Moses}, \citenamefont {Gadway}, \citenamefont {Covey},
  \citenamefont {Hazzard}, \citenamefont {Rey}, \citenamefont {Jin},\ and\
  \citenamefont {Ye}}]{yan2013observation}%
  \BibitemOpen
  \bibfield  {author} {\bibinfo {author} {\bibfnamefont {B.}~\bibnamefont
  {Yan}}, \bibinfo {author} {\bibfnamefont {S.~A.}\ \bibnamefont {Moses}},
  \bibinfo {author} {\bibfnamefont {B.}~\bibnamefont {Gadway}}, \bibinfo
  {author} {\bibfnamefont {J.~P.}\ \bibnamefont {Covey}}, \bibinfo {author}
  {\bibfnamefont {K.~R.}\ \bibnamefont {Hazzard}}, \bibinfo {author}
  {\bibfnamefont {A.~M.}\ \bibnamefont {Rey}}, \bibinfo {author} {\bibfnamefont
  {D.~S.}\ \bibnamefont {Jin}},\ and\ \bibinfo {author} {\bibfnamefont
  {J.}~\bibnamefont {Ye}},\ }\bibfield  {title} {\bibinfo {title} {Observation
  of dipolar spin-exchange interactions with lattice-confined polar
  molecules},\ }\href@noop {} {\bibfield  {journal} {\bibinfo  {journal}
  {Nature}\ }\textbf {\bibinfo {volume} {501}},\ \bibinfo {pages} {521}
  (\bibinfo {year} {2013})}\BibitemShut {NoStop}%
\bibitem [{\citenamefont {de~Paz}\ \emph {et~al.}(2013)\citenamefont {de~Paz},
  \citenamefont {Sharma}, \citenamefont {Chotia}, \citenamefont {Mar\'echal},
  \citenamefont {Huckans}, \citenamefont {Pedri}, \citenamefont {Santos},
  \citenamefont {Gorceix}, \citenamefont {Vernac},\ and\ \citenamefont
  {Laburthe-Tolra}}]{PhysRevLett.111.185305}%
  \BibitemOpen
  \bibfield  {author} {\bibinfo {author} {\bibfnamefont {A.}~\bibnamefont
  {de~Paz}}, \bibinfo {author} {\bibfnamefont {A.}~\bibnamefont {Sharma}},
  \bibinfo {author} {\bibfnamefont {A.}~\bibnamefont {Chotia}}, \bibinfo
  {author} {\bibfnamefont {E.}~\bibnamefont {Mar\'echal}}, \bibinfo {author}
  {\bibfnamefont {J.~H.}\ \bibnamefont {Huckans}}, \bibinfo {author}
  {\bibfnamefont {P.}~\bibnamefont {Pedri}}, \bibinfo {author} {\bibfnamefont
  {L.}~\bibnamefont {Santos}}, \bibinfo {author} {\bibfnamefont
  {O.}~\bibnamefont {Gorceix}}, \bibinfo {author} {\bibfnamefont
  {L.}~\bibnamefont {Vernac}},\ and\ \bibinfo {author} {\bibfnamefont
  {B.}~\bibnamefont {Laburthe-Tolra}},\ }\bibfield  {title} {\bibinfo {title}
  {Nonequilibrium quantum magnetism in a dipolar lattice gas},\ }\href
  {https://doi.org/10.1103/PhysRevLett.111.185305} {\bibfield  {journal}
  {\bibinfo  {journal} {Phys. Rev. Lett.}\ }\textbf {\bibinfo {volume} {111}},\
  \bibinfo {pages} {185305} (\bibinfo {year} {2013})}\BibitemShut {NoStop}%
\bibitem [{\citenamefont {Li}\ \emph {et~al.}(2023)\citenamefont {Li},
  \citenamefont {Matsuda}, \citenamefont {Miller}, \citenamefont {Carroll},
  \citenamefont {Tobias}, \citenamefont {Higgins},\ and\ \citenamefont
  {Ye}}]{Li2023}%
  \BibitemOpen
  \bibfield  {author} {\bibinfo {author} {\bibfnamefont {J.-R.}\ \bibnamefont
  {Li}}, \bibinfo {author} {\bibfnamefont {K.}~\bibnamefont {Matsuda}},
  \bibinfo {author} {\bibfnamefont {C.}~\bibnamefont {Miller}}, \bibinfo
  {author} {\bibfnamefont {A.~N.}\ \bibnamefont {Carroll}}, \bibinfo {author}
  {\bibfnamefont {W.~G.}\ \bibnamefont {Tobias}}, \bibinfo {author}
  {\bibfnamefont {J.~S.}\ \bibnamefont {Higgins}},\ and\ \bibinfo {author}
  {\bibfnamefont {J.}~\bibnamefont {Ye}},\ }\bibfield  {title} {\bibinfo
  {title} {Tunable itinerant spin dynamics with polar molecules},\ }\href
  {https://doi.org/10.1038/s41586-022-05479-2} {\bibfield  {journal} {\bibinfo
  {journal} {Nature (London)}\ }\textbf {\bibinfo {volume} {614}},\ \bibinfo
  {pages} {70} (\bibinfo {year} {2023})}\BibitemShut {NoStop}%
\bibitem [{\citenamefont {Christakis}\ \emph {et~al.}(2023)\citenamefont
  {Christakis}, \citenamefont {Rosenberg}, \citenamefont {Raj}, \citenamefont
  {Chi}, \citenamefont {Morningstar}, \citenamefont {Huse}, \citenamefont
  {Yan},\ and\ \citenamefont {Bakr}}]{Christakis2023}%
  \BibitemOpen
  \bibfield  {author} {\bibinfo {author} {\bibfnamefont {L.}~\bibnamefont
  {Christakis}}, \bibinfo {author} {\bibfnamefont {J.~S.}\ \bibnamefont
  {Rosenberg}}, \bibinfo {author} {\bibfnamefont {R.}~\bibnamefont {Raj}},
  \bibinfo {author} {\bibfnamefont {S.}~\bibnamefont {Chi}}, \bibinfo {author}
  {\bibfnamefont {A.}~\bibnamefont {Morningstar}}, \bibinfo {author}
  {\bibfnamefont {D.~A.}\ \bibnamefont {Huse}}, \bibinfo {author}
  {\bibfnamefont {Z.~Z.}\ \bibnamefont {Yan}},\ and\ \bibinfo {author}
  {\bibfnamefont {W.~S.}\ \bibnamefont {Bakr}},\ }\bibfield  {title} {\bibinfo
  {title} {Probing site-resolved correlations in a spin system of ultracold
  molecules},\ }\href {https://doi.org/10.1038/s41586-022-05558-4} {\bibfield
  {journal} {\bibinfo  {journal} {Nature (London)}\ }\textbf {\bibinfo {volume}
  {614}},\ \bibinfo {pages} {64} (\bibinfo {year} {2023})}\BibitemShut
  {NoStop}%
\bibitem [{\citenamefont {Yao}\ \emph {et~al.}(2018)\citenamefont {Yao},
  \citenamefont {Zaletel}, \citenamefont {Stamper-Kurn},\ and\ \citenamefont
  {Vishwanath}}]{yao2018quantum}%
  \BibitemOpen
  \bibfield  {author} {\bibinfo {author} {\bibfnamefont {N.~Y.}\ \bibnamefont
  {Yao}}, \bibinfo {author} {\bibfnamefont {M.~P.}\ \bibnamefont {Zaletel}},
  \bibinfo {author} {\bibfnamefont {D.~M.}\ \bibnamefont {Stamper-Kurn}},\ and\
  \bibinfo {author} {\bibfnamefont {A.}~\bibnamefont {Vishwanath}},\ }\bibfield
   {title} {\bibinfo {title} {A quantum dipolar spin liquid},\ }\href@noop {}
  {\bibfield  {journal} {\bibinfo  {journal} {Nature Physics}\ }\textbf
  {\bibinfo {volume} {14}},\ \bibinfo {pages} {405} (\bibinfo {year}
  {2018})}\BibitemShut {NoStop}%
\bibitem [{\citenamefont {Yi}\ \emph {et~al.}(2007)\citenamefont {Yi},
  \citenamefont {Li},\ and\ \citenamefont {Sun}}]{PhysRevLett.98.260405}%
  \BibitemOpen
  \bibfield  {author} {\bibinfo {author} {\bibfnamefont {S.}~\bibnamefont
  {Yi}}, \bibinfo {author} {\bibfnamefont {T.}~\bibnamefont {Li}},\ and\
  \bibinfo {author} {\bibfnamefont {C.~P.}\ \bibnamefont {Sun}},\ }\bibfield
  {title} {\bibinfo {title} {Novel quantum phases of dipolar bose gases in
  optical lattices},\ }\href {https://doi.org/10.1103/PhysRevLett.98.260405}
  {\bibfield  {journal} {\bibinfo  {journal} {Phys. Rev. Lett.}\ }\textbf
  {\bibinfo {volume} {98}},\ \bibinfo {pages} {260405} (\bibinfo {year}
  {2007})}\BibitemShut {NoStop}%
\bibitem [{\citenamefont {Guo}\ \emph {et~al.}(2019)\citenamefont {Guo},
  \citenamefont {B{\"o}ttcher}, \citenamefont {Hertkorn}, \citenamefont
  {Schmidt}, \citenamefont {Wenzel}, \citenamefont {B{\"u}chler}, \citenamefont
  {Langen},\ and\ \citenamefont {Pfau}}]{guo2019low}%
  \BibitemOpen
  \bibfield  {author} {\bibinfo {author} {\bibfnamefont {M.}~\bibnamefont
  {Guo}}, \bibinfo {author} {\bibfnamefont {F.}~\bibnamefont {B{\"o}ttcher}},
  \bibinfo {author} {\bibfnamefont {J.}~\bibnamefont {Hertkorn}}, \bibinfo
  {author} {\bibfnamefont {J.-N.}\ \bibnamefont {Schmidt}}, \bibinfo {author}
  {\bibfnamefont {M.}~\bibnamefont {Wenzel}}, \bibinfo {author} {\bibfnamefont
  {H.~P.}\ \bibnamefont {B{\"u}chler}}, \bibinfo {author} {\bibfnamefont
  {T.}~\bibnamefont {Langen}},\ and\ \bibinfo {author} {\bibfnamefont
  {T.}~\bibnamefont {Pfau}},\ }\bibfield  {title} {\bibinfo {title} {The
  low-energy goldstone mode in a trapped dipolar supersolid},\ }\href@noop {}
  {\bibfield  {journal} {\bibinfo  {journal} {Nature}\ }\textbf {\bibinfo
  {volume} {574}},\ \bibinfo {pages} {386} (\bibinfo {year}
  {2019})}\BibitemShut {NoStop}%
\bibitem [{\citenamefont {Tanzi}\ \emph {et~al.}(2019)\citenamefont {Tanzi},
  \citenamefont {Lucioni}, \citenamefont {Fam\`a}, \citenamefont {Catani},
  \citenamefont {Fioretti}, \citenamefont {Gabbanini}, \citenamefont {Bisset},
  \citenamefont {Santos},\ and\ \citenamefont
  {Modugno}}]{PhysRevLett.122.130405}%
  \BibitemOpen
  \bibfield  {author} {\bibinfo {author} {\bibfnamefont {L.}~\bibnamefont
  {Tanzi}}, \bibinfo {author} {\bibfnamefont {E.}~\bibnamefont {Lucioni}},
  \bibinfo {author} {\bibfnamefont {F.}~\bibnamefont {Fam\`a}}, \bibinfo
  {author} {\bibfnamefont {J.}~\bibnamefont {Catani}}, \bibinfo {author}
  {\bibfnamefont {A.}~\bibnamefont {Fioretti}}, \bibinfo {author}
  {\bibfnamefont {C.}~\bibnamefont {Gabbanini}}, \bibinfo {author}
  {\bibfnamefont {R.~N.}\ \bibnamefont {Bisset}}, \bibinfo {author}
  {\bibfnamefont {L.}~\bibnamefont {Santos}},\ and\ \bibinfo {author}
  {\bibfnamefont {G.}~\bibnamefont {Modugno}},\ }\bibfield  {title} {\bibinfo
  {title} {Observation of a dipolar quantum gas with metastable supersolid
  properties},\ }\href {https://doi.org/10.1103/PhysRevLett.122.130405}
  {\bibfield  {journal} {\bibinfo  {journal} {Phys. Rev. Lett.}\ }\textbf
  {\bibinfo {volume} {122}},\ \bibinfo {pages} {130405} (\bibinfo {year}
  {2019})}\BibitemShut {NoStop}%
\bibitem [{\citenamefont {Bland}\ \emph {et~al.}(2022)\citenamefont {Bland},
  \citenamefont {Poli}, \citenamefont {Politi}, \citenamefont {Klaus},
  \citenamefont {Norcia}, \citenamefont {Ferlaino}, \citenamefont {Santos},\
  and\ \citenamefont {Bisset}}]{PhysRevLett.128.195302}%
  \BibitemOpen
  \bibfield  {author} {\bibinfo {author} {\bibfnamefont {T.}~\bibnamefont
  {Bland}}, \bibinfo {author} {\bibfnamefont {E.}~\bibnamefont {Poli}},
  \bibinfo {author} {\bibfnamefont {C.}~\bibnamefont {Politi}}, \bibinfo
  {author} {\bibfnamefont {L.}~\bibnamefont {Klaus}}, \bibinfo {author}
  {\bibfnamefont {M.~A.}\ \bibnamefont {Norcia}}, \bibinfo {author}
  {\bibfnamefont {F.}~\bibnamefont {Ferlaino}}, \bibinfo {author}
  {\bibfnamefont {L.}~\bibnamefont {Santos}},\ and\ \bibinfo {author}
  {\bibfnamefont {R.~N.}\ \bibnamefont {Bisset}},\ }\bibfield  {title}
  {\bibinfo {title} {Two-dimensional supersolid formation in dipolar
  condensates},\ }\href {https://doi.org/10.1103/PhysRevLett.128.195302}
  {\bibfield  {journal} {\bibinfo  {journal} {Phys. Rev. Lett.}\ }\textbf
  {\bibinfo {volume} {128}},\ \bibinfo {pages} {195302} (\bibinfo {year}
  {2022})}\BibitemShut {NoStop}%
\bibitem [{\citenamefont {Chomaz}\ \emph {et~al.}(2019)\citenamefont {Chomaz},
  \citenamefont {Petter}, \citenamefont {Ilzh\"ofer}, \citenamefont {Natale},
  \citenamefont {Trautmann}, \citenamefont {Politi}, \citenamefont
  {Durastante}, \citenamefont {van Bijnen}, \citenamefont {Patscheider},
  \citenamefont {Sohmen}, \citenamefont {Mark},\ and\ \citenamefont
  {Ferlaino}}]{PhysRevX.9.021012}%
  \BibitemOpen
  \bibfield  {author} {\bibinfo {author} {\bibfnamefont {L.}~\bibnamefont
  {Chomaz}}, \bibinfo {author} {\bibfnamefont {D.}~\bibnamefont {Petter}},
  \bibinfo {author} {\bibfnamefont {P.}~\bibnamefont {Ilzh\"ofer}}, \bibinfo
  {author} {\bibfnamefont {G.}~\bibnamefont {Natale}}, \bibinfo {author}
  {\bibfnamefont {A.}~\bibnamefont {Trautmann}}, \bibinfo {author}
  {\bibfnamefont {C.}~\bibnamefont {Politi}}, \bibinfo {author} {\bibfnamefont
  {G.}~\bibnamefont {Durastante}}, \bibinfo {author} {\bibfnamefont {R.~M.~W.}\
  \bibnamefont {van Bijnen}}, \bibinfo {author} {\bibfnamefont
  {A.}~\bibnamefont {Patscheider}}, \bibinfo {author} {\bibfnamefont
  {M.}~\bibnamefont {Sohmen}}, \bibinfo {author} {\bibfnamefont {M.~J.}\
  \bibnamefont {Mark}},\ and\ \bibinfo {author} {\bibfnamefont
  {F.}~\bibnamefont {Ferlaino}},\ }\bibfield  {title} {\bibinfo {title}
  {Long-lived and transient supersolid behaviors in dipolar quantum gases},\
  }\href {https://doi.org/10.1103/PhysRevX.9.021012} {\bibfield  {journal}
  {\bibinfo  {journal} {Phys. Rev. X}\ }\textbf {\bibinfo {volume} {9}},\
  \bibinfo {pages} {021012} (\bibinfo {year} {2019})}\BibitemShut {NoStop}%
\bibitem [{\citenamefont {Ferrier-Barbut}\ \emph {et~al.}(2016)\citenamefont
  {Ferrier-Barbut}, \citenamefont {Kadau}, \citenamefont {Schmitt},
  \citenamefont {Wenzel},\ and\ \citenamefont {Pfau}}]{PhysRevLett.116.215301}%
  \BibitemOpen
  \bibfield  {author} {\bibinfo {author} {\bibfnamefont {I.}~\bibnamefont
  {Ferrier-Barbut}}, \bibinfo {author} {\bibfnamefont {H.}~\bibnamefont
  {Kadau}}, \bibinfo {author} {\bibfnamefont {M.}~\bibnamefont {Schmitt}},
  \bibinfo {author} {\bibfnamefont {M.}~\bibnamefont {Wenzel}},\ and\ \bibinfo
  {author} {\bibfnamefont {T.}~\bibnamefont {Pfau}},\ }\bibfield  {title}
  {\bibinfo {title} {Observation of quantum droplets in a strongly dipolar bose
  gas},\ }\href {https://doi.org/10.1103/PhysRevLett.116.215301} {\bibfield
  {journal} {\bibinfo  {journal} {Phys. Rev. Lett.}\ }\textbf {\bibinfo
  {volume} {116}},\ \bibinfo {pages} {215301} (\bibinfo {year}
  {2016})}\BibitemShut {NoStop}%
\bibitem [{\citenamefont {Chomaz}\ \emph {et~al.}(2016)\citenamefont {Chomaz},
  \citenamefont {Baier}, \citenamefont {Petter}, \citenamefont {Mark},
  \citenamefont {W\"achtler}, \citenamefont {Santos},\ and\ \citenamefont
  {Ferlaino}}]{PhysRevX.6.041039}%
  \BibitemOpen
  \bibfield  {author} {\bibinfo {author} {\bibfnamefont {L.}~\bibnamefont
  {Chomaz}}, \bibinfo {author} {\bibfnamefont {S.}~\bibnamefont {Baier}},
  \bibinfo {author} {\bibfnamefont {D.}~\bibnamefont {Petter}}, \bibinfo
  {author} {\bibfnamefont {M.~J.}\ \bibnamefont {Mark}}, \bibinfo {author}
  {\bibfnamefont {F.}~\bibnamefont {W\"achtler}}, \bibinfo {author}
  {\bibfnamefont {L.}~\bibnamefont {Santos}},\ and\ \bibinfo {author}
  {\bibfnamefont {F.}~\bibnamefont {Ferlaino}},\ }\bibfield  {title} {\bibinfo
  {title} {Quantum-fluctuation-driven crossover from a dilute bose-einstein
  condensate to a macrodroplet in a dipolar quantum fluid},\ }\href
  {https://doi.org/10.1103/PhysRevX.6.041039} {\bibfield  {journal} {\bibinfo
  {journal} {Phys. Rev. X}\ }\textbf {\bibinfo {volume} {6}},\ \bibinfo {pages}
  {041039} (\bibinfo {year} {2016})}\BibitemShut {NoStop}%
\bibitem [{\citenamefont {Schmitt}\ \emph {et~al.}(2016)\citenamefont
  {Schmitt}, \citenamefont {Wenzel}, \citenamefont {B{\"o}ttcher},
  \citenamefont {Ferrier-Barbut},\ and\ \citenamefont
  {Pfau}}]{schmitt2016self}%
  \BibitemOpen
  \bibfield  {author} {\bibinfo {author} {\bibfnamefont {M.}~\bibnamefont
  {Schmitt}}, \bibinfo {author} {\bibfnamefont {M.}~\bibnamefont {Wenzel}},
  \bibinfo {author} {\bibfnamefont {F.}~\bibnamefont {B{\"o}ttcher}}, \bibinfo
  {author} {\bibfnamefont {I.}~\bibnamefont {Ferrier-Barbut}},\ and\ \bibinfo
  {author} {\bibfnamefont {T.}~\bibnamefont {Pfau}},\ }\bibfield  {title}
  {\bibinfo {title} {Self-bound droplets of a dilute magnetic quantum liquid},\
  }\href@noop {} {\bibfield  {journal} {\bibinfo  {journal} {Nature}\ }\textbf
  {\bibinfo {volume} {539}},\ \bibinfo {pages} {259} (\bibinfo {year}
  {2016})}\BibitemShut {NoStop}%
\bibitem [{\citenamefont {Santos}\ \emph {et~al.}(2003)\citenamefont {Santos},
  \citenamefont {Shlyapnikov},\ and\ \citenamefont
  {Lewenstein}}]{PhysRevLett.90.250403}%
  \BibitemOpen
  \bibfield  {author} {\bibinfo {author} {\bibfnamefont {L.}~\bibnamefont
  {Santos}}, \bibinfo {author} {\bibfnamefont {G.~V.}\ \bibnamefont
  {Shlyapnikov}},\ and\ \bibinfo {author} {\bibfnamefont {M.}~\bibnamefont
  {Lewenstein}},\ }\bibfield  {title} {\bibinfo {title} {Roton-maxon spectrum
  and stability of trapped dipolar bose-einstein condensates},\ }\href
  {https://doi.org/10.1103/PhysRevLett.90.250403} {\bibfield  {journal}
  {\bibinfo  {journal} {Phys. Rev. Lett.}\ }\textbf {\bibinfo {volume} {90}},\
  \bibinfo {pages} {250403} (\bibinfo {year} {2003})}\BibitemShut {NoStop}%
\bibitem [{\citenamefont {Ritsch}\ \emph {et~al.}(2013)\citenamefont {Ritsch},
  \citenamefont {Domokos}, \citenamefont {Brennecke},\ and\ \citenamefont
  {Esslinger}}]{RevModPhys.85.553}%
  \BibitemOpen
  \bibfield  {author} {\bibinfo {author} {\bibfnamefont {H.}~\bibnamefont
  {Ritsch}}, \bibinfo {author} {\bibfnamefont {P.}~\bibnamefont {Domokos}},
  \bibinfo {author} {\bibfnamefont {F.}~\bibnamefont {Brennecke}},\ and\
  \bibinfo {author} {\bibfnamefont {T.}~\bibnamefont {Esslinger}},\ }\bibfield
  {title} {\bibinfo {title} {Cold atoms in cavity-generated dynamical optical
  potentials},\ }\href {https://doi.org/10.1103/RevModPhys.85.553} {\bibfield
  {journal} {\bibinfo  {journal} {Rev. Mod. Phys.}\ }\textbf {\bibinfo {volume}
  {85}},\ \bibinfo {pages} {553} (\bibinfo {year} {2013})}\BibitemShut
  {NoStop}%
\bibitem [{\citenamefont {Defenu}\ \emph {et~al.}(2023)\citenamefont {Defenu},
  \citenamefont {Donner}, \citenamefont {Macr\`{\i}}, \citenamefont {Pagano},
  \citenamefont {Ruffo},\ and\ \citenamefont
  {Trombettoni}}]{RevModPhys.95.035002}%
  \BibitemOpen
  \bibfield  {author} {\bibinfo {author} {\bibfnamefont {N.}~\bibnamefont
  {Defenu}}, \bibinfo {author} {\bibfnamefont {T.}~\bibnamefont {Donner}},
  \bibinfo {author} {\bibfnamefont {T.}~\bibnamefont {Macr\`{\i}}}, \bibinfo
  {author} {\bibfnamefont {G.}~\bibnamefont {Pagano}}, \bibinfo {author}
  {\bibfnamefont {S.}~\bibnamefont {Ruffo}},\ and\ \bibinfo {author}
  {\bibfnamefont {A.}~\bibnamefont {Trombettoni}},\ }\bibfield  {title}
  {\bibinfo {title} {Long-range interacting quantum systems},\ }\href
  {https://doi.org/10.1103/RevModPhys.95.035002} {\bibfield  {journal}
  {\bibinfo  {journal} {Rev. Mod. Phys.}\ }\textbf {\bibinfo {volume} {95}},\
  \bibinfo {pages} {035002} (\bibinfo {year} {2023})}\BibitemShut {NoStop}%
\bibitem [{\citenamefont {Mivehvar}\ \emph {et~al.}(2021)\citenamefont
  {Mivehvar}, \citenamefont {Piazza}, \citenamefont {Donner},\ and\
  \citenamefont {Ritsch}}]{Ritsch_2021}%
  \BibitemOpen
  \bibfield  {author} {\bibinfo {author} {\bibfnamefont {F.}~\bibnamefont
  {Mivehvar}}, \bibinfo {author} {\bibfnamefont {F.}~\bibnamefont {Piazza}},
  \bibinfo {author} {\bibfnamefont {T.}~\bibnamefont {Donner}},\ and\ \bibinfo
  {author} {\bibfnamefont {H.}~\bibnamefont {Ritsch}},\ }\bibfield  {title}
  {\bibinfo {title} {Cavity qed with quantum gases: new paradigms in many-body
  physics},\ }\href {https://doi.org/10.1080/00018732.2021.1969727} {\bibfield
  {journal} {\bibinfo  {journal} {Adv. Phys}\ }\textbf {\bibinfo {volume}
  {70}},\ \bibinfo {pages} {1} (\bibinfo {year} {2021})}\BibitemShut {NoStop}%
\bibitem [{\citenamefont {Forn-D\'{\i}az}\ \emph {et~al.}(2019)\citenamefont
  {Forn-D\'{\i}az}, \citenamefont {Lamata}, \citenamefont {Rico}, \citenamefont
  {Kono},\ and\ \citenamefont {Solano}}]{RevModPhys.91.025005}%
  \BibitemOpen
  \bibfield  {author} {\bibinfo {author} {\bibfnamefont {P.}~\bibnamefont
  {Forn-D\'{\i}az}}, \bibinfo {author} {\bibfnamefont {L.}~\bibnamefont
  {Lamata}}, \bibinfo {author} {\bibfnamefont {E.}~\bibnamefont {Rico}},
  \bibinfo {author} {\bibfnamefont {J.}~\bibnamefont {Kono}},\ and\ \bibinfo
  {author} {\bibfnamefont {E.}~\bibnamefont {Solano}},\ }\bibfield  {title}
  {\bibinfo {title} {Ultrastrong coupling regimes of light-matter
  interaction},\ }\href {https://doi.org/10.1103/RevModPhys.91.025005}
  {\bibfield  {journal} {\bibinfo  {journal} {Rev. Mod. Phys.}\ }\textbf
  {\bibinfo {volume} {91}},\ \bibinfo {pages} {025005} (\bibinfo {year}
  {2019})}\BibitemShut {NoStop}%
\bibitem [{\citenamefont {Deng}\ \emph {et~al.}(2014)\citenamefont {Deng},
  \citenamefont {Cheng}, \citenamefont {Jing},\ and\ \citenamefont
  {Yi}}]{PhysRevLett.112.143007}%
  \BibitemOpen
  \bibfield  {author} {\bibinfo {author} {\bibfnamefont {Y.}~\bibnamefont
  {Deng}}, \bibinfo {author} {\bibfnamefont {J.}~\bibnamefont {Cheng}},
  \bibinfo {author} {\bibfnamefont {H.}~\bibnamefont {Jing}},\ and\ \bibinfo
  {author} {\bibfnamefont {S.}~\bibnamefont {Yi}},\ }\bibfield  {title}
  {\bibinfo {title} {Bose-einstein condensates with cavity-mediated spin-orbit
  coupling},\ }\href {https://doi.org/10.1103/PhysRevLett.112.143007}
  {\bibfield  {journal} {\bibinfo  {journal} {Phys. Rev. Lett.}\ }\textbf
  {\bibinfo {volume} {112}},\ \bibinfo {pages} {143007} (\bibinfo {year}
  {2014})}\BibitemShut {NoStop}%
\bibitem [{\citenamefont {Kroeze}\ \emph {et~al.}(2018)\citenamefont {Kroeze},
  \citenamefont {Guo}, \citenamefont {Vaidya}, \citenamefont {Keeling},\ and\
  \citenamefont {Lev}}]{PhysRevLett.121.163601}%
  \BibitemOpen
  \bibfield  {author} {\bibinfo {author} {\bibfnamefont {R.~M.}\ \bibnamefont
  {Kroeze}}, \bibinfo {author} {\bibfnamefont {Y.}~\bibnamefont {Guo}},
  \bibinfo {author} {\bibfnamefont {V.~D.}\ \bibnamefont {Vaidya}}, \bibinfo
  {author} {\bibfnamefont {J.}~\bibnamefont {Keeling}},\ and\ \bibinfo {author}
  {\bibfnamefont {B.~L.}\ \bibnamefont {Lev}},\ }\bibfield  {title} {\bibinfo
  {title} {Spinor self-ordering of a quantum gas in a cavity},\ }\href
  {https://doi.org/10.1103/PhysRevLett.121.163601} {\bibfield  {journal}
  {\bibinfo  {journal} {Phys. Rev. Lett.}\ }\textbf {\bibinfo {volume} {121}},\
  \bibinfo {pages} {163601} (\bibinfo {year} {2018})}\BibitemShut {NoStop}%
\bibitem [{\citenamefont {Kroeze}\ \emph {et~al.}(2019)\citenamefont {Kroeze},
  \citenamefont {Guo},\ and\ \citenamefont {Lev}}]{PhysRevLett.123.160404}%
  \BibitemOpen
  \bibfield  {author} {\bibinfo {author} {\bibfnamefont {R.~M.}\ \bibnamefont
  {Kroeze}}, \bibinfo {author} {\bibfnamefont {Y.}~\bibnamefont {Guo}},\ and\
  \bibinfo {author} {\bibfnamefont {B.~L.}\ \bibnamefont {Lev}},\ }\bibfield
  {title} {\bibinfo {title} {Dynamical spin-orbit coupling of a quantum gas},\
  }\href {https://doi.org/10.1103/PhysRevLett.123.160404} {\bibfield  {journal}
  {\bibinfo  {journal} {Phys. Rev. Lett.}\ }\textbf {\bibinfo {volume} {123}},\
  \bibinfo {pages} {160404} (\bibinfo {year} {2019})}\BibitemShut {NoStop}%
\bibitem [{\citenamefont {Zhang}\ \emph {et~al.}(2021)\citenamefont {Zhang},
  \citenamefont {Chen}, \citenamefont {Wu}, \citenamefont {Wang}, \citenamefont
  {Fan}, \citenamefont {j.~Deng},\ and\ \citenamefont
  {Wu}}]{doi:10.1126/science.abd4385}%
  \BibitemOpen
  \bibfield  {author} {\bibinfo {author} {\bibfnamefont {X.-T.}\ \bibnamefont
  {Zhang}}, \bibinfo {author} {\bibfnamefont {Y.}~\bibnamefont {Chen}},
  \bibinfo {author} {\bibfnamefont {Z.-M.}\ \bibnamefont {Wu}}, \bibinfo
  {author} {\bibfnamefont {J.}~\bibnamefont {Wang}}, \bibinfo {author}
  {\bibfnamefont {J.-J.}\ \bibnamefont {Fan}}, \bibinfo {author} {\bibfnamefont
  {S.}~\bibnamefont {j.~Deng}},\ and\ \bibinfo {author} {\bibfnamefont {H.-B.}\
  \bibnamefont {Wu}},\ }\bibfield  {title} {\bibinfo {title} {Observation of a
  superradiant quantum phase transition in an intracavity degenerate fermi
  gas},\ }\href {https://doi.org/10.1126/science.abd4385} {\bibfield  {journal}
  {\bibinfo  {journal} {Science}\ }\textbf {\bibinfo {volume} {373}},\ \bibinfo
  {pages} {1359} (\bibinfo {year} {2021})}\BibitemShut {NoStop}%
\bibitem [{\citenamefont {Baumann}\ \emph {et~al.}(2010)\citenamefont
  {Baumann}, \citenamefont {Guerlin}, \citenamefont {Brennecke},\ and\
  \citenamefont {Esslinger}}]{Baumann2010}%
  \BibitemOpen
  \bibfield  {author} {\bibinfo {author} {\bibfnamefont {K.}~\bibnamefont
  {Baumann}}, \bibinfo {author} {\bibfnamefont {C.}~\bibnamefont {Guerlin}},
  \bibinfo {author} {\bibfnamefont {F.}~\bibnamefont {Brennecke}},\ and\
  \bibinfo {author} {\bibfnamefont {T.}~\bibnamefont {Esslinger}},\ }\bibfield
  {title} {\bibinfo {title} {Dicke quantum phase transition with a superfluid
  gas in an optical cavity},\ }\href {https://doi.org/10.1038/nature09009}
  {\bibfield  {journal} {\bibinfo  {journal} {Nature (London)}\ }\textbf
  {\bibinfo {volume} {464}},\ \bibinfo {pages} {1301} (\bibinfo {year}
  {2010})}\BibitemShut {NoStop}%
\bibitem [{\citenamefont {Mottl}\ \emph {et~al.}(2012)\citenamefont {Mottl},
  \citenamefont {Brennecke}, \citenamefont {Baumann}, \citenamefont {Landig},
  \citenamefont {Donner},\ and\ \citenamefont
  {Esslinger}}]{doi:10.1126/science.1220314}%
  \BibitemOpen
  \bibfield  {author} {\bibinfo {author} {\bibfnamefont {R.}~\bibnamefont
  {Mottl}}, \bibinfo {author} {\bibfnamefont {F.}~\bibnamefont {Brennecke}},
  \bibinfo {author} {\bibfnamefont {K.}~\bibnamefont {Baumann}}, \bibinfo
  {author} {\bibfnamefont {R.}~\bibnamefont {Landig}}, \bibinfo {author}
  {\bibfnamefont {T.}~\bibnamefont {Donner}},\ and\ \bibinfo {author}
  {\bibfnamefont {T.}~\bibnamefont {Esslinger}},\ }\bibfield  {title} {\bibinfo
  {title} {Roton-type mode softening in a quantum gas with cavity-mediated
  long-range interactions},\ }\href {https://doi.org/10.1126/science.1220314}
  {\bibfield  {journal} {\bibinfo  {journal} {Science}\ }\textbf {\bibinfo
  {volume} {336}},\ \bibinfo {pages} {1570} (\bibinfo {year}
  {2012})}\BibitemShut {NoStop}%
\bibitem [{\citenamefont {Deng}\ and\ \citenamefont
  {Yi}(2023)}]{PhysRevResearch.5.013002}%
  \BibitemOpen
  \bibfield  {author} {\bibinfo {author} {\bibfnamefont {Y.-G.}\ \bibnamefont
  {Deng}}\ and\ \bibinfo {author} {\bibfnamefont {S.}~\bibnamefont {Yi}},\
  }\bibfield  {title} {\bibinfo {title} {Self-ordered supersolid phase beyond
  dicke superradiance in a ring cavity},\ }\href
  {https://doi.org/10.1103/PhysRevResearch.5.013002} {\bibfield  {journal}
  {\bibinfo  {journal} {Phys. Rev. Res.}\ }\textbf {\bibinfo {volume} {5}},\
  \bibinfo {pages} {013002} (\bibinfo {year} {2023})}\BibitemShut {NoStop}%
\bibitem [{\citenamefont {L\'{e}onard}\ \emph {et~al.}(2017)\citenamefont
  {L\'{e}onard}, \citenamefont {Morales}, \citenamefont {Zupancic},
  \citenamefont {Esslinger},\ and\ \citenamefont {Donner}}]{Leonard2017}%
  \BibitemOpen
  \bibfield  {author} {\bibinfo {author} {\bibfnamefont {J.}~\bibnamefont
  {L\'{e}onard}}, \bibinfo {author} {\bibfnamefont {A.}~\bibnamefont
  {Morales}}, \bibinfo {author} {\bibfnamefont {P.}~\bibnamefont {Zupancic}},
  \bibinfo {author} {\bibfnamefont {T.}~\bibnamefont {Esslinger}},\ and\
  \bibinfo {author} {\bibfnamefont {T.}~\bibnamefont {Donner}},\ }\bibfield
  {title} {\bibinfo {title} {Supersolid formation in a quantum gas breaking a
  continuous translational symmetry},\ }\href
  {https://doi.org/10.1038/nature21067} {\bibfield  {journal} {\bibinfo
  {journal} {Nature (London)}\ }\textbf {\bibinfo {volume} {543}},\ \bibinfo
  {pages} {87} (\bibinfo {year} {2017})}\BibitemShut {NoStop}%
\bibitem [{\citenamefont {L{\'e}onard}\ \emph {et~al.}(2017)\citenamefont
  {L{\'e}onard}, \citenamefont {Morales}, \citenamefont {Zupancic},
  \citenamefont {Donner},\ and\ \citenamefont
  {Esslinger}}]{leonard2017monitoring}%
  \BibitemOpen
  \bibfield  {author} {\bibinfo {author} {\bibfnamefont {J.}~\bibnamefont
  {L{\'e}onard}}, \bibinfo {author} {\bibfnamefont {A.}~\bibnamefont
  {Morales}}, \bibinfo {author} {\bibfnamefont {P.}~\bibnamefont {Zupancic}},
  \bibinfo {author} {\bibfnamefont {T.}~\bibnamefont {Donner}},\ and\ \bibinfo
  {author} {\bibfnamefont {T.}~\bibnamefont {Esslinger}},\ }\bibfield  {title}
  {\bibinfo {title} {Monitoring and manipulating higgs and goldstone modes in a
  supersolid quantum gas},\ }\href@noop {} {\bibfield  {journal} {\bibinfo
  {journal} {Science}\ }\textbf {\bibinfo {volume} {358}},\ \bibinfo {pages}
  {1415} (\bibinfo {year} {2017})}\BibitemShut {NoStop}%
\bibitem [{\citenamefont {Finger}\ \emph {et~al.}(2024)\citenamefont {Finger},
  \citenamefont {Rosa-Medina}, \citenamefont {Reiter}, \citenamefont
  {Christodoulou}, \citenamefont {Donner},\ and\ \citenamefont
  {Esslinger}}]{PhysRevLett.132.093402}%
  \BibitemOpen
  \bibfield  {author} {\bibinfo {author} {\bibfnamefont {F.}~\bibnamefont
  {Finger}}, \bibinfo {author} {\bibfnamefont {R.}~\bibnamefont {Rosa-Medina}},
  \bibinfo {author} {\bibfnamefont {N.}~\bibnamefont {Reiter}}, \bibinfo
  {author} {\bibfnamefont {P.}~\bibnamefont {Christodoulou}}, \bibinfo {author}
  {\bibfnamefont {T.}~\bibnamefont {Donner}},\ and\ \bibinfo {author}
  {\bibfnamefont {T.}~\bibnamefont {Esslinger}},\ }\bibfield  {title} {\bibinfo
  {title} {Spin- and momentum-correlated atom pairs mediated by photon exchange
  and seeded by vacuum fluctuations},\ }\href
  {https://doi.org/10.1103/PhysRevLett.132.093402} {\bibfield  {journal}
  {\bibinfo  {journal} {Phys. Rev. Lett.}\ }\textbf {\bibinfo {volume} {132}},\
  \bibinfo {pages} {093402} (\bibinfo {year} {2024})}\BibitemShut {NoStop}%
\bibitem [{\citenamefont {Muniz}\ \emph {et~al.}(2020)\citenamefont {Muniz},
  \citenamefont {Barberena}, \citenamefont {Lewis-Swan}, \citenamefont {Young},
  \citenamefont {Cline}, \citenamefont {Rey},\ and\ \citenamefont
  {Thompson}}]{muniz2020exploring}%
  \BibitemOpen
  \bibfield  {author} {\bibinfo {author} {\bibfnamefont {J.~A.}\ \bibnamefont
  {Muniz}}, \bibinfo {author} {\bibfnamefont {D.}~\bibnamefont {Barberena}},
  \bibinfo {author} {\bibfnamefont {R.~J.}\ \bibnamefont {Lewis-Swan}},
  \bibinfo {author} {\bibfnamefont {D.~J.}\ \bibnamefont {Young}}, \bibinfo
  {author} {\bibfnamefont {J.~R.}\ \bibnamefont {Cline}}, \bibinfo {author}
  {\bibfnamefont {A.~M.}\ \bibnamefont {Rey}},\ and\ \bibinfo {author}
  {\bibfnamefont {J.~K.}\ \bibnamefont {Thompson}},\ }\bibfield  {title}
  {\bibinfo {title} {Exploring dynamical phase transitions with cold atoms in
  an optical cavity},\ }\href@noop {} {\bibfield  {journal} {\bibinfo
  {journal} {Nature}\ }\textbf {\bibinfo {volume} {580}},\ \bibinfo {pages}
  {602} (\bibinfo {year} {2020})}\BibitemShut {NoStop}%
\bibitem [{\citenamefont {Landig}\ \emph {et~al.}(2016)\citenamefont {Landig},
  \citenamefont {Hruby}, \citenamefont {Dogra}, \citenamefont {Landini},
  \citenamefont {Mottl}, \citenamefont {Donner},\ and\ \citenamefont
  {Esslinger}}]{landig2016quantum}%
  \BibitemOpen
  \bibfield  {author} {\bibinfo {author} {\bibfnamefont {R.}~\bibnamefont
  {Landig}}, \bibinfo {author} {\bibfnamefont {L.}~\bibnamefont {Hruby}},
  \bibinfo {author} {\bibfnamefont {N.}~\bibnamefont {Dogra}}, \bibinfo
  {author} {\bibfnamefont {M.}~\bibnamefont {Landini}}, \bibinfo {author}
  {\bibfnamefont {R.}~\bibnamefont {Mottl}}, \bibinfo {author} {\bibfnamefont
  {T.}~\bibnamefont {Donner}},\ and\ \bibinfo {author} {\bibfnamefont
  {T.}~\bibnamefont {Esslinger}},\ }\bibfield  {title} {\bibinfo {title}
  {Quantum phases from competing short-and long-range interactions in an
  optical lattice},\ }\href@noop {} {\bibfield  {journal} {\bibinfo  {journal}
  {Nature}\ }\textbf {\bibinfo {volume} {532}},\ \bibinfo {pages} {476}
  (\bibinfo {year} {2016})}\BibitemShut {NoStop}%
\bibitem [{\citenamefont {Defenu}\ \emph {et~al.}(2024)\citenamefont {Defenu},
  \citenamefont {Lerose},\ and\ \citenamefont {Pappalardi}}]{defenu2024out}%
  \BibitemOpen
  \bibfield  {author} {\bibinfo {author} {\bibfnamefont {N.}~\bibnamefont
  {Defenu}}, \bibinfo {author} {\bibfnamefont {A.}~\bibnamefont {Lerose}},\
  and\ \bibinfo {author} {\bibfnamefont {S.}~\bibnamefont {Pappalardi}},\
  }\bibfield  {title} {\bibinfo {title} {Out-of-equilibrium dynamics of quantum
  many-body systems with long-range interactions},\ }\href@noop {} {\bibfield
  {journal} {\bibinfo  {journal} {Physics Reports}\ }\textbf {\bibinfo {volume}
  {1074}},\ \bibinfo {pages} {1} (\bibinfo {year} {2024})}\BibitemShut
  {NoStop}%
\bibitem [{\citenamefont {Young}\ \emph {et~al.}(2024)\citenamefont {Young},
  \citenamefont {Chu}, \citenamefont {Song}, \citenamefont {Barberena},
  \citenamefont {Wellnitz}, \citenamefont {Niu}, \citenamefont {Sch{\"a}fer},
  \citenamefont {Lewis-Swan}, \citenamefont {Rey},\ and\ \citenamefont
  {Thompson}}]{young2024observing}%
  \BibitemOpen
  \bibfield  {author} {\bibinfo {author} {\bibfnamefont {D.~J.}\ \bibnamefont
  {Young}}, \bibinfo {author} {\bibfnamefont {A.}~\bibnamefont {Chu}}, \bibinfo
  {author} {\bibfnamefont {E.~Y.}\ \bibnamefont {Song}}, \bibinfo {author}
  {\bibfnamefont {D.}~\bibnamefont {Barberena}}, \bibinfo {author}
  {\bibfnamefont {D.}~\bibnamefont {Wellnitz}}, \bibinfo {author}
  {\bibfnamefont {Z.}~\bibnamefont {Niu}}, \bibinfo {author} {\bibfnamefont
  {V.~M.}\ \bibnamefont {Sch{\"a}fer}}, \bibinfo {author} {\bibfnamefont
  {R.~J.}\ \bibnamefont {Lewis-Swan}}, \bibinfo {author} {\bibfnamefont
  {A.~M.}\ \bibnamefont {Rey}},\ and\ \bibinfo {author} {\bibfnamefont {J.~K.}\
  \bibnamefont {Thompson}},\ }\bibfield  {title} {\bibinfo {title} {Observing
  dynamical phases of bcs superconductors in a cavity qed simulator},\
  }\href@noop {} {\bibfield  {journal} {\bibinfo  {journal} {Nature}\ }\textbf
  {\bibinfo {volume} {625}},\ \bibinfo {pages} {679} (\bibinfo {year}
  {2024})}\BibitemShut {NoStop}%
\bibitem [{\citenamefont {Nairn}\ \emph {et~al.}(2025)\citenamefont {Nairn},
  \citenamefont {Giannelli}, \citenamefont {Morigi}, \citenamefont {Slama},
  \citenamefont {Olmos},\ and\ \citenamefont
  {J\"ager}}]{PhysRevLett.134.083603}%
  \BibitemOpen
  \bibfield  {author} {\bibinfo {author} {\bibfnamefont {M.}~\bibnamefont
  {Nairn}}, \bibinfo {author} {\bibfnamefont {L.}~\bibnamefont {Giannelli}},
  \bibinfo {author} {\bibfnamefont {G.}~\bibnamefont {Morigi}}, \bibinfo
  {author} {\bibfnamefont {S.}~\bibnamefont {Slama}}, \bibinfo {author}
  {\bibfnamefont {B.}~\bibnamefont {Olmos}},\ and\ \bibinfo {author}
  {\bibfnamefont {S.~B.}\ \bibnamefont {J\"ager}},\ }\bibfield  {title}
  {\bibinfo {title} {Spin self-organization in an optical cavity facilitated by
  inhomogeneous broadening},\ }\href
  {https://doi.org/10.1103/PhysRevLett.134.083603} {\bibfield  {journal}
  {\bibinfo  {journal} {Phys. Rev. Lett.}\ }\textbf {\bibinfo {volume} {134}},\
  \bibinfo {pages} {083603} (\bibinfo {year} {2025})}\BibitemShut {NoStop}%
\bibitem [{\citenamefont {Deng}\ and\ \citenamefont {Yi}(2015)}]{Deng2015}%
  \BibitemOpen
  \bibfield  {author} {\bibinfo {author} {\bibfnamefont {Y.}~\bibnamefont
  {Deng}}\ and\ \bibinfo {author} {\bibfnamefont {S.}~\bibnamefont {Yi}},\
  }\bibfield  {title} {\bibinfo {title} {Spinor bose-einstein condensates of
  rotating polar molecules},\ }\href
  {https://doi.org/10.1103/PhysRevA.92.033624} {\bibfield  {journal} {\bibinfo
  {journal} {Phys. Rev. A}\ }\textbf {\bibinfo {volume} {92}},\ \bibinfo
  {pages} {033624} (\bibinfo {year} {2015})}\BibitemShut {NoStop}%
\bibitem [{\citenamefont {Yi}\ and\ \citenamefont
  {Pu}(2006)}]{PhysRevLett.97.020401}%
  \BibitemOpen
  \bibfield  {author} {\bibinfo {author} {\bibfnamefont {S.}~\bibnamefont
  {Yi}}\ and\ \bibinfo {author} {\bibfnamefont {H.}~\bibnamefont {Pu}},\
  }\bibfield  {title} {\bibinfo {title} {Spontaneous spin textures in dipolar
  spinor condensates},\ }\href {https://doi.org/10.1103/PhysRevLett.97.020401}
  {\bibfield  {journal} {\bibinfo  {journal} {Phys. Rev. Lett.}\ }\textbf
  {\bibinfo {volume} {97}},\ \bibinfo {pages} {020401} (\bibinfo {year}
  {2006})}\BibitemShut {NoStop}%
\bibitem [{\citenamefont {Vengalattore}\ \emph {et~al.}(2008)\citenamefont
  {Vengalattore}, \citenamefont {Leslie}, \citenamefont {Guzman},\ and\
  \citenamefont {Stamper-Kurn}}]{PhysRevLett.100.170403}%
  \BibitemOpen
  \bibfield  {author} {\bibinfo {author} {\bibfnamefont {M.}~\bibnamefont
  {Vengalattore}}, \bibinfo {author} {\bibfnamefont {S.~R.}\ \bibnamefont
  {Leslie}}, \bibinfo {author} {\bibfnamefont {J.}~\bibnamefont {Guzman}},\
  and\ \bibinfo {author} {\bibfnamefont {D.~M.}\ \bibnamefont {Stamper-Kurn}},\
  }\bibfield  {title} {\bibinfo {title} {Spontaneously modulated spin textures
  in a dipolar spinor bose-einstein condensate},\ }\href
  {https://doi.org/10.1103/PhysRevLett.100.170403} {\bibfield  {journal}
  {\bibinfo  {journal} {Phys. Rev. Lett.}\ }\textbf {\bibinfo {volume} {100}},\
  \bibinfo {pages} {170403} (\bibinfo {year} {2008})}\BibitemShut {NoStop}%
\bibitem [{\citenamefont {Lin}\ \emph {et~al.}(2009)\citenamefont {Lin},
  \citenamefont {Compton}, \citenamefont {Jim{\'e}nez-Garc{\'\i}a},
  \citenamefont {Porto},\ and\ \citenamefont {Spielman}}]{lin2009synthetic}%
  \BibitemOpen
  \bibfield  {author} {\bibinfo {author} {\bibfnamefont {Y.-J.}\ \bibnamefont
  {Lin}}, \bibinfo {author} {\bibfnamefont {R.~L.}\ \bibnamefont {Compton}},
  \bibinfo {author} {\bibfnamefont {K.}~\bibnamefont
  {Jim{\'e}nez-Garc{\'\i}a}}, \bibinfo {author} {\bibfnamefont {J.~V.}\
  \bibnamefont {Porto}},\ and\ \bibinfo {author} {\bibfnamefont {I.~B.}\
  \bibnamefont {Spielman}},\ }\bibfield  {title} {\bibinfo {title} {Synthetic
  magnetic fields for ultracold neutral atoms},\ }\href@noop {} {\bibfield
  {journal} {\bibinfo  {journal} {Nature}\ }\textbf {\bibinfo {volume} {462}},\
  \bibinfo {pages} {628} (\bibinfo {year} {2009})}\BibitemShut {NoStop}%
\bibitem [{SM()}]{SM}%
  \BibitemOpen
  \bibfield  {title} {\bibinfo {title} {See supplemental material for
  derivations of the cavity-molecule hamiltonian, the dipolar interaction for
  polar molecules, and the bogoliubov-de-gennes excitation spectrum},\
  }\href@noop {} {\ }\BibitemShut {NoStop}%
\bibitem [{\citenamefont {Paik}\ \emph {et~al.}(2011)\citenamefont {Paik},
  \citenamefont {Schuster}, \citenamefont {Bishop}, \citenamefont {Kirchmair},
  \citenamefont {Catelani}, \citenamefont {Sears}, \citenamefont {Johnson},
  \citenamefont {Reagor}, \citenamefont {Frunzio}, \citenamefont {Glazman},
  \citenamefont {Girvin}, \citenamefont {Devoret},\ and\ \citenamefont
  {Schoelkopf}}]{PhysRevLett.107.240501}%
  \BibitemOpen
  \bibfield  {author} {\bibinfo {author} {\bibfnamefont {H.}~\bibnamefont
  {Paik}}, \bibinfo {author} {\bibfnamefont {D.~I.}\ \bibnamefont {Schuster}},
  \bibinfo {author} {\bibfnamefont {L.~S.}\ \bibnamefont {Bishop}}, \bibinfo
  {author} {\bibfnamefont {G.}~\bibnamefont {Kirchmair}}, \bibinfo {author}
  {\bibfnamefont {G.}~\bibnamefont {Catelani}}, \bibinfo {author}
  {\bibfnamefont {A.~P.}\ \bibnamefont {Sears}}, \bibinfo {author}
  {\bibfnamefont {B.~R.}\ \bibnamefont {Johnson}}, \bibinfo {author}
  {\bibfnamefont {M.~J.}\ \bibnamefont {Reagor}}, \bibinfo {author}
  {\bibfnamefont {L.}~\bibnamefont {Frunzio}}, \bibinfo {author} {\bibfnamefont
  {L.~I.}\ \bibnamefont {Glazman}}, \bibinfo {author} {\bibfnamefont {S.~M.}\
  \bibnamefont {Girvin}}, \bibinfo {author} {\bibfnamefont {M.~H.}\
  \bibnamefont {Devoret}},\ and\ \bibinfo {author} {\bibfnamefont {R.~J.}\
  \bibnamefont {Schoelkopf}},\ }\bibfield  {title} {\bibinfo {title}
  {Observation of high coherence in josephson junction qubits measured in a
  three-dimensional circuit qed architecture},\ }\href
  {https://doi.org/10.1103/PhysRevLett.107.240501} {\bibfield  {journal}
  {\bibinfo  {journal} {Phys. Rev. Lett.}\ }\textbf {\bibinfo {volume} {107}},\
  \bibinfo {pages} {240501} (\bibinfo {year} {2011})}\BibitemShut {NoStop}%
\bibitem [{\citenamefont {Norcia}\ \emph {et~al.}(2018)\citenamefont {Norcia},
  \citenamefont {Lewis-Swan}, \citenamefont {Cline}, \citenamefont {Zhu},
  \citenamefont {Rey},\ and\ \citenamefont {Thompson}}]{norcia2018cavity}%
  \BibitemOpen
  \bibfield  {author} {\bibinfo {author} {\bibfnamefont {M.~A.}\ \bibnamefont
  {Norcia}}, \bibinfo {author} {\bibfnamefont {R.~J.}\ \bibnamefont
  {Lewis-Swan}}, \bibinfo {author} {\bibfnamefont {J.~R.}\ \bibnamefont
  {Cline}}, \bibinfo {author} {\bibfnamefont {B.}~\bibnamefont {Zhu}}, \bibinfo
  {author} {\bibfnamefont {A.~M.}\ \bibnamefont {Rey}},\ and\ \bibinfo {author}
  {\bibfnamefont {J.~K.}\ \bibnamefont {Thompson}},\ }\bibfield  {title}
  {\bibinfo {title} {Cavity-mediated collective spin-exchange interactions in a
  strontium superradiant laser},\ }\href@noop {} {\bibfield  {journal}
  {\bibinfo  {journal} {Science}\ }\textbf {\bibinfo {volume} {361}},\ \bibinfo
  {pages} {259} (\bibinfo {year} {2018})}\BibitemShut {NoStop}%
\bibitem [{\citenamefont {Deng}\ \emph {et~al.}(2012)\citenamefont {Deng},
  \citenamefont {Cheng}, \citenamefont {Jing}, \citenamefont {Sun},\ and\
  \citenamefont {Yi}}]{PhysRevLett.108.125301}%
  \BibitemOpen
  \bibfield  {author} {\bibinfo {author} {\bibfnamefont {Y.}~\bibnamefont
  {Deng}}, \bibinfo {author} {\bibfnamefont {J.}~\bibnamefont {Cheng}},
  \bibinfo {author} {\bibfnamefont {H.}~\bibnamefont {Jing}}, \bibinfo {author}
  {\bibfnamefont {C.-P.}\ \bibnamefont {Sun}},\ and\ \bibinfo {author}
  {\bibfnamefont {S.}~\bibnamefont {Yi}},\ }\bibfield  {title} {\bibinfo
  {title} {Spin-orbit-coupled dipolar bose-einstein condensates},\ }\href
  {https://doi.org/10.1103/PhysRevLett.108.125301} {\bibfield  {journal}
  {\bibinfo  {journal} {Phys. Rev. Lett.}\ }\textbf {\bibinfo {volume} {108}},\
  \bibinfo {pages} {125301} (\bibinfo {year} {2012})}\BibitemShut {NoStop}%
\bibitem [{\citenamefont {B\"ottcher}\ \emph {et~al.}(2019)\citenamefont
  {B\"ottcher}, \citenamefont {Schmidt}, \citenamefont {Wenzel}, \citenamefont
  {Hertkorn}, \citenamefont {Guo}, \citenamefont {Langen},\ and\ \citenamefont
  {Pfau}}]{PhysRevX.9.011051}%
  \BibitemOpen
  \bibfield  {author} {\bibinfo {author} {\bibfnamefont {F.}~\bibnamefont
  {B\"ottcher}}, \bibinfo {author} {\bibfnamefont {J.-N.}\ \bibnamefont
  {Schmidt}}, \bibinfo {author} {\bibfnamefont {M.}~\bibnamefont {Wenzel}},
  \bibinfo {author} {\bibfnamefont {J.}~\bibnamefont {Hertkorn}}, \bibinfo
  {author} {\bibfnamefont {M.}~\bibnamefont {Guo}}, \bibinfo {author}
  {\bibfnamefont {T.}~\bibnamefont {Langen}},\ and\ \bibinfo {author}
  {\bibfnamefont {T.}~\bibnamefont {Pfau}},\ }\bibfield  {title} {\bibinfo
  {title} {Transient supersolid properties in an array of dipolar quantum
  droplets},\ }\href {https://doi.org/10.1103/PhysRevX.9.011051} {\bibfield
  {journal} {\bibinfo  {journal} {Phys. Rev. X}\ }\textbf {\bibinfo {volume}
  {9}},\ \bibinfo {pages} {011051} (\bibinfo {year} {2019})}\BibitemShut
  {NoStop}%
\bibitem [{\citenamefont {Lima}\ and\ \citenamefont
  {Pelster}(2011)}]{PhysRevA.84.041604}%
  \BibitemOpen
  \bibfield  {author} {\bibinfo {author} {\bibfnamefont {A.~R.~P.}\
  \bibnamefont {Lima}}\ and\ \bibinfo {author} {\bibfnamefont {A.}~\bibnamefont
  {Pelster}},\ }\bibfield  {title} {\bibinfo {title} {Quantum fluctuations in
  dipolar bose gases},\ }\href {https://doi.org/10.1103/PhysRevA.84.041604}
  {\bibfield  {journal} {\bibinfo  {journal} {Phys. Rev. A}\ }\textbf {\bibinfo
  {volume} {84}},\ \bibinfo {pages} {041604} (\bibinfo {year}
  {2011})}\BibitemShut {NoStop}%
\bibitem [{\citenamefont {Wang}\ \emph {et~al.}(2020)\citenamefont {Wang},
  \citenamefont {Guo}, \citenamefont {Yi},\ and\ \citenamefont
  {Shi}}]{PhysRevResearch.2.043074}%
  \BibitemOpen
  \bibfield  {author} {\bibinfo {author} {\bibfnamefont {Y.}~\bibnamefont
  {Wang}}, \bibinfo {author} {\bibfnamefont {L.}~\bibnamefont {Guo}}, \bibinfo
  {author} {\bibfnamefont {S.}~\bibnamefont {Yi}},\ and\ \bibinfo {author}
  {\bibfnamefont {T.}~\bibnamefont {Shi}},\ }\bibfield  {title} {\bibinfo
  {title} {Theory for self-bound states of dipolar bose-einstein condensates},\
  }\href {https://doi.org/10.1103/PhysRevResearch.2.043074} {\bibfield
  {journal} {\bibinfo  {journal} {Phys. Rev. Res.}\ }\textbf {\bibinfo {volume}
  {2}},\ \bibinfo {pages} {043074} (\bibinfo {year} {2020})}\BibitemShut
  {NoStop}%
\bibitem [{\citenamefont {Deng}\ \emph {et~al.}(2023)\citenamefont {Deng},
  \citenamefont {Chen}, \citenamefont {Luo}, \citenamefont {Zhang},
  \citenamefont {Yi},\ and\ \citenamefont {Shi}}]{PhysRevLett.130.183001}%
  \BibitemOpen
  \bibfield  {author} {\bibinfo {author} {\bibfnamefont {F.}~\bibnamefont
  {Deng}}, \bibinfo {author} {\bibfnamefont {X.-Y.}\ \bibnamefont {Chen}},
  \bibinfo {author} {\bibfnamefont {X.-Y.}\ \bibnamefont {Luo}}, \bibinfo
  {author} {\bibfnamefont {W.}~\bibnamefont {Zhang}}, \bibinfo {author}
  {\bibfnamefont {S.}~\bibnamefont {Yi}},\ and\ \bibinfo {author}
  {\bibfnamefont {T.}~\bibnamefont {Shi}},\ }\bibfield  {title} {\bibinfo
  {title} {Effective potential and superfluidity of microwave-shielded polar
  molecules},\ }\href {https://doi.org/10.1103/PhysRevLett.130.183001}
  {\bibfield  {journal} {\bibinfo  {journal} {Phys. Rev. Lett.}\ }\textbf
  {\bibinfo {volume} {130}},\ \bibinfo {pages} {183001} (\bibinfo {year}
  {2023})}\BibitemShut {NoStop}%
\bibitem [{\citenamefont {Jin}\ \emph {et~al.}(2024)\citenamefont {Jin},
  \citenamefont {Deng}, \citenamefont {Yi},\ and\ \citenamefont
  {Shi}}]{jin2024bose}%
  \BibitemOpen
  \bibfield  {author} {\bibinfo {author} {\bibfnamefont {W.-J.}\ \bibnamefont
  {Jin}}, \bibinfo {author} {\bibfnamefont {F.}~\bibnamefont {Deng}}, \bibinfo
  {author} {\bibfnamefont {S.}~\bibnamefont {Yi}},\ and\ \bibinfo {author}
  {\bibfnamefont {T.}~\bibnamefont {Shi}},\ }\bibfield  {title} {\bibinfo
  {title} {Bose-einstein condensates of microwave-shielded polar molecules},\
  }\href@noop {} {\bibfield  {journal} {\bibinfo  {journal} {arXiv preprint
  arXiv:2406.06412}\ } (\bibinfo {year} {2024})}\BibitemShut {NoStop}%
\bibitem [{\citenamefont {Abo-Shaeer}\ \emph {et~al.}(2001)\citenamefont
  {Abo-Shaeer}, \citenamefont {Raman}, \citenamefont {Vogels},\ and\
  \citenamefont {Ketterle}}]{Ketterle2001}%
  \BibitemOpen
  \bibfield  {author} {\bibinfo {author} {\bibfnamefont {J.~R.}\ \bibnamefont
  {Abo-Shaeer}}, \bibinfo {author} {\bibfnamefont {C.}~\bibnamefont {Raman}},
  \bibinfo {author} {\bibfnamefont {J.~M.}\ \bibnamefont {Vogels}},\ and\
  \bibinfo {author} {\bibfnamefont {W.}~\bibnamefont {Ketterle}},\ }\bibfield
  {title} {\bibinfo {title} {Observation of vortex lattices in bose-einstein
  condensates},\ }\href {https://doi.org/10.1126/science.1060182} {\bibfield
  {journal} {\bibinfo  {journal} {Science}\ }\textbf {\bibinfo {volume}
  {292}},\ \bibinfo {pages} {476} (\bibinfo {year} {2001})}\BibitemShut
  {NoStop}%
\bibitem [{\citenamefont {Engels}\ \emph {et~al.}(2003)\citenamefont {Engels},
  \citenamefont {Coddington}, \citenamefont {Haljan}, \citenamefont
  {Schweikhard},\ and\ \citenamefont {Cornell}}]{Cornell2003}%
  \BibitemOpen
  \bibfield  {author} {\bibinfo {author} {\bibfnamefont {P.}~\bibnamefont
  {Engels}}, \bibinfo {author} {\bibfnamefont {I.}~\bibnamefont {Coddington}},
  \bibinfo {author} {\bibfnamefont {P.~C.}\ \bibnamefont {Haljan}}, \bibinfo
  {author} {\bibfnamefont {V.}~\bibnamefont {Schweikhard}},\ and\ \bibinfo
  {author} {\bibfnamefont {E.~A.}\ \bibnamefont {Cornell}},\ }\bibfield
  {title} {\bibinfo {title} {Observation of long-lived vortex aggregates in
  rapidly rotating bose-einstein condensates},\ }\href
  {https://doi.org/10.1103/PhysRevLett.90.170405} {\bibfield  {journal}
  {\bibinfo  {journal} {Phys. Rev. Lett.}\ }\textbf {\bibinfo {volume} {90}},\
  \bibinfo {pages} {170405} (\bibinfo {year} {2003})}\BibitemShut {NoStop}%
\bibitem [{\citenamefont {Schweikhard}\ \emph {et~al.}(2004)\citenamefont
  {Schweikhard}, \citenamefont {Coddington}, \citenamefont {Engels},
  \citenamefont {Mogendorff},\ and\ \citenamefont {Cornell}}]{Cornell2004}%
  \BibitemOpen
  \bibfield  {author} {\bibinfo {author} {\bibfnamefont {V.}~\bibnamefont
  {Schweikhard}}, \bibinfo {author} {\bibfnamefont {I.}~\bibnamefont
  {Coddington}}, \bibinfo {author} {\bibfnamefont {P.}~\bibnamefont {Engels}},
  \bibinfo {author} {\bibfnamefont {V.~P.}\ \bibnamefont {Mogendorff}},\ and\
  \bibinfo {author} {\bibfnamefont {E.~A.}\ \bibnamefont {Cornell}},\
  }\bibfield  {title} {\bibinfo {title} {Rapidly rotating bose-einstein
  condensates in and near the lowest landau level},\ }\href
  {https://doi.org/10.1103/PhysRevLett.92.040404} {\bibfield  {journal}
  {\bibinfo  {journal} {Phys. Rev. Lett.}\ }\textbf {\bibinfo {volume} {92}},\
  \bibinfo {pages} {040404} (\bibinfo {year} {2004})}\BibitemShut {NoStop}%
\bibitem [{\citenamefont {Fleischhauer}\ and\ \citenamefont
  {Lukin}(2000)}]{PhysRevLett.84.5094}%
  \BibitemOpen
  \bibfield  {author} {\bibinfo {author} {\bibfnamefont {M.}~\bibnamefont
  {Fleischhauer}}\ and\ \bibinfo {author} {\bibfnamefont {M.~D.}\ \bibnamefont
  {Lukin}},\ }\bibfield  {title} {\bibinfo {title} {Dark-state polaritons in
  electromagnetically induced transparency},\ }\href
  {https://doi.org/10.1103/PhysRevLett.84.5094} {\bibfield  {journal} {\bibinfo
   {journal} {Phys. Rev. Lett.}\ }\textbf {\bibinfo {volume} {84}},\ \bibinfo
  {pages} {5094} (\bibinfo {year} {2000})}\BibitemShut {NoStop}%
\bibitem [{\citenamefont {Micheli}\ \emph {et~al.}(2006)\citenamefont
  {Micheli}, \citenamefont {Brennen},\ and\ \citenamefont
  {Zoller}}]{micheli2006toolbox}%
  \BibitemOpen
  \bibfield  {author} {\bibinfo {author} {\bibfnamefont {A.}~\bibnamefont
  {Micheli}}, \bibinfo {author} {\bibfnamefont {G.~K.}\ \bibnamefont
  {Brennen}},\ and\ \bibinfo {author} {\bibfnamefont {P.}~\bibnamefont
  {Zoller}},\ }\bibfield  {title} {\bibinfo {title} {A toolbox for lattice-spin
  models with polar molecules},\ }\href@noop {} {\bibfield  {journal} {\bibinfo
   {journal} {Nature Physics}\ }\textbf {\bibinfo {volume} {2}},\ \bibinfo
  {pages} {341} (\bibinfo {year} {2006})}\BibitemShut {NoStop}%
\bibitem [{\citenamefont {Gregory}\ \emph {et~al.}(2021)\citenamefont
  {Gregory}, \citenamefont {Blackmore}, \citenamefont {Bromley}, \citenamefont
  {Hutson},\ and\ \citenamefont {Cornish}}]{gregory2021robust}%
  \BibitemOpen
  \bibfield  {author} {\bibinfo {author} {\bibfnamefont {P.~D.}\ \bibnamefont
  {Gregory}}, \bibinfo {author} {\bibfnamefont {J.~A.}\ \bibnamefont
  {Blackmore}}, \bibinfo {author} {\bibfnamefont {S.~L.}\ \bibnamefont
  {Bromley}}, \bibinfo {author} {\bibfnamefont {J.~M.}\ \bibnamefont
  {Hutson}},\ and\ \bibinfo {author} {\bibfnamefont {S.~L.}\ \bibnamefont
  {Cornish}},\ }\bibfield  {title} {\bibinfo {title} {Robust storage qubits in
  ultracold polar molecules},\ }\href@noop {} {\bibfield  {journal} {\bibinfo
  {journal} {Nature Physics}\ }\textbf {\bibinfo {volume} {17}},\ \bibinfo
  {pages} {1149} (\bibinfo {year} {2021})}\BibitemShut {NoStop}%
\bibitem [{\citenamefont {Capogrosso-Sansone}\ \emph
  {et~al.}(2010)\citenamefont {Capogrosso-Sansone}, \citenamefont {Trefzger},
  \citenamefont {Lewenstein}, \citenamefont {Zoller},\ and\ \citenamefont
  {Pupillo}}]{PhysRevLett.104.125301}%
  \BibitemOpen
  \bibfield  {author} {\bibinfo {author} {\bibfnamefont {B.}~\bibnamefont
  {Capogrosso-Sansone}}, \bibinfo {author} {\bibfnamefont {C.}~\bibnamefont
  {Trefzger}}, \bibinfo {author} {\bibfnamefont {M.}~\bibnamefont
  {Lewenstein}}, \bibinfo {author} {\bibfnamefont {P.}~\bibnamefont {Zoller}},\
  and\ \bibinfo {author} {\bibfnamefont {G.}~\bibnamefont {Pupillo}},\
  }\bibfield  {title} {\bibinfo {title} {Quantum phases of cold polar molecules
  in 2d optical lattices},\ }\href
  {https://doi.org/10.1103/PhysRevLett.104.125301} {\bibfield  {journal}
  {\bibinfo  {journal} {Phys. Rev. Lett.}\ }\textbf {\bibinfo {volume} {104}},\
  \bibinfo {pages} {125301} (\bibinfo {year} {2010})}\BibitemShut {NoStop}%
\bibitem [{\citenamefont {Bao}\ \emph {et~al.}(2023)\citenamefont {Bao},
  \citenamefont {Yu}, \citenamefont {Anderegg}, \citenamefont {Chae},
  \citenamefont {Ketterle}, \citenamefont {Ni},\ and\ \citenamefont
  {Doyle}}]{doi:10.1126/science.adf8999}%
  \BibitemOpen
  \bibfield  {author} {\bibinfo {author} {\bibfnamefont {Y.-C.}\ \bibnamefont
  {Bao}}, \bibinfo {author} {\bibfnamefont {S.~S.}\ \bibnamefont {Yu}},
  \bibinfo {author} {\bibfnamefont {L.}~\bibnamefont {Anderegg}}, \bibinfo
  {author} {\bibfnamefont {E.}~\bibnamefont {Chae}}, \bibinfo {author}
  {\bibfnamefont {W.}~\bibnamefont {Ketterle}}, \bibinfo {author}
  {\bibfnamefont {K.-K.}\ \bibnamefont {Ni}},\ and\ \bibinfo {author}
  {\bibfnamefont {J.~M.}\ \bibnamefont {Doyle}},\ }\bibfield  {title} {\bibinfo
  {title} {Dipolar spin-exchange and entanglement between molecules in an
  optical tweezer array},\ }\href {https://doi.org/10.1126/science.adf8999}
  {\bibfield  {journal} {\bibinfo  {journal} {Science}\ }\textbf {\bibinfo
  {volume} {382}},\ \bibinfo {pages} {1138} (\bibinfo {year}
  {2023})}\BibitemShut {NoStop}%
\bibitem [{\citenamefont {Holland}\ \emph {et~al.}(2023)\citenamefont
  {Holland}, \citenamefont {Lu},\ and\ \citenamefont
  {Cheuk}}]{doi:10.1126/science.adf4272}%
  \BibitemOpen
  \bibfield  {author} {\bibinfo {author} {\bibfnamefont {C.~M.}\ \bibnamefont
  {Holland}}, \bibinfo {author} {\bibfnamefont {Y.-K.}\ \bibnamefont {Lu}},\
  and\ \bibinfo {author} {\bibfnamefont {L.~W.}\ \bibnamefont {Cheuk}},\
  }\bibfield  {title} {\bibinfo {title} {On-demand entanglement of molecules in
  a reconfigurable optical tweezer array},\ }\href
  {https://doi.org/10.1126/science.adf4272} {\bibfield  {journal} {\bibinfo
  {journal} {Science}\ }\textbf {\bibinfo {volume} {382}},\ \bibinfo {pages}
  {1143} (\bibinfo {year} {2023})}\BibitemShut {NoStop}%
\bibitem [{\citenamefont {You}\ \emph {et~al.}(2025)\citenamefont {You},
  \citenamefont {Yi},\ and\ \citenamefont {Deng}}]{You25}%
  \BibitemOpen
  \bibfield  {author} {\bibinfo {author} {\bibfnamefont {J.}~\bibnamefont
  {You}}, \bibinfo {author} {\bibfnamefont {S.}~\bibnamefont {Yi}},\ and\
  \bibinfo {author} {\bibfnamefont {Y.}~\bibnamefont {Deng}},\ }\bibfield
  {title} {\bibinfo {title} {Spin-momentum-mixing interactions with
  cavity-mediated supersolid in spinor condensates},\ }\href@noop {} {\bibfield
   {journal} {\bibinfo  {journal} {Photonics Research}\ }\textbf {\bibinfo
  {volume} {13}},\ \bibinfo {pages} {987} (\bibinfo {year} {2025})}\BibitemShut
  {NoStop}%
\bibitem [{\citenamefont {Luo}\ \emph {et~al.}(2024)\citenamefont {Luo},
  \citenamefont {Zhang}, \citenamefont {Koh}, \citenamefont {Wilson},
  \citenamefont {Chu}, \citenamefont {Holland}, \citenamefont {Rey},\ and\
  \citenamefont {Thompson}}]{luo2024momentum}%
  \BibitemOpen
  \bibfield  {author} {\bibinfo {author} {\bibfnamefont {C.}~\bibnamefont
  {Luo}}, \bibinfo {author} {\bibfnamefont {H.}~\bibnamefont {Zhang}}, \bibinfo
  {author} {\bibfnamefont {V.~P.}\ \bibnamefont {Koh}}, \bibinfo {author}
  {\bibfnamefont {J.~D.}\ \bibnamefont {Wilson}}, \bibinfo {author}
  {\bibfnamefont {A.}~\bibnamefont {Chu}}, \bibinfo {author} {\bibfnamefont
  {M.~J.}\ \bibnamefont {Holland}}, \bibinfo {author} {\bibfnamefont {A.~M.}\
  \bibnamefont {Rey}},\ and\ \bibinfo {author} {\bibfnamefont {J.~K.}\
  \bibnamefont {Thompson}},\ }\bibfield  {title} {\bibinfo {title}
  {Momentum-exchange interactions in a bragg atom interferometer suppress
  doppler dephasing},\ }\href@noop {} {\bibfield  {journal} {\bibinfo
  {journal} {Science}\ }\textbf {\bibinfo {volume} {384}},\ \bibinfo {pages}
  {551} (\bibinfo {year} {2024})}\BibitemShut {NoStop}%
\bibitem [{\citenamefont {Aldegunde}\ \emph {et~al.}(2008)\citenamefont
  {Aldegunde}, \citenamefont {Rivington}, \citenamefont {\ifmmode~\dot{Z}\else
  \.{Z}\fi{}uchowski},\ and\ \citenamefont {Hutson}}]{Huston2008}%
  \BibitemOpen
  \bibfield  {author} {\bibinfo {author} {\bibfnamefont {J.}~\bibnamefont
  {Aldegunde}}, \bibinfo {author} {\bibfnamefont {B.~A.}\ \bibnamefont
  {Rivington}}, \bibinfo {author} {\bibfnamefont {P.~S.}\ \bibnamefont
  {\ifmmode~\dot{Z}\else \.{Z}\fi{}uchowski}},\ and\ \bibinfo {author}
  {\bibfnamefont {J.~M.}\ \bibnamefont {Hutson}},\ }\bibfield  {title}
  {\bibinfo {title} {Hyperfine energy levels of alkali-metal dimers:
  Ground-state polar molecules in electric and magnetic fields},\ }\href
  {https://doi.org/10.1103/PhysRevA.78.033434} {\bibfield  {journal} {\bibinfo
  {journal} {Phys. Rev. A}\ }\textbf {\bibinfo {volume} {78}},\ \bibinfo
  {pages} {033434} (\bibinfo {year} {2008})}\BibitemShut {NoStop}%
\bibitem [{\citenamefont {Ran}\ \emph {et~al.}(2010)\citenamefont {Ran},
  \citenamefont {Aldegunde},\ and\ \citenamefont {Hutson}}]{ran2010hyperfine}%
  \BibitemOpen
  \bibfield  {author} {\bibinfo {author} {\bibfnamefont {H.}~\bibnamefont
  {Ran}}, \bibinfo {author} {\bibfnamefont {J.}~\bibnamefont {Aldegunde}},\
  and\ \bibinfo {author} {\bibfnamefont {J.~M.}\ \bibnamefont {Hutson}},\
  }\bibfield  {title} {\bibinfo {title} {Hyperfine structure in the microwave
  spectra of ultracold polar molecules},\ }\href@noop {} {\bibfield  {journal}
  {\bibinfo  {journal} {New Journal of Physics}\ }\textbf {\bibinfo {volume}
  {12}},\ \bibinfo {pages} {043015} (\bibinfo {year} {2010})}\BibitemShut
  {NoStop}%
\bibitem [{\citenamefont {Aldegunde}\ and\ \citenamefont
  {Hutson}(2017)}]{PhysRevA.96.042506}%
  \BibitemOpen
  \bibfield  {author} {\bibinfo {author} {\bibfnamefont {J.}~\bibnamefont
  {Aldegunde}}\ and\ \bibinfo {author} {\bibfnamefont {J.~M.}\ \bibnamefont
  {Hutson}},\ }\bibfield  {title} {\bibinfo {title} {Hyperfine structure of
  alkali-metal diatomic molecules},\ }\href
  {https://doi.org/10.1103/PhysRevA.96.042506} {\bibfield  {journal} {\bibinfo
  {journal} {Phys. Rev. A}\ }\textbf {\bibinfo {volume} {96}},\ \bibinfo
  {pages} {042506} (\bibinfo {year} {2017})}\BibitemShut {NoStop}%
\bibitem [{\citenamefont {Fischer}(2006)}]{Fischer2006}%
  \BibitemOpen
  \bibfield  {author} {\bibinfo {author} {\bibfnamefont {U.~R.}\ \bibnamefont
  {Fischer}},\ }\bibfield  {title} {\bibinfo {title} {Stability of
  quasi-two-dimensional bose-einstein condensates with dominant dipole-dipole
  interactions},\ }\href {https://doi.org/10.1103/PhysRevA.73.031602}
  {\bibfield  {journal} {\bibinfo  {journal} {Phys. Rev. A}\ }\textbf {\bibinfo
  {volume} {73}},\ \bibinfo {pages} {031602} (\bibinfo {year}
  {2006})}\BibitemShut {NoStop}%
\end{thebibliography}

\end{document}